\documentclass[fontsize=11pt]{arxiv-article}
\bibliography{References,not-arxiv}

\DeclareMathSizes{10.95}{10}{7}{5}

\usepackage{comment}

\usepackage{Definitions}

\usepackage{hyperref}
\usepackage{local}

\allowdisplaybreaks[2]

\newcommand{\OfficialTitle}{
  Spinning correlators in large-charge CFTs
}
\title{\setstretch{1.4}
  {\color{Thoughtless}\OfficialTitle}
}

\hypersetup{pdfauthor={Nicola Dondi, Ioannis Kalogerakis, Rafael Moser, Domenico Orlando, Susanne Reffert},pdftitle={\OfficialTitle},%
  colorlinks=true,linkcolor=ThoughtYouWere,citecolor=ThoughtYouWere,urlcolor=ThoughtYouWere,ocgcolorlinks}

\author{%
  \begin{minipage}{.94\textwidth}
    \vspace{1cm}
    \begin{center} \dosserif%
      {\small
      	\textbf{Nicola~Dondi}\textsuperscript{\ding{73}}, 
      	\textbf{Ioannis~Kalogerakis}\textsuperscript{\ding{73}},
      	\textbf{Rafael~Moser}\textsuperscript{\ding{73}},\\
        \textbf{Domenico~Orlando}\textsuperscript{\ding{72}\ding{73}} and
        \textbf{Susanne~Reffert}\textsuperscript{\ding{73}} 
         }
    \end{center}
    \authorBlock{\ding{73}}{\dosserif{}%
      Albert Einstein Center for Fundamental Physics\\
      Institute for Theoretical Physics, University of Bern,\\
      Sidlerstrasse 5, CH-3012 Bern, Switzerland}
    \authorBlock{\ding{72}}{\dosserif{}%
      INFN sezione di Torino.\\
      via Pietro Giuria 1, 10125 Torino, Italy}
  \end{minipage}
}

\date{}

\numberwithin{equation}{section}

\begin{document}

\begin{titlepage}

  \maketitle

  \thispagestyle{empty}

  \vfill\dosserif{}

  \abstract{\normalfont{}\noindent{}%
    We systematically study correlators of a generic conformal field theory with a global $O(2)$ symmetry in a sector of large global charge.
    We focus in particular on three- and four-point correlators with conserved current insertions sandwiched between spinful excited states corresponding to phonons over the large-charge vacuum.
    We also discuss loop corrections to the scaling dimensions and observe the presence of multiple logarithms in even dimensions. 
  }

\end{titlepage}

\setstretch{1.1}
\setcounter{tocdepth}{\subsectiontocdepth}
\tableofcontents

\title{Spinning correlators in large-charge CFTs}

\section{Introduction}

Working in sectors of large charge results in important simplifications~\cite{Gaume:2020bmp,Hellerman:2015nra,Alvarez-Gaume:2016vff,Alvarez-Gaume:2019biu,Orlando:2021usz,Moser:2021bes}. In the case of strongly coupled \acp{cft}, it allows in particular the calculation of the \emph{conformal data}, \emph{i.e.} scaling dimensions and three-point function coefficients, which encode all the relevant information for solving the theory. 

In the existing literature, the main stress was put on the calculation of conformal dimensions of large-charge operators~\cite{Badel:2019oxl,Giombi:2020enj,Antipin:2020abu,Antipin:2020rdw,Jack:2021lja,Jack:2021ziq}, which were independently verified via lattice calculations~\cite{Banerjee:2017fcx,Banerjee:2019jpw,Banerjee:2021bbw,Singh:2022akp}.
A number of (mostly three- and four-point) correlators have appeared scattered in the literature~\cite{Monin:2016jmo,Jafferis:2017zna,Arias-Tamargo:2019kfr,Cuomo:2020rgt,Cuomo:2020thesis,Komargodski:2021qp}.
In this note, we aim to close a gap in the literature by systematically collecting three- and four-point correlators involving currents in the large charge sector of a generic \ac{cft} in $d$ dimensions with a global $O(2)$ symmetry.
We not only organize and express existing results in a self-contained, consistent way and unified language, but go beyond the state of the art by giving current correlators inserted not only between the scalar ground state at large charge but between states with phonons, \emph{i.e.} spinning states, representing excitations of the large-charge ground state. While we do not directly use conformal symmetry to obtain our results, they are consistent with the expected form of conformal correlators. 

Technically, we make use of the fact that the \ac{eft} at large charge is to leading order a free theory, which allows us to perform explicit computations also for a strongly-coupled theory.
The leading term in the \ac{eft} is fixed by scale invariance and receives corrections subleading in $Q$ from curvature terms. 
The quantum fluctuations arising from the leading term in the \ac{eft} are of order $Q^0$. 
We write the correlators as a sum of semiclassical terms with non-negative \(Q\) scaling plus a quantum \(Q^0\) correction, neglecting terms suppressed by negative powers of $Q$. 
Since in general the ground state is homogeneous, any non-trivial position dependence in our correlators must be due to the quantum fluctuations.
In odd dimensions, the tree-level expressions do not have any $Q^0$ contribution, so all terms at this order are due to quantum fluctuations and are universal. In even dimensions, the universal $Q^0$ term is replaced by a $Q^0 \log Q$ term~\cite{Cuomo:2020rgt}.
Studying the structure of the higher-order loop corrections we find in even $D$ logarithmic $l$-loop contributions of the form
\begin{equation}
    \Delta_l \supset \frac{1}{Q^{(l-1)D/(D-1)}}  \left( \alpha_0 + \alpha_1 \log Q + ... + \alpha_l (\log Q)^l \right).
\end{equation}
This result is especially relevant for applications in the context of resurgent asymptotics as in odd dimensions, large-$Q$ expansions are expected to be log-free transseries with non-perturbative corrections related to worldline instantons~\cite{Dondi:2021buw,Antipin:2022dsm}. 

\medskip

In this note, we focus on the non-supersymmetric case with a single large-charge vacuum. Correlators of supersymmetric theories with a moduli space must be approached differently and have appeared in~\cite{Hellerman:2017veg,Hellerman:2017sur,Bourget:2018obm,Hellerman:2018xpi,Hellerman:2020sqj,Beccaria:2018xxl,Hellerman:2021yqz,Hellerman:2021duh,Giombi:2021zfb,Giombi:2022anm}.

\bigskip
The plan of this paper is as follows. In \autoref{sec:O(2)_review}, we review the basics of the $O(2)$ sector at large charge, focusing on canonical quantization. In \autoref{sec:pathIntergral}, we instead turn to path-integral methods, computing in \autoref{sec:twoPointFn} the basic two-point functions of the ground state $\braket{Q|Q}$ and the one-phonon state $\braket{\myatop{Q}{\ell_2 m_2} | \myatop{Q}{\ell_1 m_1} } $.
In \autoref{sec:loopCorrections}, we discuss loop corrections, in particular the one- and two-loop corrections to the scaling dimensions of the primary operator corresponding to the large-charge ground state.
In \autoref{sec:ConformalAlgebraAndChargeCorrelators}, we present the main results of this paper, namely correlators of the conserved currents $J$ of the $O(2)$ symmetry and $T$ inserted between one-phonon states.
In \autoref{sec:HLH} we give an overview of correlators in which a small-charge state is inserted between large-charge states. These results have however already appeared in the literature.
We close in \autoref{sec:Conclusions} with concluding remarks and an outlook.
In \autoref{sec:Ylm-identities}, we collect properties and identities of the hyperspherical harmonics used throughout this paper.
In \autoref{sec:constraints}, we give the form of conformal correlators due to the constraints of conformal invariance.
The Casimir energy of the fluctuations is computed in various dimensions in \autoref{sec:Casimir_appendix}. In \autoref{sec:Loop_appendix}, we give details of the loop calculations.
Finally, in \autoref{sec:methods}, we give some calculational details for the correlators appearing in \autoref{sec:ConformalAlgebraAndChargeCorrelators}.

\section{The O(2) sector at large charge}\label{sec:O(2)_review}

In this section we collect some review material about the large-charge expansion for a generic \ac{cft} with a global $O(2)$ symmetry. 

\subsection{Effective field theory}
\label{sec:effect-field-theory}

We consider a \ac{cft} in $D$-dimensional flat space with an $O(2)$ internal symmetry which can generically be a subgroup of a larger global symmetry. In particular, consider the state $\ket{Q}$ generated by the scalar primary $\Opp$
with $O(2)$ charge $Q$. Since flat space is conformally equivalent to the cylinder $\setR \times S^{D-1}$ we will work for future convenience in the cylinder frame. We are interested in correlators of such
primaries at long distances, which can be expressed on the cylinder as
\begin{equation}
  \braket{ Q, \infty | Q , - \infty} = \lim_{\beta \longrightarrow \infty} \braket {Q | e^{-\beta  {H}_{\text{cyl}} } |Q}.
  \label{eq:corr}
\end{equation}
There is strong indication~\cite{Hellerman:2015nra,Monin:2016jmo,Gaume:2020bmp} that as $Q$
becomes very large, this correlator on the cylinder has a description
in terms of a weakly coupled \ac{eft} based  on the coset model
\begin{align}%
  \label{eq:coset}
  \frac{SO(D+1,1) \times U(1)_Q}{SO(D) \times U(1)_{D+\mu Q}} \, ,  && \text{valid for energy scales} \quad \quad \frac{1}{R} \ll E \ll \mu \sim \frac{Q^{1/(D-1)}}{R},
\end{align}
where $R$ is the cylinder radius. The parameter $\mu(Q)$ can be interpreted as the chemical potential dual to the quantum number $Q$ which is the fixed control parameter. The symmetry-breaking pattern of
this coset model is known as the \emph{conformal superfluid phase}.%
\footnote{The state $\ket{Q} $ is not a superfluid state, as it is an
  eigenstate of the charge operator. More precisely one assumes that the
  two-point function in Eq.~\eqref{eq:corr} is dominated by a saddle
  corresponding to a superfluid state.} The corresponding \ac{eft}
in Euclidean spacetime%
\footnote{Our convention for Euclidean space is $\tau=i t$, so that
  $i \del_\tau{} = \del_t{} $.} has been computed in terms of a Goldstone
field
$\chi = -i\mu \tau + \pi(\tau,\n)$~\cite{Hellerman:2015nra,Monin:2016jmo}, where $\pi(\tau,\n)$ are the fluctuations over the fixed-charge ground state $\chi^{\saddle} = -i\mu \tau$. The result is
\begin{equation}
  S= - c_1 \int_{\setR \times S^{D-1}} \dd{\tau}\dd{S}  \left( - \del_\mu \chi \del^\mu \chi \right)^{D/2} + \text{curvature couplings},
  \label{eq:action}
\end{equation}
where $c_1$ an unknown Wilsonian coefficient which depends on the \ac{uv} theory (\emph{i.e.} the starting $\ac{cft}_{D}$) and $\dd{S} = R^{D-1} \dd{\Omega}$. This is to be
interpreted as an action for the fluctuation $\pi(\tau,\n)$ with cutoff $\Lambda\sim \mu$, so that a hierarchy is generated, and it is controlled by the dimensionless ratio $(R\mu) \gg 1$. Every observable in the \ac{eft} is expressed as an expansion in inverse powers of $\mu$. In particular, the ground-state action takes the form
\begin{equation}\label{eq:S_expansion}
 S^{\saddle} = \left( \frac{\tau_2-\tau_1}{R} \right) \sum_{r=0}^{\infty} \alpha_r (R\mu)^{D-2r},
\end{equation}
where the coefficients $\alpha_r$ depend on $c_1$ and all other Wilsonian coefficients associated to curvature terms. 
Other than the scaling behavior in the quantum number $Q$, there is a number of universal predictions that do not depend on the Wilsonian parameters of the \ac{eft}~\cite{Alvarez-Gaume:2016vff,delaFuente:2018qwv,Cuomo:2020rgt}. 

We will now review the classical and quantum treatment of the action~\eqref{eq:action}, from which we will be able to compute some important $\ac{cft}$ correlators and corrections to the scaling dimension of the primary $\Opp$.

\subsection{Classical treatment}
\label{sec:classical_treatment}
Neglecting curvature couplings and expanding to quadratic order in $\pi(\tau,\n)$, the \ac{eft} Lagrangian reads
\begin{equation}\label{eq:Quadratic_Lagrangian}
\mathscr{L} = - c_1 \mu^D - i c_1 \mu^{D-1} D \dot{\pi} + c_1 \mu^{D-2} \frac{D(D-1)}{2} \left( \dot\pi^2 + \frac{1}{D-1} (\del_i \pi)^2\right) +\order{\mu^{D-3}}.
\end{equation}
The conjugate momentum to \(\pi\) is defined in the usual manner from the quadratic Lagrangian
\begin{equation}
    \Pi = \eval*{ i \fdv{\mathscr{L}}{\dot{\pi}} }_{\text{lin}} =  c_1 D \mu^{D-1} + i c_1 D(D-1) \mu^{D-2} \dot{\pi}.
\end{equation}
At leading order, this gives rise to the usual canonical Poisson brackets. We will for now neglect interactions and study the spectrum of the quadratic Lagrangian.

The fields $\pi$ and $\Pi$ can be decomposed into a complete set of solutions of the \ac{eom}~\cite{Alvarez-Gaume:2016vff}: 
\begin{align}\label{eq:FieldDecompositionInLadders}
  &\begin{multlined}[][.9\linewidth]
    \pi(\tau, \n) = \pi_0 -  \frac{i \Pi_0 \tau}{c_1 \Omega_D R^{D-1}  D(D-1) \mu^{D-2} } \\
    + \frac{1 }{\sqrt{c_1 R^{D-1} D(D-1) \mu^{D-2} }} \sum_{\ell \geq 1 , m}\left( \frac{a_{\ell m}}{\sqrt{2\omega_{\ell}}} e^{-\omega_{\ell} \tau} Y_{\ell m} (\n)  +  \frac{ a^*_{\ell m} }{ \sqrt{2\omega_{\ell}} } e^{\omega_{\ell} \tau} Y_{\ell m}^* (\n) \right) ,
  \end{multlined}
  \\
  &\begin{multlined}[][.9\linewidth]
    \Pi(\tau,\n) = c_1 D \mu^{D-1} + \frac{\Pi_0}{\Omega_D R^{D-1}} \\
    + i \sqrt{\frac{c_1 D(D-1)\mu^{D-2}}{R^{D-1}}} \sum_{\ell,m} \left( - a_{\ell m} \sqrt{\frac{\omega_\ell}{2}} e^{-\omega_\ell \tau} Y_{\ell m} (\n) + a^*_{\ell m} \sqrt{\frac{\omega_\ell}{2}} e^{\omega_\ell \tau} Y_{\ell m}^* (\n) \right) ,
  \end{multlined}
\end{align}
where $\pi_0$ and $\Pi_0$ are constant zero modes of the fields, $\Omega_D = \frac{2 \pi^{D/2}}{\Gamma(D/2)}$ is the volume of the $D-1$-sphere and the \(Y_{\ell m}\) are hyperspherical harmonics.\footnote{The index \(m\) is a vector with \(D-2\) components. For conventions and properties see Appendix~\ref{sec:Ylm-identities}.} The dispersion relation for the oscillator modes reads
\begin{equation}\label{eq:DispersionRelation}
  R \omega_\ell= \sqrt{\frac{\ell(\ell+D-2)}{(D-1)} } \ .
\end{equation}
Adding higher-curvature terms in the \ac{eft} will add subleading corrections in \(1/Q\) to this expression, which will depend on the Wilsonian coefficients and are not universal.

The complex Fourier coefficients $a_{\ell m}$ can be extracted as follows:
\begin{equation}\label{eq:ExtractingTheOscillatorModes}
	a_{\ell m} = \sqrt{ \frac{c_1 D (D-1) \mu^{D-2}}{2 \omega_{\ell} \, R^{D-1} } } \int \dd S \bqty*{\pi(\tau, \n) \del_\tau \pqty*{Y^*_{\ell m}(\n) e^{\omega_{\ell} \tau}} - \del_\tau \pi(\tau, \n) Y^*_{\ell m}(\n) e^{\omega_{\ell} \tau}}.
\end{equation}
The canonical Poisson bracket between $\pi$ and $\Pi$ corresponds to the Fourier mode brackets $\{ a_{\ell m},  a_{\ell' m'}^\dagger \} = \delta_{\ell \ell'} \delta_{m m'}$.
The classical $O(2)$ current and conserved charge are
\begin{align}
  J^{\mu} &= \fdv{ \mathscr{L}}{\del_{\mu} \chi} \ , & Q &= \int \dd{S}  J^{\tau} = c_1 D \Omega_D(R\mu)^{D-1} + \Pi_0 .
\end{align}
The leading contribution to the charge comes from the homogeneous term (zero mode on the sphere) corresponding to the ground state.
This relates the \ac{eft} scale $\mu$ to the ground state charge $Q_0$ as
\begin{equation}\label{eq:mu_vs_Q}
  \mu = \left[ \frac{Q_0}{c_1 D R^{D-1} \Omega_D} \right]^{1/(D-1)} \ .
\end{equation}
As mentioned in the previous section, the $O(2)$ charge is our controlling parameter since $\Lambda R \sim \mu R \sim Q_0^{1/(D-1)}$ and the validity of the \ac{eft} is controlled by \(1/(\mu R)\).
At leading order in the fluctuations, the charge $Q$ of a generic solution of the \ac{eom} depends only additively on the zero mode $\Pi_0$,
\begin{equation}
   Q = Q_0 + \Pi_0 \ .
\end{equation}

Using the state-operator correspondence, we can compute the scaling dimension of the operator \(\Opp\) from the cylinder Hamiltonian.
A generic solution of the \ac{eom} corresponds to an operator with scaling dimension%
\begin{equation}\label{eq:classical_H}
    \Delta = R E_{\text{cyl}} = \Delta_0 + \frac{\del \Delta_0}{\del Q_0} \Pi_0 + \frac{1}{2} \frac{\del^2 \Delta_0}{\del Q_0 \del Q_0} \Pi_0^2 + R \sum_{\ell \geq 1 , m}\omega_{\ell}  a_{\ell m}^* a_{\ell m }, 
\end{equation}
where we have defined
\begin{align}\label{eq:Delta0}
  \Delta_0 = c_1 (D-1) \Omega_D(\mu R)^D + \order*{(R\mu)^{D-2}}, &&
  \frac{\del \Delta_0}{\del Q_0} = R \mu, &&
  \frac{\del^2 \Delta_0}{\del Q_0 \del Q_0} = \frac{1}{c_1 D(D-1) \Omega_D (R\mu)^{D-2}} \ .
\end{align}
The quantity $\Delta_0$ corresponds to the leading (classical) contribution to the action in Eq.~\eqref{eq:S_expansion}.
Note that the Hamiltonian $H_\chi$ for the $\chi$-field is shifted \emph{w.r.t.} the one for \(\pi\) as $H_\chi = H_\pi + \mu Q$.
Thus, for the fluctuation $\pi$ the effective time evolution is generated by $H_\pi + \mu Q$, as expected for a superfluid Goldstone fluctuation.

\subsection{Canonical quantization}

Canonical quantization in the cylinder frame is obtained by $\tau$-slicing, associating a Hilbert space $\mathscr{H}_Q$ to each fixed \(\tau\). This poses no conceptual problems since the cylinder is a direct product of the time direction and a curved manifold. 
The mode coefficients in the decompositions~\eqref{eq:FieldDecompositionInLadders} are promoted to field operators with non-vanishing commutators,
\begin{align}\label{eq:CanonicalCommutators1}
  \comm{ {\pi}_0}{ {\Pi}_0} &= i, & \comm{ a_{\ell m }}{ a^{\dagger}_{\ell' m'}} = \delta_{\ell\ell'} \delta_{mm'}.
\end{align}
These are equivalent to the canonical equal-$\tau$ commutator $\comm{ {\pi}(\tau,\n)}{ {\Pi}(\tau, \n')} = i \delta_{S^{D-1}}(\n,\n')$, where $\delta_{S^{D-1}}(\n,\n')$ is the invariant delta function on $S^{D-1}$.
To build a representation of the Heisenberg algebra we start with a vacuum $\ket{Q}$ which satisfies
\begin{equation}\label{eq:QisVacuumOfTheAs}
   a_{\ell m} \ket{Q} =  {\Pi}_0 \ket{Q} = 0.
\end{equation}
As we are in finite volume, the $O(2)$ charge is a well-defined operator acting on $\mathscr{H}_Q$ as
\begin{align}
   \Qop &= \int \dd{ S}  {\Pi}(\tau,\n) = Q_0  {\Id} +  {\Pi}_0 \ , &  \Qop\ket{Q} &= Q_0 \ket{Q}.
\end{align}

The non-zero charge of the vacuum can be increased by acting with the mode $ {\pi}_0$, which is the only one carrying non-zero charge,
\begin{align}
  \comm{ \Qop }{ {\pi}_0} &= -i \ , & \comm{ \Qop}{  a_{\ell m}} &= \comm{ \Qop}{ a_{\ell m}^\dagger } = 0.
\end{align}
However, this does not lead to a degeneracy in the spectrum.
Starting from the vacuum $\ket{Q}$ one can obtain a state annihilated by $ a_{\ell m}$ with charge $Q_0+q$ and scaling dimension $\Delta_0(Q_0 + q)$:\footnote{$\Delta_0 = \Delta_0(Q)$ is defined via equations \eqref{eq:mu_vs_Q} and \eqref{eq:Delta0}.}
\begin{equation}
  \ket{ Q + q} = e^{i  {\pi}_0 q} \ket{Q} = \exp*[\frac{i q}{\Omega_D R^{D-1}} \int \dd{ S}  {\pi}(\tau,\n)] \ket{Q}.
\end{equation}
While these states are all annihilated by the ladder operators, since $\comm{ a_{\ell m}}{ {\pi}_0} = 0$, they are not zero modes of $ {\Pi}_0$. 
Thus, they do not represent degenerate vacua, but they have a gap
\begin{equation}
  \Delta_0(Q_0+q) - \Delta_0(Q_0) \sim q (R \mu) \ .
\end{equation}
Since $ {\pi}_0$ is the only operator on which the $O(2)$-charge acts non-trivially, it has to be compact, $ {\pi}_0 \sim  {\pi}_0 + 2\pi  {\Id}$, which implies that $q \in \setZ$. This shows that the states with charge $Q_0 + q$ live at the \ac{eft} cutoff and will not be discussed any further.

\bigskip

The quantized quadratic Hamiltonian corresponding to the classical expression of the scaling dimension in~\eqref{eq:classical_H} can be written as the sum of a normal-ordered\footnote{Normal order refers to the vacuum $\ket{Q}$ where $\braket{ Q | \normord{H} | Q } = \Delta_0/R$.} operator $\normord{H}$ and a vacuum contribution,
\begin{equation}\label{eq:Delta1}
   {D} = R \normord{ {H}} {}+ \Delta_1  {\Id}, \quad\quad \text{where} \quad\quad \Delta_1 \coloneqq \frac{1}{2} \sum_{\ell \ge  1, m} (R\omega_\ell) .
\end{equation}
The vacuum contribution needs regulation and has physical consequences. This is computed in various dimensions in \autoref{sec:Casimir_appendix} and first appeared in $D=3$ in~\cite{Monin:2016jmo} and $D=4,5,6$ in~\cite{Cuomo:2020rgt}. 
From the point of view of the large-charge expansion, the one-loop correction comes at order \(\order{Q^0}\). 
This means that we need to keep track in the tree-level computation also of all the curvature terms up to this order. 
For example in \(D = 3\) we know that
\begin{equation}
	\Delta_0 = d_{3/2} Q^{3/2} + d_{1/2} Q^{1/2} + \order{Q^{-1/2}} ,
\end{equation}
and, in general there will be \(\ceil{(D+1)/2} \) terms with positive \(Q \)-scaling, each controlled by a Wilsonian coefficient~\cite{Gaume:2020bmp}.

The commutators between $ {D}$ and the various modes show which ones generate excited states when acting on the vacuum:
\begin{align}\label{eq:CanonicalCommutators2}
  \comm{ {D}}{ a_{\ell m}} &= - R \omega_\ell   a_{\ell m} ,
  &\comm{ {D}}{ a_{\ell m}^\dagger } &=  R \omega_\ell   a_{\ell m}^\dagger, \\
   \comm{ {D}}{ {\pi}_0} &= -i \frac{\partial \Delta_0}{\partial Q_0} - i \frac{\partial^2 \Delta_0}{\partial Q_0^2}  {\Pi}_0  ,
  &\comm{ {D}}{ {\Pi}_0} &=0.
\end{align}
The Hilbert space $\mathscr{H}_Q$ of the theory is described as the Fock space generated by states of the form
\begin{equation}\label{eq:generic_state}
a_{\ell_1 m_1}^\dagger \dots a_{\ell_k m_k}^\dagger | Q \rangle
\end{equation}
with charge $Q_0$ and scaling dimension 
\begin{equation}\label{eq:scalingDimPhonon}
	\Delta = \Delta_0 + \Delta_1+ \sum_{i =1}^k (R \omega_{\ell_k}).
\end{equation}
 These states are also known as \emph{superfluid phonon} states in the literature. Some comments are in order:
\begin{itemize}
    \item From the $\ac{cft}_D$ perspective, these states correspond to primary operators with different quantum numbers than  $\Opp$ (corresponding to the vacuum $\ket{Q}$) but same $O(2)$ charge. The only exception are states including at least one $a_{1m}^\dagger$ which are descendants since their energy is $R \omega_1 =1$.
    \item The $SO(D)$ part of the isometry group of the cylinder is realized in terms of some unitary operator $ {U}$ on $\mathscr{H}_Q$, under which the mode operators transform as
    \begin{equation}
     {U}(R)  a_{\ell m}^\dagger  {U}^\dagger(R) = \sum_{m'} D^{\ell}_{mm'}(R^{-1}) \, a_{\ell m'}^\dagger, \quad R \in SO(D).
    \end{equation}
    This follows from the decomposition \eqref{eq:FieldDecompositionInLadders} and the properties of hyperspherical harmonics, where $D^{\ell}_{mm'}$ is a finite-dimensional irrep of $SO(D)$ (generalizing Wigner's D-symbol in \(D>3\)).
    In the $\ac{cft}_D$ this is the group of Euclidean rotations, so that states $a_{\ell_1 m_1}^\dagger \dots a_{\ell_k m_k}^\dagger \ket{ Q }$ will generically correspond to spinning primaries in the appropriate reducible representation.
    \item Not all phonon states can be described within the \ac{eft}. When the $\ell$-quantum number becomes too large, their contribution $R\omega_\ell$ can compete with the leading $\Delta_0$ term, breaking the large-$Q$ expansion. We have seen that $\Delta_0 \sim Q^{D/(D-1)}$, but higher-curvature terms in~\eqref{eq:action} will introduce lower order corrections up to $Q^{1/(D-1)}$. Phonon states with comparable energy $\omega_\ell$ should be excluded from the \ac{eft}. This sets a cutoff for the $\ell$-quantum number as
    \begin{equation}\label{eq:ell_cutoff}
    \ell_{\text{cutoff}} \sim Q^{1/(D-1)}.
    \end{equation}
    Operators with such high spin should be described by new coset models with more complicated breaking pattern, resembling the ground states found in non-Abelian models of~\cite{Hellerman:2017efx,Hellerman:2018sjf,Banerjee:2019jpw}. Work in this direction has been carried out in~\cite{Cuomo:2017vzg,Cuomo:2019ejv}.
\end{itemize}
It is worth stressing that the structure of the spectrum and the existence of the above-mentioned charged spinning primaries is a direct prediction of the superfluid hypothesis for generic a $O(2)-\ac{cft}_D$. Canonical quantization is the appropriate framework for this discussion, but one will expect corrections to scaling dimensions and the spectrum structure coming from interactions in Eq.~\eqref{eq:Quadratic_Lagrangian}, corresponding to subleading terms in large $Q$. These are best discussed within a path integral formulation, so that ordinary loop expansions techniques can be employed. This will be the subject of the next section.

\section{Path integral methods}\label{sec:pathIntergral}

An equivalent basis of the fixed-$\tau$ Hilbert space $\mathscr{H}_Q$ is given by the field/momentum eigenstates
\begin{align}
   {\chi}(\n) \ket{\chi} &= \chi(\n) \ket{\chi}, &  {\Pi}(\n) \ket{\Pi} &= \Pi(\n) \ket{\Pi}.
\end{align}
Their bracket is fixed by the canonical commutation relations,
\begin{equation}
  \braket{ \chi | \Pi} = e^{i \int \dd{S} \chi \Pi}.
\end{equation}
Generically, the vacuum $\ket{Q}$ is a superposition of momentum eigenstates without the $\Pi_0$-component:
\begin{equation}
  \ket{Q} = \mathscr{N}_Q \int \DD{\Pi} \delta( \Pi_0 )  \Psi_Q(\Pi) \ket{\Pi},
\end{equation}
where $\mathscr{N}_Q$ is a normalization factor. In the limit of large separation, $\tau \to \infty$, correlators will not depend on the specifics of the vacuum wave function $\Psi_Q$, which will only affect the overall normalization. Without loss of generality, we can take $\Psi_Q= 1$.
The overlap of $\ket{ Q}$ with field eigenstates is then given by
\begin{equation}%
\label{eq:overlap_chiQ}
	\braket{\chi | Q } = \begin{cases}
		\mathscr{N}_Q \exp\left\{\frac{i Q}{\Omega_D R^{D-1}} \int \dd{S} \chi\right\} & \text{if \(\chi \) is constant,} \\
		0 & \text{otherwise}
	\end{cases}
\end{equation}
Generically, on a $\tau$-slice, the zero-modes of any field configuration can be separated by integrating on the sphere:
\begin{equation}
	\chi_0 = \int \dd{S} \chi .
\end{equation}
This bracket sets the correct boundary conditions for any correlators in the path integral representation of the form $\braket{ Q | \dots | Q} $. These boundary conditions are a generalization of open boundary condition on a segment in the special case of $D=1$.
We are mostly interested in correlators in which the vacuum $|Q\rangle$ is inserted at large separation on the cylinder, namely at $\tau = \pm \infty$. In this case the details on the boundary conditions the vacuum imposes are irrelevant. We can now construct path integrals for the norm of the states \eqref{eq:generic_state} which correspond to two-points functions of the corresponding primaries in the $\ac{cft}_D$ at large cylinder-time separation. 

\subsection{Two-point functions}\label{sec:twoPointFn}

\subsubsection[\texorpdfstring%
{$\braket{ Q  | Q}$ correlator}%
{<Q|Q>}]%
{$\braket{\myatop{Q}{\ell_2 m_2} | J  | \myatop{Q}{\ell_1 m_1} } $ correlator}%
\label{sec:QQ_corr}

The vacuum correlator with cylinder times $\tau_2 > \tau_1$ can be written using \eqref{eq:overlap_chiQ} as follows:
\begin{equation} \label{eq:QQcorrelator}   
   \braket{Q | e^{-\frac{(\tau_2 - \tau_1)}{R}   D} | Q} = \abs{\mathscr{N}_Q}^2\int \DD{\chi} \exp*[-S[\chi] - \frac{i Q}{\Omega_D R^{D-1}} \int_{\tau_1}^{\tau_2} \dd{\tau} \int \dd{S}   \dot \chi ] \coloneq \Anew ,
\end{equation}
where we have introduced the notation $\Anew$ for future convenience.
This path integral can be taken as the working definition of the correlator, without referring any more to the canonically quantized picture and taking the \ac{eft} action~\eqref{eq:action} as a starting point.

The path integral can be computed as a saddle-point expansion around a field configuration $\chi^{\saddle}(\tau, \n)$ which is a solution to the minimization problem
\begin{equation}
  \delta S[\chi] = \int_{\tau_1}^{\tau_2} \dd{\tau} \dd{S} \left( - \del_\mu \frac{ \del \mathscr{L}}{\del (\del_\mu \chi)} \right) \delta \chi + \eval*{ \int \dd{S}  \left(  \frac{ \del\mathscr{L}}{\del (\del_\tau \chi)} + \frac{i Q}{\Omega_D R^{D-1}} \right) \delta  \chi }_{\tau_1}^{\tau_2} .
\end{equation}
The bulk \ac{eom} requires the (Euclidean) $O(2)$ conserved current
\begin{equation}
  \frac{\del \mathscr{L}}{\del(\del^\mu \chi)} = c_1 D (-\del_\mu \chi \del_\mu \chi)^{D/2-1} \del_\mu \chi = J_\mu
\end{equation}
to be divergence-free. The general solution compatible with the boundary conditions is the homogeneous configuration $\chi^{\saddle}(\tau, \n) = - i \mu \tau + \pi_0$~\cite{Hellerman:2015nra}, with $\pi_0$ constant and the parameter $\mu$ fixed by the boundary condition to
\begin{equation}
  c_1 D \mu^{D-1} = \frac{Q}{\Omega_D R^{D-1}},
\end{equation}
which we had already seen in Eq.~\eqref{eq:mu_vs_Q}. The action expansion for this ground-state fluctuation $\chi(\tau,\n) = \chi^{\saddle}(\tau, \n)+\pi( \tau,\n)$ is, to quadratic order,
\begin{equation}\label{eq:Action_quadratic}
  S = \Delta_0 \frac{\tau_2-\tau_1}{R} + c_1 \mu^{D-2} \frac{D(D-1)}{2} \int_{\tau_1}^{\tau_2} \dd \tau \int \dd{S} \left( \dot{\pi}^2 + \frac{1}{(D-1)R^2} (\del_i \pi)^2 \right) + \Op(\mu^{D-3}).
\end{equation}
The boundary term eliminates the linear term in~\eqref{eq:Quadratic_Lagrangian} and correspondingly the zero-mode terms of~\eqref{eq:classical_H}, as expected.
As we will see later, this ground state is a good starting point for a loop expansion controlled by \(\mu R\). The normalization $\mathscr{N}_Q$ is chosen such that the correlator takes the form
\begin{equation}
  \Anew = R^{-2 (\Delta_0 + \Delta_1 +\dots)}  \exp*\left\{- \frac{(\tau_2 - \tau_1)}{R} \Big[ \Delta_0+\Delta_1 +\dots\Big] \right\} ,
\end{equation}
which corresponds to the two-point function in $\mathbb{R}^D$ normalized to unity. The correction $\Delta_1$ introduced in~\eqref{eq:Delta1} is the Casimir energy of the fluctuation $\pi$ around the homogeneous ground state $\chi^{\saddle}$.

In the state-operator correspondence, the reference states \(\ket{Q}\) and \(\bra{Q}\) correspond to insertions of scalar primaries at $\tau = \pm \infty$:
\begin{align}
  \ket{Q} &\coloneq \Opp(-\infty) \ket{0}, & \bra{Q} &\coloneq \bra{0} \Opp(\infty)^{\dagger},
\end{align}
recalling that conjugation on the cylinder is $\Opp(\tau, \n)^\dagger = \Opp*(-\tau,\n)$. The Weyl map to $\setR^D$ can then be performed as
\begin{equation}
  \begin{aligned}
   \braket{\Opp*(x_2) \Opp(x_1)}_{\setR^D } &= \left( \frac{|x_1|}{R} \right)^{-\Delta_Q} \left( \frac{|x_2|}{R} \right)^{-\Delta_Q} \braket{\Opp*(\tau_2, \n_2) \Opp(\tau_1, \n_1)}_{\text{cyl}} .
  \end{aligned}
\end{equation}
\subsubsection[\texorpdfstring%
{$\braket{\myatop{Q}{\ell_2 m_2} | \myatop{Q}{\ell_1 m_1} } $ correlators}%
{<phonon|phonon>}]{$\braket{\myatop{Q}{\ell_2 m_2} | \myatop{Q}{\ell_1 m_1} } $ correlators}%
\label{sec:phonon2pt}

The next class of two-point functions we study are correlators of one-phonon states obtained by acting with a single creation operator $ a^\dagger_{\ell m}$ on the vacuum $|Q \rangle$:
\begin{align}\label{eq:one_phonon}
	\ket{\myatop{Q}{\ell m}} &=  a_{\ell m}^\dagger \ket{Q}, &&  \text{where} && \ket{\myatop{Q}{0 0}} =\ket{Q}.
\end{align}
In canonical quantization, using the commutation relations of the $ a_{\ell m}$ and $ a_{\ell m}^\dagger$ the two-point function is found to be
\begin{equation}\label{eq:one_phonon_correlator_quad}
\begin{aligned}
	\braket{ \myatop{Q}{\ell_2 m_2} | \myatop{Q}{\ell_1 m_1}} &= \braket{ Q |  a_{\ell_2 m_2} \, e^{ - (\tau_2 - \tau_1)   D / R }  a_{\ell_1 m_1}^\dagger |Q} = \Anew e^{-(\tau_2 -\tau_1) \omega_\ell} \delta_{\ell_1 \ell_2} \delta_{m_1 m_2} \\
	&= R^{\Delta} e^{-\Delta (\tau_2 - \tau_1)/R} \delta_{\ell_1 \ell_2} \delta_{m_1 m_2} ,
\end{aligned}
\end{equation}
where \(\Delta \) is the conformal dimension that we have found in Eq.~\eqref{eq:scalingDimPhonon}, consistently with the general structure of a conformal two-point function on the cylinder given in Eq.~\eqref{eq:TwoPointCylinder}.

This is true to quadratic order in the Hamiltonian, however we expect loop corrections to shift the spectrum in a complicated way.
It is convenient to formulate the correlator as a path integral. This can be done in a straightforward manner by expressing $a_{\ell m}$ in terms of the fields as in Eq.~\eqref{eq:ExtractingTheOscillatorModes}, so that one finds
\begin{multline}
  \braket{ \myatop{Q}{\ell_2 m_2} | \myatop{Q}{\ell_1 m_1}} = { \frac{ c_1 D (D-1) \mu^{D-2}}{2 R^{D-1}\sqrt{ \omega_{ \ell_2} \omega_{ \ell_1} } } } \int \dd{S(\n_2)} \int \dd{S(\n_1)} Y^*_{\ell_2 m_2} (\n_2) \, Y_{\ell_1 m_1} (\n_1) \\
  \Anew \lim_{ \substack{ \tau \to \tau_1 \\ \tau' \to \tau_2}} \left( \omega_{\ell_2} - \del_{\tau'}{} \right) \left( \omega_{ \ell_1} + \del_{ \tau} {} \right) \braket{ \pi (\tau', \n_2)  \pi (\tau, \n_1) },
\end{multline}
where the two-point function of the Goldstone fluctuations is defined as 
\begin{equation}
\braket{ \pi (\tau_2, \n_2)  \pi (\tau_1, \n_1) }  = \frac{1}{\braket{ Q , \tau_2 | Q , \tau_1}} \int \DD{\pi} \pi(\tau_2, \n_2 ) \pi(\tau_1, \n_1 ) \, e^{- S[\pi]} ,
\end{equation}
where \(S[\pi]\) is the action~\eqref{eq:Action_quadratic}.
As expected, the information about the spectrum is contained in the full $\pi$-fluctuation two-point function.
In this formalism, the result in Eq.~\eqref{eq:one_phonon_correlator_quad} is found by using the tree-level propagator, which on the cylinder reads
\begin{equation}
   \braket{ \pi (\tau_2, \n_2 ) \pi (\tau_1, \n_1 )} = \frac{1}{c_1 D(D-1) (\mu R)^{D-2} } \pqty*{ \sum_{\ell=1}^\infty \sum_m e^{-\omega_\ell |\tau_2-\tau_1|} \frac{Y_{\ell m}(\n_2)^* Y_{\ell m}(\n_1)}{2 R \omega_\ell } - \frac{\abs{ \tau_2 - \tau_1} }{2 R \Omega_D} }.
\end{equation}

As discussed previously, the state \(\ket{\myatop{Q}{\ell m}}\) defines a spin-$\ell$ symmetric and traceless tensor operator inserted in the infinite past on the cylinder,
\begin{equation}
   \ket*{\myatop{Q}{\ell m}} \coloneq \Vpp[Q][\ell m](-\infty) \ket{0}.
\end{equation}

\medskip

The computation of $\braket{ \myatop{Q}{\ell_2 m_2} | \myatop{Q}{\ell_1 m_1}}$ in canonical quantization is easily generalized to states with more phonon excitations. For example, for two phonon excitations we get
\begin{multline}
        \braket*{ \myatop{Q}{(\ell_2 m_2) \otimes (\ell'_2 m'_2)} | \myatop{Q}{(\ell_1 m_1) \otimes (\ell'_1 m'_1)}} =  \braket{ Q |  a_{\ell_2 m_2 } a_{\ell'_2 m'_2} \, e^{ - (\tau_2 - \tau_1)   D / R }  a_{\ell'_1 m'_1}^\dagger  a_{\ell_1 m_1}^\dagger |Q} \\
        = \Anew  e^{- (\tau_2 -\tau_1) \left( \omega_{\ell_2} + \omega_{\ell'_2} \right) } \left( \delta_{\ell_1 \ell_2} \delta_{m_1 m_2} \delta_{\ell'_1 \ell'_2} \delta_{m'_1 m'_2} + \delta_{\ell_1 \ell'_2} \delta_{m_1 m'_2} \delta_{\ell'_1 \ell_2} \delta_{m'_1 m_2} \right).
\end{multline}
For states with higher numbers of phonon excitations the energy is just corrected accordingly and there is a sum over all possible permutations of Kronecker deltas.
These states are primary (as long as none of the $\ell$s is equal to one) but they will not be in irreducible representations like the one-phonon states.
For example in \(D = 3\), by virtue of the Clebsch--Gordan decomposition, we have
\begin{equation}
    \ell \otimes \ell' = (\ell + \ell') \oplus (\ell + \ell'-2) \oplus \dots \oplus \abs{\ell-\ell'}.
\end{equation}

\subsection{Loop corrections}\label{sec:loopCorrections}

The perturbation theory for the \ac{eft} action~\eqref{eq:action} can be conveniently set up on the thermal circle $S^1_\beta \times S^{D-1}$. 
One then recovers the original \ac{cft} predictions in the zero-temperature limit $\beta \rightarrow \infty$. 
The fluctuations $\pi$ can be decomposed into modes as
\begin{align}
\pi(\tau,\n) &= \sqrt{\frac{\beta}{R}} \sum_{n \in \mathbb{Z}} \sum_{\ell \geq 1 , m} Y_{\ell m} (\n) e^{i \nu_n \tau} \pi_{n\ell m}, & \pi_{n\ell m}^* &= (-1)^{m_{D-2}} \pi_{-n,\ell,m^*} ,
\end{align}
where the notation for the $m$-type quantum numbers follows from the standard tree convention for the hyperspherical harmonics, see Appendix~\ref{sec:Ylm-identities}. We have also introduced the Matsubara frequencies $\nu_n = 2\pi n /\beta$.%
\footnote{Standard references for thermal field theory methods are~\cite{Kapusta:2006pm,Laine:2016hma}.}
On $S^1_\beta \times S^{D-1}$ there is a unique zero mode, which can be excluded as it never appears in the derivative-only interactions.

The propagator in mode space can be computed from the quadratic part of the action:
\begin{equation}
	\braket{ \pi_{n \ell m } \pi_{n' \ell' m'} } = \frac{1}{c_1 D(D-1) (\mu R)^{D-2} } \frac{1}{\beta^2} \underbrace{\frac{1}{\nu_n^2 + \omega_{\ell}^2 }}_{\coloneq D_{n \ell}} \delta_{n,-n'} \delta_{\ell \ell'} (-1)^{\abs{m}} \delta_{m, -m'} .
\end{equation}
The zero mode does not mix with the other modes and has 
  $\braket{ \pi_0 \pi_0 } = \text{constant}$, and cannot be corrected at any order in perturbation theory because all vertices contain derivatives of the fields.
  
 For generic spacetime dimension $D$ the \ac{eft} action has all possible $k$-point vertices
  \begin{equation}\label{eq:S_interaction}
 S_{\text{int}} = \sum_{k=3}^{\infty} \mu^{D-k} S^{(k)} .
  \end{equation}
  The two-loop corrections to the $\Opp$ primary scaling dimension, $\Delta_2$, is computed via diagrams involving only three-point and four-point vertices:
\begin{align}
	S^{(3)} &= \frac{i}{6} c_1 D(D-1)(D-2) \int_0^\beta \dd \tau \int_{S^{D-1}} \dd S \dot{\pi} \left\{ \dot{\pi}^2 + \frac{3}{D-1} \frac{1}{R^2} (\del_i \pi)^2 \right\}, \\
	S^{(4)}&= -\frac{1}{24} c_1 D(D-1)(D-2) \int_0^\beta \dd \tau \underset{S^{D-1} \,}{ \int \dd S} \left\{ \frac{3}{R^4(D-1)} (\del_i \pi)^4+ \frac{6}{R^2}\left(  \frac{D-3}{D-1} \right) \dot{\pi}^2 (\del_i \pi)^2 + (D-3) \dot{\pi}^4    \right\}.
\end{align}
The corrections to the partition function are then computed as 
\begin{equation}\label{eq:Z_int}
\log \mathcal{Z} = \log \mathcal{Z}_0 - \mu^{D-4} \braket{ S^{(4)}}_c + \frac{1}{2} \mu^{2D-6} \braket{ S^{(3)} S^{(3)}}_c , 
\end{equation}
where we indicated with $\braket{ \dots }_c$ connected contractions only. 
Since propagators scale as $\sim \mu^{2-D}$ it is clear that both contributions enter at order $\mu^{-D} \sim Q^{-\frac{D}{D-1}}$ modulo $Q^0 \log Q$ powers. 
Following the same counting, an $\ell$-loop diagram produces the term
\begin{equation}
\Delta \supset Q^{-\frac{(\ell-1)D}{D-1}}
\end{equation}
in the $\Delta$ expansion. In what follows we will concentrate on the two-loop partition function \eqref{eq:Z_int}.

\subsubsection{Tadpole (sub)diagrams are vanishing}
\label{sec:tadpole}

One can easily verify that in the two-loop contribution $\Delta_2$ all tadpole-type diagrams vanish identically. More generally, no vertex appearing in \eqref{eq:S_interaction} can generate non-vanishing tadpole subdiagrams. This guarantees the ground state $\chi^{\saddle}(\tau,\n)$ introduced in \autoref{sec:QQ_corr} to be stable under quantum corrections. This is a direct consequence of the $SO(D) \times O(2)_{\mathrm{shift}}$ symmetry.  
 
A semi-diagrammatic proof of this statement goes as follows.  Any tadpole sub-diagram is generated by contractions from a vertex with an odd number of legs, $\braket{ S^{(2k-1)}}_c$ for some $k \in \mathbb{N}$. 
By $SO(D)$ invariance, the quantum-corrected propagator is $ \braket{ \pi_{n\ell m} \pi_{n' \ell' m'}} \propto \delta_{m-m'}$. Generically, each term will contain $k$ pairs $YY^*$ and one unpaired $Y$:
\begin{equation}
    \braket{ S^{(2k-1)} } \supset \sum_{m , m_1 \dotsm_{k} } \int_{S^{D-1}} \dd S Y_{\ell m} Y_{\ell_1 m_1} Y_{\ell_1 m_1}^* \dots Y_{\ell_k m_k} Y_{\ell_k m_k}^* .
\end{equation}
The sum $\sum_m Y_{\ell m}Y^*_{\ell m}$ is a constant, see Appendix~\ref{sec:Ylm-identities}. So
\begin{equation}
     \sum_{m , m_1 \dotsm_{k} } \int_{S^{D-1}} \dd S Y_{\ell m} Y_{\ell_1 m_1} Y_{\ell_1 m_1}^* \dots Y_{\ell_k m_k} Y_{\ell_k m_k}^* \propto \sum_m \int_{S^{D-1}} \dd S Y_{\ell m} =0 .
\end{equation}
This holds since the $O(2)$ shift symmetry guarantees that the zero mode $\ell=0$ never appears in perturbation theory. When derivatives on the hyperspherical harmonics $Y_{\ell m}$ are present, the same argument applies if one uses the identities in Eq.~\eqref{eq:Y_orthogonality}.

\subsubsection{One-loop scaling dimension $\Delta_1$ and regularization}

Let us briefly review the computation of the one-loop scaling dimension $\Delta_1$ for the primary $\Opp$. 
On $S^1_\beta \times S^{D-1}$, this is computed as
\begin{align}
  \Delta_1 = -\lim_{\beta \rightarrow \infty} \frac{\del}{\del \beta} \log \mathcal{Z}_0 =  \frac{1}{2} \sum_{\ell >0} M_\ell (R \omega_\ell),
\end{align}
where the Matsubara sum has been performed as in Appendix~\ref{sec:Loop_appendix}. It coincides with the expression found in Eq.~\eqref{eq:Delta1}.
We indicated with $M_\ell$ the degeneracy of the $\ell$-th eigenvalue $\lambda_\ell$ of $\Delta_{S^{D-1}}$, see Eq.~\eqref{eq:laplacian_degenerancy}. The sum appearing above is obviously divergent and needs regularization.
For future convenience we introduce the regularization
\begin{equation}\label{eq:regulated_sphere_sum}
\Sigma(s) = \lim_{\Lambda \to \infty} \sum_{\ell >0} M_\ell (R \omega_\ell)^{s} e^{-\omega_\ell^2/ \Lambda^2}.
\end{equation}
The one-loop scaling dimension reads $\Delta_1 = \frac{1}{2} \Sigma(1)$ and is computed in detail in Appendix~\ref{sec:Casimir_appendix} for various spacetime dimensions.
The use of a momentum-dependent regulator is natural.
Our controlling parameter is the charge \(Q\), which is used to define  the \ac{eft}  \ac{uv} scale, via $\Lambda R \sim \mu R \sim Q^{1/(D-1)}$, which cuts the phonon states running in internal lines as discussed in Eq.~\eqref{eq:ell_cutoff}.

\subsubsection{Two-loop scaling dimension $\Delta_2$}

In this section we compute the two-loop correction to the scaling dimension $\Delta_2$. 
In higher-loop computations, the regularization procedure needs to take into account also the Matsubara modes.
This is because the derivative couplings lead to sums of the form 
\begin{equation}
    \sum_{n \in \mathbb{Z}} \sum_{\ell >0} \frac{\nu_n^2 \omega_\ell^2}{\nu_n^2 + \omega_\ell^2}
\end{equation}
in which the Matsubara sum is divergent.
A convenient choice~\cite{Dietz:1982uc} is a regularization procedure which is linear,
\begin{equation}\label{eq:reg_lin}
    \Reg \bqty*{ \sum_{n \in \mathbb{Z}} \sum_{\ell >0} \left[ \alpha f(n ,\ell) + \beta g(n, \ell) \right] } =  \alpha \Reg \bqty*{ \sum_{n \in \mathbb{Z}} \sum_{\ell >0}  f(n ,\ell) }+ \beta \Reg \bqty*{ \sum_{n \in \mathbb{Z}} \sum_{\ell >0}  g(n ,\ell) },
\end{equation}
symmetric under $n \leftrightarrow -n$, and $m$-independent. 
We use a smooth cutoff regularization~\cite{Bilal:2013iva,Monin:2016bwf} which now involves also the thermal circle:
\begin{equation}\label{eq:regulation}
    \Reg \bqty*{ \sum_{n \in \mathbb{Z}} \sum_{\ell >0} f(n ,\ell) } = \sum_{n \in \mathbb{Z}} \sum_{\ell >0} f(n ,\ell)e^{-(\nu_n^2 + \omega_\ell^2)/\Lambda^2}.
\end{equation}
The zero temperature limit for these regulated sums is then taken with $\beta \rightarrow \infty$ keeping $(R \Lambda)$ fixed, and will eventually produce regulated sums of the type \eqref{eq:regulated_sphere_sum}. 
Usual Feynman-integral methods can be used on $S^1_\beta \times S^{D-1}$, and the contribution from the quartic action is found to be
\begin{equation}\label{eq:fourpoint_diag}
    \braket{ S^{(4)}} = - \frac{c_1 D(D-2)}{24} \left\{ \frac{3}{R^4} \mathinner{\vcenter{\hbox{\includegraphics[angle=90]{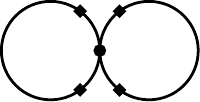}}}}  + \frac{6(D-3)}{R^2} \mathinner{\vcenter{\hbox{\includegraphics[angle=90]{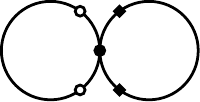}}}} + (D-1)(D-3) \mathinner{\vcenter{\hbox{\includegraphics[angle=90]{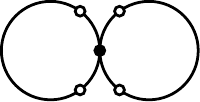}}}}  \right\}.
\end{equation}
We indicate with square dots spatial derivatives $\del_i$ acting on the corresponding legs, while white dots stand for time derivatives $\del_\tau$. In this pictorial notation we have suppressed the permutations of these derivatives on the legs, which need to be included as independent Wick contractions. In Appendix~\ref{sec:Loop_appendix} we provide separate computations for each contribution and find
\begin{equation}
   \mu^{D-4} \braket{ S^{(4)} }_c = - \frac{(D-2)}{8 c_1 D(D-1) \Omega_D (\mu R)^D} \left( \frac{R}{\beta} \right) \left\{ (D-3) \left[ \sum_{n \ell} M_\ell \right]^2 -4 \left[ \sum_{n\ell} D_{n\ell} \omega_\ell^2 M_\ell\right]^2  \right\} ,
\end{equation}
where each sum is regulated by Eq.~\eqref{eq:regulation}. Performing the Matsubara sum in the $\beta \rightarrow \infty$ limit we find the contribution to the scaling dimension coming from the quartic vertices,
\begin{equation}\label{eq:Delta_vertex4}
    \Delta_2^{(4)} = - \frac{(D-2)}{8c_1 D(D-1) \Omega_D (\mu R)^D} \left\{ (D-3)\frac{(\Lambda R)^2}{4\pi} \Sigma(0)^2 - \Sigma(1)^2\right\}.
\end{equation}
The contributions coming from the 4-point vertices are purely 2-loop and are 
\begin{equation}\label{eq:threepoint_diag}
\braket{ S^{(3)} S^{(3)}}_c = - \frac{c_1^2}{36}D^2(D-1)^2(D-2)^2 \left\{ \mathinner{\vcenter{\hbox{\includegraphics{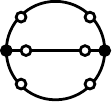}}}} + \frac{6}{R^2(D-1)} \mathinner{\vcenter{\hbox{\includegraphics{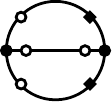}}}} + \frac{9}{R^4(D-1)^2} \mathinner{\vcenter{\hbox{\includegraphics{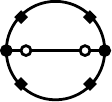}}}} \right\},
\end{equation}
where again each graph stands for all permutations of different derivatives on the internal legs. As discussed in Section~\ref{sec:tadpole}, there are no tadpole contractions. 

Each graph is computed separately in Appendix~\ref{sec:Loop_appendix}. We can express the final result in a completely symmetrized form as:
\begin{multline}%
\label{eq:sum-of-two-loop-graphs}
    \mu^{2D-6}\braket{ S^{(3)} S^{(3)} }_c = - \frac{(D-2)}{12 c_1 D(D-1) \Omega_D (\mu R)^D} \left( \frac{R}{\beta} \right) \sum_{n_a + n_b + n_c = 0} \sum_{\ell_a \ell_b \ell_c} D_{n_a \ell_a} D_{n_b \ell_b} D_{n_c \ell_c} S_{\ell_a \ell_b \ell_c} \\
    \times \Bigg\{ 2\nu_{n_a}^2 \nu_{n_b}^2 \nu_{n_c}^2 -  \nu_{n_a} \nu_{n_b} \nu_{n_c} \left[ \nu_{n_a} (\omega_{\ell_b} + \omega_{\ell_c})^2 + (\text{cyclic perm.} ) \right] \\
     + \frac{1}{2}  \nu_{n_a}^2 (\omega_{\ell_b}^2 + \omega_{\ell_c}^2 -\omega_{\ell_a}^2 )^2 + (\text{cyclic perm.} )\\
    -   \nu_{n_a} \nu_{n_b} \left[ \omega_{\ell_c}^4- (\omega_{\ell_a}^2 -\omega_{\ell_b}^2)^2 \right] + (\text{cyclic perm.} ) \Bigg\}.
\end{multline}
The symmetric structure $S_{\ell_a \ell_b \ell_c}$ is defined in~\autoref{sec:Loop_appendix} in  Eq.~\eqref{eq:def_Sabc}.
The symbol $\vartri_{\ell_a \ell_b \ell_c}$ enforces the $SO(D)$ quantum numbers $\ell_a$, $\ell_b $, $\ell_c$ to satisfy a triangle inequality, which corresponds to momentum conservation on $S^{D-1}$:
\begin{equation}\label{eq:def_triangle}
    \vartri_{\ell_a \ell_b \ell_c} = \begin{cases} 1 & \text{if} \ \abs{\ell_b - \ell_a } \leq \ell_c \leq \ell_b+\ell_a \quad \text{and} \quad \ell_c - \ell_a - \ell_b \ \text{even}, \\
    0 & \text{otherwise.} \end{cases}
\end{equation}

The corresponding contribution to the scaling dimension can be obtained in the $\beta \rightarrow \infty$ limit, computing the Matsubara sums along the same lines as before. The end result for the two-loop scaling dimension for the operator $\Opp$ reads 
\begin{multline}\label{eq:Delta2_final}
    \Delta_2 = \frac{1}{16 c_1 D(D-1)\Omega_D (\mu R)^D} \Bigg\{ \frac{(R \Lambda)^2}{6 \pi} (D-2) \left[ 2D-4 + (D-5) \Sigma(0) \right] \Sigma(0)  \\
    - \frac{(R \Lambda)}{\sqrt{\pi}} (D-2)^2 \left( 1-\Sigma(0) \right) \Sigma(1)  \\
    - \frac{D-2}{3} \big[ (D-2)( \Sigma(2) + 6 \Sigma(0) \Sigma(2) + 2 \Sigma(-1) \Sigma(3) ) - (5D-16) \Sigma(1)^2 - 8\Sigma^{(2\ell)} \big] \Bigg\},
\end{multline}
where $\Sigma^{(2\ell)}$ is a sum which cannot be readily reduced to a combination of $\Sigma(s)$ sums:
\begin{equation}
    \Sigma^{(2\ell)} = \sum_{\ell_a \ell_b \ell_c} S_{\ell_a \ell_b \ell_c} \vartri_{\ell_a \ell_b \ell_c} \frac{\omega_{\ell_a} \omega_{\ell_b} \omega_{\ell_c}}{\omega_{\ell_a} + \omega_{\ell_b} + \omega_{\ell_c}}.
\end{equation}
This new sum above can be regulated and computed, but for the present work we limit ourselves to noting that it has two divergent regimes: when $\ell_a \sim \ell_b \gg 1$ it grows as $\Sigma(1)$, and when $\ell_a \sim \ell_b \sim \ell_c \gg 1$ it grows as $\Sigma(2)$. 
The former corresponds to a degenerate triangle described by three integers $\ell_a$, $\ell_b$, $\ell_c$, while the latter corresponds to an equilateral triangle. 
These are the discrete versions of different collinear divergences in ordinary loop integrals. 

When $D$ is even, the result~\eqref{eq:Delta2_final} has the structure
\begin{equation}
    \Delta_2 \supset \frac{1}{Q^{D/(D-1)}}  \left( \alpha_0 + \alpha_1 \log Q + \alpha_2 (\log Q)^2 \right)
\end{equation}
and a (non-universal) $Q^0 \log Q^2$ term appears. This result is expected to generalize to any loop $l$, where a term $\Sigma(1)^l$ is expected to appear. Thus, in general, we expect in even $D$ to have an $l$-loop contribution of the form
\begin{equation}
    \Delta_l \supset \frac{1}{Q^{(l-1)D/(D-1)}}  \left( \alpha_0 + \alpha_1 \log Q + ... + \alpha_l (\log Q)^l \right).
\end{equation}
This result is especially relevant for applications in the context of resurgent asymptotics. In odd dimensions, large-$Q$ expansions are expected to be log-free transseries with non-perturbative corrections related to worldline instantons~\cite{Dondi:2021buw,Antipin:2022dsm}. This is an important simplification with respect to transseries which contain logarithm terms, which can be been found as consequence of quasi-zero mode integration in quantum mechanics problems~\cite{Zinn-Justin:1982hva,Jentschura:2011zza,Zinn-Justin:2002ecy}. It would be interesting to investigate whether a similar phenomenon is behind the appearance of $Q^0 \log Q$ terms in even-$D$ large-$Q$ expansions.

\section{Correlators with current insertions}\label{sec:ConformalAlgebraAndChargeCorrelators}

The advantage of working with an \ac{eft} at large charge is that the physics at the fixed point is captured by a free theory. We are therefore able to explicitly compute three- and four-point functions for a strongly coupled system using only the operator algebra \eqref{eq:CanonicalCommutators1}.

This section contains one of the main results of this paper, namely three- and pour-point correlators with current insertions between spinful large-charge primaries $\Vpp[Q][\ell m]$. A handful of these correlators have already appeared in the literature in the scalar case $\ell = 0$~\cite{Monin:2016jmo,Komargodski:2021qp,Jafferis:2017zna,Cuomo:2020thesis}. 
Even though we do not use conformal invariance to compute the correlators, we find that our results respect as expected the structure of conformal correlators on the cylinder in the limit of large separation collected in Appendix~\ref{sec:constraints}.
Note that our results have to be understood as an expansion in $Q$ and contain only the classical contribution plus the leading-order quantum correction.

\subsection{Conserved currents and Ward identities in the EFT}\label{sec:currents}

The classical conserved currents in the model~\eqref{eq:action} are
\begin{align}
    J_\mu &= c_1 D (-\del_\mu \chi \del^\mu \chi)^{D/2-1} \del_\mu \chi ,\\
    T_{\mu\nu} &= c_1 \left\{ D  (-\del_\mu \chi \del^\mu \chi)^{D/2-1} \del_\mu \chi \del_\mu \chi + g_{\mu\nu}  (-\del_\mu \chi \del^\mu \chi)^{D/2} \right\} .
\end{align}
Their integrals are related to the conserved charges of $\mathbb{R}^d$ as already discussed in \autoref{sec:classical_treatment}. 

On the cylinder, the currents are expanded in fluctuations around $\chi^{\saddle}(\tau,\n)$ up to quadratic order as  
\begin{subequations}\label{eq:ExpansionsOfTsAndQsInTheField}
\begin{align}
  & J_\tau = - i \frac{Q}{\Omega_D R^{D-1}} \left\{ 1 + \frac{i}{\mu}  (D-1)  \dot{\pi} - \frac{  (D-2) (D-1) }{2 \mu^2} \left[ \dot{\pi}^2 + \frac{ ( \del_i \pi )^2 }{R^2 (D-1)} \right] + \order{\mu^{-3}} \right\} \label{eq:Jtau}, \\
  & J_i = \frac{Q}{\Omega_D R^{D-1}} \left\{\frac{1}{\mu R} \del_i \pi + \frac{i }{\mu} \frac{(D-2)}{\mu R} \dot{\pi} \del_i \pi + \order{\mu^{-3}}\right\},\label{eq:Ji} \\
  & T_{\tau\tau} = - \frac{ \Delta_0 }{ \Omega_D R^{D}} \left\{1 + i \frac{ D}{ \mu}  \dot{\pi} - \frac{ D (D-1) }{ 2 \mu^2} \left[ \dot{\pi}^2 + \frac{(D-3) \, (\del_i \pi)^2 }{R^2 (D-1)^2} \right] + \order{\mu^{-3}} \right\}, \\
   & T_{\tau i} = - i \frac{\Delta_0}{\Omega_D R^{D}} \left[ \frac{1}{\mu R} \frac{D}{D-1} \partial_i \pi + \frac{i}{\mu} \frac{D}{\mu R} \dot{\pi}\partial_i \pi + \order{\mu^{-3}}  \right] \\
  &
    \begin{multlined}[][.9\linewidth]
      T_{ij} = \frac{ \Delta_0 }{ \Omega_D R^{D} } \frac{ h_{ij}}{(D-1)} \left\{ 1 + i \frac{ D }{ \mu } \dot{\pi} - \frac{ D (D-1) }{ 2 \mu^2} \left[ \dot{\pi}^2 + \frac{ (\del_i \pi)^2 }{R^2 (D-1)} \right]  +\order{\mu^{-3}} \right\} \\
      + \frac{ \Delta_0 }{   \Omega_D R^{D}} \frac{1}{(\mu R)^2}\frac{D}{(D-1)}  \left\{ \del_i \pi \del_j \pi + \order{\mu^{-3}} \right\},
    \end{multlined}
\end{align}
\end{subequations}
where $h_{ij}$ is the metric on the $D-1$-sphere. Homogeneity of $\chi^{\saddle}$ guarantees that $T_{\tau i} = T_{ij} = 0$ at leading order in the fluctuations. 

The above expressions were calculated using only the leading term in the effective action~\eqref{eq:action}. As observed before, this gives rise to contributions of $\order{Q^0}$, while the effect of the sub-leading curvature terms on the fluctuations is suppressed at large charge. 

We will discuss correlators of these currents in the canonically quantized setting of~\autoref{sec:O(2)_review}, which is sufficient for leading-order results. Integrating $J_\tau$ or $T_{\tau\tau}$ over spatial slices  gives rise to the charge operators $ {D}$ and $\Qop$. 
When inserted at time $\tau$ they measure the scaling dimension and the $O(2)$-charge of any operator insertion contained in the half-cylinder $(\tau, -\infty) \times S^{D-1}$. This fact is expressed by the Ward identities
\begin{align}\label{eq:WardIdentities}
\braket{\Qop(\tau) \prod_{i} \mathscr{O}_i(\tau_i, \n_i) } &=  \sum_{ \tau_i < \tau} Q_i \braket{\prod_{i} \mathscr{O}_i(\tau_i, \n_i ) }, \\
\braket{  {D}(\tau) \prod_{i} \mathscr{O}_i(\tau_i, \n_i) } &=  \sum_{ \tau_i < \tau} \Delta_i \braket{ \prod_{i} \mathscr{O}_i(\tau_i, \n_i ) }.
\end{align}
These identities hold order by order in a loop expansion and can be used to constrain correlators with insertions of the currents in Eq.~\eqref{eq:ExpansionsOfTsAndQsInTheField}.

\subsection[\texorpdfstring%
{$\braket{\myatop{Q}{\ell_2 m_2} | J  | \myatop{Q}{\ell_1 m_1} } $}%
{<QJQ>}]{$\braket{\myatop{Q}{\ell_2 m_2} | J  | \myatop{Q}{\ell_1 m_1} } $ correlators}%
\label{sec:VJV}
We first compute the correlator of two spinning primaries $\Vpp[Q][\ell m] |0 \rangle=  a^\dagger_{\ell m} |Q\rangle$ inserted at times $\tau_1 , \tau_2$ with an insertion of $J_\mu(\tau,x)$ at time $\tau_1 < \tau < \tau_2$.\footnote{Recall that in the special case $\ell=1$ the operator $\Vpp[Q][\ell m]$ is not a primary, but a descendant.} To leading order one finds
\begin{align}
  &\begin{aligned}\
  &\braket{\Vpp* J_{\tau}(\tau,\n) \Vpp} = - i \frac{Q}{\Omega_D R^{D-1}} \Bigg\{ \Anew[\Delta_Q + R \omega_{\ell_2}] \delta_{\ell_1 \ell_2}  \delta_{m_1 m_2}  \\
  & + \Anew[\Delta_Q + R \omega_{\ell_1}][\Delta_Q + R \omega_{\ell_2}][\tau] (D-1)(D-2) \Omega_D \frac{ R \sqrt{ \omega_{ \ell_2} \omega_{ \ell_1} } }{ 2 D  \Delta_0 } \Big[ Y_{\ell_2 m_2}^* (\n) Y_{\ell_1 m_1} (\n) - \frac{ \partial_i Y_{\ell_2 m_2}^* (\n) \, \partial_i Y_{\ell_1 m_1} (\n) }{ R^2 (D-1) \omega_{\ell_2} \omega_{\ell_1} } \Big]  \Bigg\} , 
  \end{aligned} \\
  &\braket{\Vpp* J_i(\tau,\n) \Vpp} =  i \frac{ Q (D-2)}{2 \Delta_0 R^{Ð-1} D } \Anew[ \Delta_Q + R \omega_{ \ell_{1}} ][ \Delta_Q + R \omega_{ \ell_{2}} ][ \tau]  \bigg[ \sqrt{ \frac{ \omega_{ \ell_2} }{ \omega_{ \ell_1} } } Y_{\ell_2 m_2}^* (\n) \del_i Y_{\ell_1 m_1} (\n) - (1 \leftrightarrow 2 )^* \bigg] .
\end{align}
For later convenience we have introduced 
\begin{align}
    \Anew[ \Delta_Q + R \omega_{\ell_2}]  &\coloneqq \Anew  e^{- (\tau_2 - \tau_1) \, \omega_{ \ell_2} } , & \Anew[\Delta_1][\Delta_2][\tau] &\coloneqq e^{- \Delta_2 ( \tau_2 - \tau)/R - \Delta_1 (\tau - \tau_1)/R} .
\end{align}
This generalizes \(\Anew = \Anew[\Delta]\) defined by the two-point function \(\braket{Q|Q}\) in Eq.~\eqref{eq:QQcorrelator}. The $\ell=0$ case of the $J_\tau$ correlator appeared first in~\cite{Monin:2016jmo}.

If we integrate $J_\tau$ over the sphere, we obtain the conserved charge, so the integral over the three-point function with a $J_\tau$ is fixed by the Ward identity~\eqref{eq:WardIdentities}, giving us a consistency check:
\begin{equation}
    \int \dd{S(\n)} \braket{\Vpp* J_{\tau}(\tau,\n) \Vpp} = - i Q  \Anew[\Delta_Q + R \omega_{\ell_2}] \delta_{\ell_1 \ell_2}  \delta_{m_1 m_2} .
\end{equation}

Let us examine the structure of the $J_\tau$ correlator. The current is a sum of a classical piece and quantum corrections, see Eq.~\eqref{eq:Jtau}. The classical part is homogeneous and, by charge conservation, time-independent. It follows that the classical contribution to the three-point function must be proportional to the two-point function
\begin{equation}
	\braket*{ \myatop{Q}{\ell_2 m_2} | \myatop{Q}{\ell_1 m_1}} = \Anew[\Delta_Q + R \omega_{\ell_2}] \delta_{\ell_1 \ell_2}  \delta_{m_1 m_2}.
\end{equation}
The quantum piece will in general give a contribution that has the same tensor structure as the \ac{lhs} and since it is not homogeneous, can be decomposed into spherical harmonics. Moreover, as we have seen above, by charge conservation its integral must vanish.

In the same way, the classical piece of $J_i$ is zero (see Eq.~\eqref{eq:Ji}), so we only have the inhomogeneous quantum contribution in the $J_i$ correlator. The separation into a homogeneous classical part and a space-dependent quantum contribution applies to any physical observable.

The special case of $\ell_i = 0$ corresponds to the (scalar) ground state $\ket{Q}$ and Ward identities guarantee that $\braket{ Q | J_i |Q} = 0$ to all orders.

From the above relations one can extract the corresponding \ac{ope} coefficient:
\begin{equation}
    C\indices{_{\Vpp[Q][\ell m] J_\tau}^{\Vpp[Q][\ell m]}} =\frac{ \braket{\Vpp[-Q][\ell m] J_{\tau}(\tau,\n) \Vpp[Q][\ell m]} }{ \braket{\Vpp[-Q][\ell m]  \Vpp[Q][\ell m] }} = - i \frac{Q}{\Omega_D R^{D-1}}.
\end{equation}

\medskip

These correlators can also be computed for higher-phonon states. For example, for two-phonon states we have
\begin{multline}
 \braket*{ \myatop{Q}{(\ell_2 m_2) \otimes (\ell'_2 m'_2)} | J_\tau (\tau, \n) | \myatop{Q}{(\ell_1 m_1) \otimes (\ell'_1 m'_1)} } = - i \frac{Q }{ \Omega_D R^{D-1}} \Anew[\Delta_Q + R \omega_{\ell_2} + R \omega_{\ell'_2}]  \\
 \Bigg\{ \Big( \delta_{\ell_1 \ell_2} \delta_{m_1 m_2} \delta_{\ell'_1 \ell'_2} \delta_{m'_1 m'_2} + \delta_{\ell_1 \ell'_2} \delta_{m_1 m'_2} \delta_{\ell'_1 \ell_2} \delta_{m'_1 m_2} \Big) \\
 + \Omega_D \frac{ (D-2) (D-1) }{2D \, \Delta_0 } \bigg[ \frac{ R \sqrt{ \omega_{\ell'_2} \omega_{\ell'_1}} }{ e^{ (\tau - \tau_1) ( \omega_{\ell'_1} - \omega_{\ell'_2}) } } \Big[  Y_{\ell'_1 m'_1}(\n) Y_{\ell'_2 m'_2}^*(\n) - \frac{ \partial_i Y_{\ell'_1 m'_1}(\n) \partial_i Y_{\ell'_2 m'_2}^*(\n) }{R^2 (D-1) {\omega_{\ell'_2} \omega_{\ell'_1}} } \Big]  \delta_{\ell_2 \ell_1} \delta_{m_2 m_1} \\
 + \frac{ R \sqrt{ \omega_{\ell'_2} \omega_{\ell_1}} }{ e^{ (\tau - \tau_1) ( \omega_{\ell_1} - \omega_{\ell'_2} ) } } \Big[ Y_{\ell_1 m_1} (\n) Y_{\ell'_2m'_2}^*(\n) - \frac{ \partial_i Y_{\ell_1 m_1}(\n) \partial_i Y_{\ell'_2 m'_2}^*(\n) }{R^2 (D-1) {\omega_{\ell'_2} \omega_{\ell_1}} } \Big] \delta_{\ell_2 \ell'_1} \delta_{m_2 m'_1} \\
 + \frac{ R \sqrt{ \omega_{\ell_2} \omega_{\ell'_1}} }{ e^{ (\tau - \tau_1) ( \omega_{\ell'_1} - \omega_{\ell_2} ) } } \Big[ Y_{\ell'_1 m'_1}(\n) Y_{\ell_2 m_2}^*(\n) - \frac{ \partial_i Y_{\ell'_1 m'_1}(\n) \partial_i Y_{\ell_2 m_2}^*(\n) }{R^2 (D-1) {\omega_{\ell_2} \omega_{\ell'_1}} } \Big] \delta_{\ell'_2 \ell_1} \delta_{m'_2 m_1} \\
 + \frac{ R \sqrt{ \omega_{\ell_2} \omega_{\ell_1}} }{ e^{ (\tau - \tau_1) ( \omega_{\ell_1} - \omega_{\ell_2} ) } } \Big[  Y_{\ell_1 m_1} (\n) Y_{\ell_2 m_2}^* (\n) - \frac{ \partial_i Y_{\ell_1 m_1}(\n) \partial_i Y_{\ell_2 m_2}^*(\n) }{R^2 (D-1) {\omega_{\ell_2} \omega_{\ell_1}} } \Big] \delta_{\ell'_2 \ell'_1} \delta_{m'_2 m'_1} \bigg] \Bigg\} .
\end{multline}

In the computations above we have neglected the linear terms in the fluctuation $\pi$, which appears in the various conserved charges in Eq.~\eqref{eq:ExpansionsOfTsAndQsInTheField}.
These terms cannot be neglected if we consider matrix elements between states with different number of phonons.
For example, for correlators between the ground state and one-phonon states one finds the following leading-order expression:
\begin{align}
    \braket*{ \Opp* | J_\tau (\tau, \n) | \myatop{Q}{\ell m} } %
    &= -\frac{ Q (D-1)}{ \Omega_D R^{D-1} } \sqrt{ \frac{ \Omega_D }{ 2D } \frac{ R \omega_{\ell} }{ \Delta_0 } } \Anew[\Delta_Q + R \omega_{\ell}][\Delta_Q ][\tau] Y_{\ell m} (\n) ,  \\
    \braket*{ \Opp* | J_i (\tau, \n) | \myatop{Q}{\ell m} } %
    &= \frac{ Q}{ \Omega_D R^{D-1} } \sqrt{ \frac{ R \, \Omega_D }{ 2D \, \Delta_0 R\omega_{\ell} } } \, \Anew[\Delta_Q + R \omega_{\ell}][\Delta_Q ][\tau] \, \partial_i Y_{\ell m} (\n). 
\end{align}
The same holds for all the correlators with insertions of $T$ or $J$.
We will omit this special case in what follows as it represents a straightforward modification to the formulas presented there.

\subsection[\texorpdfstring%
{$\braket{\myatop{Q}{\ell_2 m_2} | JJ  | \myatop{Q}{\ell_1 m_1}}$}%
{<OJJO>}]%
{$\braket{\myatop{Q}{\ell_2 m_2} | JJ  | \myatop{Q}{\ell_1 m_1} }$ correlators}%
\label{sec:VJJV}
We next compute the correlators with two insertions of $J_\mu$ at cylinder times $\tau < \tau'$ between insertions of $\Vpp[Q][\ell m]$ at $\tau_1$ and $\tau_2$ such that $\tau_2 > \tau > \tau' > \tau_1$. Two insertions of $J_\tau$ result in
\begin{multline}
        \braket*{\Vpp* J_\tau(\tau,\n) J_\tau(\tau',\n') \Vpp } = - \Anew[\Delta_Q + R \omega_{ \ell_2}] \frac{ Q^2 }{ \Omega_D^2 R^{2D-2}} \, \delta_{\ell_1 \ell_2} \delta_{m_1m_2} \\
        \shoveright{ \times \Bigg\{ 1 + \frac{ (D-1)^2  }{ 2 D \, \Delta_1 } \sum_{ \ell} e^{-|\tau -\tau'| \omega_{\ell} } R \omega_{\ell} \frac{ (D+2\ell -2)}{ (D-2) } C^{ \frac{D }{2} -1}_{\ell} (\n \cdot \n') \Bigg\} } \\
        {+ \Bigg\{ \Anew[\Delta_Q + R \omega_{ \ell_1}][\Delta_Q + R \omega_{ \ell_2}][\tau] \frac{ Q^2 (D-1)^2 }{ 2 \Omega_D  R^{2D-2} D  } \frac{ R\sqrt{ \omega_{ \ell_1} \omega_{ \ell_2} } }{ \Delta_0 } \,\bigg( - \frac{ Y_{\ell_2 m_2}^* (\n) Y_{\ell_1 m_1} (\n') }{ e^{- (\tau-\tau') \omega_{\ell_1}} } } \\
        + \frac{ (D-2) }{ (D-1)} \Big[ \frac{ \del_i Y_{ \ell_1 m_1} (\n) \del_i Y_{\ell_2 m_2}^* (\n) }{ (D-1)\, R^2 {\omega_{ \ell_1} \omega_{ \ell_2}} } - Y_{\ell_1 m_1} (\n) Y_{\ell_2 m_2}^* (\n) \Big] \bigg)  +  \Big((\tau, \n) \leftrightarrow (\tau', \n') \Big)  \Bigg\} ,
\end{multline}
where we have introduced the Gegenbauer polynomials $C^{{D}/{2}-1}_{\ell}$, defined in Eq.~\eqref{eq:DefGegenbauerPolynomials}. The $\ell=0$ case of this correlator has appeared first in~\cite{Cuomo:2020thesis}.

Integrating this result over the sphere centered at the insertion point $(\tau, \n)$ gives again the conserved charge as in the Ward identity~\eqref{eq:WardIdentities} and hence must eliminate the $\tau$-dependence, again providing a consistency check of our result:
\begin{equation}
        \int \dd{S(\n)} \braket*{\Vpp* J_\tau(\tau,\n) J_\tau(\tau',\n') \Vpp } = -i Q \braket{\Vpp* J_{\tau}(\tau',\n') \Vpp} .
\end{equation}

The remaining components of the $JJ$ correlators read
\begin{gather}
  	\braket{\Vpp* J_\tau(\tau, \n) J_i (\tau', \n') \Vpp } = 0 , \\
   \begin{multlined}[][.9\linewidth]
  \braket*{\Vpp* J_i (\tau, \n) J_j (\tau',\n') \Vpp } = \Anew[\Delta_Q + R \omega_{\ell_1}][\Delta_Q + R \omega_{\ell_2}][\tau] \frac{ Q^2 }{2 \Omega_D R^{2D -2} \Delta_0 D } \\
  \Bigg[ \del_i \del'_j \sum_{ \ell} \frac{ e^{-|\tau -\tau'| \omega_{\ell} } }{ R \omega_{\ell} } \frac{ (D+2\ell -2)}{ (D-2) \Omega_D} C^{\frac{D }{2} - 1}_{\ell} (\n \cdot \n') \, \delta_{ \ell_2 \ell_1} \delta_{m_2 m_1}\\
   + \frac{\del_j Y_{\ell_2 m_2}^* (\n') \del_i Y_{\ell_1 m_1} (\n) }{ e^{ (\tau - \tau') \omega_{\ell_2}} R \sqrt{ \omega_{ \ell_1} \omega_{ \ell_2} } } 
  + \frac{ \del_i Y_{\ell_2 m_2}^* (\n) \del_j Y_{\ell_1 m_1}(\n') }{ e^{- (\tau- \tau') \omega_{ \ell_1}} R \sqrt{\omega_{\ell_1} \omega_{\ell_2}}} \Bigg] .
\end{multlined}
\end{gather}
For the last correlator the tree-level contribution vanishes, but it is not symmetry protected so that subleading corrections may appear. 
\medskip

In the special case of \(\ell_i = 0 \) these correlators reduce to
\begin{align}
 &\begin{multlined}[][.9\linewidth]
    \braket{ \Opp* J_\tau(\tau,\n) J_\tau(\tau',\n') \Opp } = -   \Anew\frac{ Q}{ (\Omega_D R^{D -1})^2}  \\
    \times \bigg[ Q + \frac{(D-1) }{ 2 \mu } \sum_{\ell} \omega_{ \ell}e^{-|\tau -\tau'| \omega_{\ell} }  \frac{ (D+2\ell -2)}{ (D-2)} C^{\frac{D}{2}-1}_{\ell} (\n \cdot \n') \bigg] ,
\end{multlined}\\
    &\braket{ \Opp* J_i(\tau,\n) J_j(\tau', \n') \Opp } =  \frac{ Q \Anew }{ 2 \mu \Omega_D (D-1) R^{2D-1}} \del_i \del'_j  \sum_{\ell} \frac{ (D+2\ell -2) \, C^{ \frac{D }{2}-1}_{\ell} (\n \cdot \n') }{ (D-2) \Omega_D  \omega_{\ell} e^{ |\tau -\tau'| \omega_{ \ell} } } ,\\
    &\braket{ \Opp* J_\tau(\tau, \n) J_i (\tau', \n') \Opp } = 0 .
\end{align}
The mixed correlator $J_\tau J_i$ vanishes exactly on the homogeneous ground state associated to $\Opp$ by rotational invariance. It is straightforward to show that these correlators satisfy the Ward identity~\eqref{eq:WardIdentities} by integrating over the sphere insertion of $J_\tau$.

\subsection[\texorpdfstring%
{$\braket{\myatop{Q}{\ell_2 m_2} | T  | \myatop{Q}{\ell_1 m_1}}$}%
{<QTQ>}]%
{$\braket{\myatop{Q}{\ell_2 m_2} | T  | \myatop{Q}{\ell_1 m_1} }$ correlators}%
\label{sec:VTV}
Now we compute correlators with an insertion of the stress-energy tensor $T$ at cylinder time $\tau$ with spinning operators $\Vpp[Q][\ell m]$ at $\tau_1,\tau_2$ such $\tau_2 > \tau > \tau_1$. The insertion of the $T_{\tau\tau}$ component leads to
\begin{multline}\label{eq:3pt_Ttautau}
    \braket*{\Vpp* T_{\tau\tau} (\tau, \n)  \Vpp } = - \Anew[\Delta_Q + R \omega_{\ell_{1}}][\Delta_Q + R \omega_{\ell_{2}}][\tau] \frac{1 }{ \Omega_D R^D}  \Bigg\{ (\Delta_0+ \Delta_1) \delta_{\ell_2 \ell_1} \delta_{m_2 m_1} \\
    + \frac{ \Omega_D }{2} R\sqrt{ \omega_{ \ell_1} \omega_{ \ell_2} } \, \bigg[ (D-1) Y_{ \ell_2 m_2}^* (\n) Y_{\ell_1 m_1} (\n) 
    - \frac{(D-3) }{ (D-1)} \frac{ \del_i Y_{\ell_2 m_2}^* (\n) \del_i Y_{\ell_1 m_1} (\n) }{ R^2 { \omega_{ \ell_1} \omega_{\ell_2}}} \bigg] \Bigg\} .
\end{multline}
For insertions at large separation \(\tau_1,\tau_2 \rightarrow \pm \infty\) and $\ell_2=\ell_1$, the three-point function does not depend on the $\tau$-slice of the $T$ insertion, see also Appendix~\ref{sec:constraints}. Moreover, integrating this result over the sphere insertion point $\n$ eliminates the $\tau$-dependence according to the Ward identity~\eqref{eq:WardIdentities}:
\begin{equation}
        \int \dd{S(\n)} \braket*{\Vpp* T_{\tau\tau} (\tau, \n)  \Vpp } = - \Anew[ \Delta_Q + R \omega_{ \ell}] \frac{1}{R} \pqty*{ \Delta_0 + \Delta_1 + R \omega_{\ell_2}} \, \delta_{\ell_1 \ell_2} \delta_{m_1 m_2} .
\end{equation}
Correlators involving the other components of the stress-energy tensor read
\begin{equation}
	\braket*{\Vpp*  T_{\tau i}(\tau, \n) \Vpp } = \Anew[\Delta_Q + R \omega_{\ell_{1}}][\Delta_Q + R \omega_{\ell_{2}}][\tau] \frac{1}{ 2 R^{D}}\bigg[ \sqrt{ \frac{ \omega_{ \ell_2} }{ \omega_{ \ell_1} } } Y_{\ell_2 m_2}^* (\n) \del_i Y_{\ell_1 m_1} (\n) - (1 \leftrightarrow 2 )^* \bigg] , 
\end{equation}
\begin{multline}
    \braket*{\Vpp*  T_{ij}(\tau, \n) \Vpp } = \Anew[ \Delta_Q + R \omega_{ \ell_1}][ \Delta_Q + R \omega_{ \ell_2}][\tau] \frac{ 1 }{ (D-1)  \Omega_D R^D} \\
     \Bigg\{  h_{ij} \Big[ \left( \Delta_0 + \Delta_1\right)  \delta_{ \ell_2 \ell_1} \delta_{m_2 m_1} \\
     + \frac{\Omega_D R \sqrt{\omega_{\ell_1} \omega_{\ell_2}} }{2} \Big((D-1)  Y_{ \ell_2 m_2}^* (\n) Y_{\ell_1 m_1} (\n) 
    - \frac{ \del_i Y_{ \ell_2 m_2}^* (\n) \del_i  Y_{\ell_1 m_1} (\n) }{ R^2 \omega_{ \ell_1 } \omega_{ \ell_2 } } \Big) \Big]  \\
    + R\sqrt{ \omega_{ \ell_1} \omega_{ \ell_2} } \Omega_D \frac{ \del_{ ( i} Y_{ \ell_2 m_2 }^* (\n)  \del_{j) }  Y_{\ell_1  m_1} (\n) }{ R^2 \omega_{ \ell_1} \omega_{ \ell_2}} \Bigg\} .
\end{multline}

Note that an insertion of the trace of the energy-momentum tensor $T_{\tau \tau} + h^{ij} T_{ij}$ vanishes on any phonon state by conformal invariance.
We find the expression
\begin{multline}
    \braket*{\Vpp* h^{ij} T_{ij}(\tau, \n) \Vpp } = \Anew[ \Delta_Q + R \omega_{ \ell_1}][ \Delta_Q + R \omega_{ \ell_2}][\tau] \frac{ 1 }{ (D-1) \, \Omega_D R^D} \Bigg\{ (D-1) \left( \Delta_0 + \Delta_1\right)  \delta_{ \ell_2 \ell_1} \delta_{m_2 m_1}  \\
    + \frac{\Omega_D R \sqrt{\omega_{\ell_1} \omega_{\ell_2}} }{2 } \bigg( (D-1)^2 \, Y_{ \ell_2 m_2}^* (\n) Y_{\ell_1 m_1} (\n) - (D-3) \, \frac{ \del_i Y_{ \ell_2 m_2}^* (\n) \del_i  Y_{\ell_1 m_1} (\n) }{ R^2 \omega_{ \ell_1 } \omega_{ \ell_2 } } \bigg) \Bigg\} ,
\end{multline}
which sums to zero with~\eqref{eq:3pt_Ttautau}. 
For \(\ell_1 = \ell_2 = 0 \) one finds the expressions
\begin{align}
  &\braket{ \Opp* T_{\tau\tau}(\tau, \n)  \Opp } = - \Anew \frac{ \Delta_0 + \Delta_1 }{ \Omega_D R^D} , \\
  &\braket{ \Opp* T_{\tau i}(\tau, \n) \Opp } = 0 , \\
  &\braket{\Opp* T_{ij}(\tau, \n) \Opp} = \frac{\Anew }{(D-1)} \frac{\Delta_0 + \Delta_1 }{ \Omega_D R^D} h_{ij} .
\end{align}
The three-point function with a single insertion of $T_{\tau i}$ vanishes by rotational symmetry. Moreover, since the ground state is homogeneous, the correlator with an insertion $T_{ij}$ can only be proportional to the sphere metric.

\subsection[\texorpdfstring%
{$\braket{\myatop{Q}{\ell_2 m_2} | T T | \myatop{Q}{\ell_1 m_1}}$}%
{<QTTQ>}]%
{$\braket{\myatop{Q}{\ell_2 m_2} | T T  | \myatop{Q}{\ell_1 m_1} }$ correlators.}%
\label{sec:VTTV}
We next compute the correlators with two insertions of the stress-energy tensor at $\tau>\tau'$ between spinning operators $\Vpp[Q][\ell m]$ at $\tau_1,\tau_2$ such that $\tau_2 > \tau > \tau' > \tau_1$.
There is a total of six correlators. These read
\begin{multline}
	\braket*{\Vpp* T_{\tau\tau}(\tau, \n) T_{\tau\tau}(\tau', \n') \Vpp } = \Anew[\Delta_Q + R \omega_{ \ell_1}][\Delta_Q + R \omega_{ \ell_2}][\tau] \frac{ \Delta_0 }{ \Omega_D^2 R^{2D}}  \\
	\Bigg\{  \bigg[ { \Delta_0} + {2 \Delta_1} + \frac{ D }{ 2} \sum_{\ell} e^{-|\tau -\tau'| \omega_{\ell} } R \omega_{\ell} \frac{ (D+2\ell -2)}{ D-2 } C^{ \frac{D }{2} -1}_{\ell} (\n \cdot \n') \bigg] \delta_{\ell_1 \ell_2} \delta_{m_1 m_2} \\
	+ \frac{ D \Omega_D}{ 2} R \sqrt{ \omega_{\ell_1} \omega_{\ell_2} } \bigg[   Y^*_{\ell_2 m_2} (\n) Y_{\ell_1 m_1} (\n') \, e^{ (\tau -\tau') \omega_{ \ell_1} } + Y^*_{\ell_2 m_2} (\n') Y_{\ell_1 m_1} (\n) \, e^{- (\tau -\tau') \omega_{ \ell_2} } \bigg] \Bigg\} \\
	+ \Bigg\{ \, \Anew[\Delta_Q + R \omega_{ \ell_1}][\Delta_Q + R \omega_{ \ell_2}][\tau] \frac{\Omega_D \Delta_0 R\sqrt{ \omega_{\ell_1} \omega_{\ell_2} } }{ 2 \Omega_D^2 R^{2D} } \bigg( \, (D-1) \, Y_{\ell_1 m_1} (\n) Y^*_{\ell_2 m_2} (\n) \\
    - \frac{(D-3)}{(D-1)} \frac{ \del_i Y_{\ell_1 m_1} (\n) \del_i Y^*_{\ell_2 m_2} (\n) }{ R^2 {\omega_{ \ell_1} \omega_{ \ell_2}} } \bigg)\, + \Big((\tau, \n) \leftrightarrow (\tau', \n') \Big)  \Bigg\} .
\end{multline}
This correlator is symmetric under $(\tau, \n) \leftrightarrow (\tau', \n')$. The $\ell=0$ special case of this correlator has already appeared in~\cite{Komargodski:2021qp}.
\begin{multline}
         \braket*{\Vpp* T_{ij} (\tau, \n)  T_{kn} (\tau', \n') \Vpp } = \Anew[\Delta_Q + R \omega_{ \ell_1}][\Delta_Q + R \omega_{ \ell_2}][\tau] \frac{ \Delta_0 }{ (D-1)^2 \Omega_D^2 R^{2D} }  \\
         \Bigg\{  \bigg[ \Delta_0 + 2 \Delta_1  + \frac{ D }{ 2} \sum_{\ell} e^{-|\tau -\tau'| \omega_{\ell} } R \omega_{\ell} \frac{ (D+2\ell -2)}{ D-2} C^{ \frac{D }{2} -1}_{\ell} (\n \cdot \n') \bigg] h_{ij} h_{kn}   \delta_{\ell_2 \ell_1} \delta_{m_2 m_1} \\
         + \frac{ D \, \Omega_D }{ 2} R \sqrt{ \omega_{\ell_2} \omega_{ \ell_1} } \left( Y_{\ell_2  m_2}^* (\n) Y_{\ell_1 m_1} (\n') \, e^{ (\tau - \tau') \omega_{ \ell_1}} + Y_{\ell_2  m_2}^* (\n') Y_{\ell_1 m_1} (\n) \, e^{ -(\tau -\tau') \omega_{ \ell_2} } \right) h_{ij} h_{kn}  \Bigg\} \\
        + \Bigg\{ \Anew[\Delta_Q + R \omega_{ \ell_1}][\Delta_Q + R \omega_{ \ell_2}][\tau] \frac{\Omega_D \Delta_0 R \sqrt{ \omega_{ \ell_1 } \omega_{\ell_2} } }{ 2 (D-1) \, \Omega_D^2 R^{2D} } \bigg[ 2 \frac{ \del_{( i}  Y_{ \ell_2  m_2}^* (\n) \del_{j)} Y_{\ell_1 m_1} (\n) }{ R^2 (D-1) \, { \omega_{ \ell_1} \omega_{ \ell_2} } } + Y_{\ell_2 m_2}^* (\n)   Y_{\ell_1 m_1} (\n) h_{ij} \\
        - \frac{ \del_i Y_{\ell_2 m_2}^* (\n) \del_i Y_{\ell_1 m_1} (\n) }{ R^2 (D-1) \, { \omega_{\ell_1} \omega_{\ell_2}} } h_{ij} \bigg] \, h_{kn} +  \Big((\tau, \n, {ij} ) \leftrightarrow (\tau', \n', {kn} ) \Big)  \Bigg\} .
\end{multline}
\begin{multline}
        \braket*{\Vpp* T_{\tau i} (\tau,\n)  T_{\tau j} (\tau',\n') \Vpp } = - \Anew[ \Delta_Q + R \omega_{ \ell_1}][ \Delta_Q + R \omega_{ \ell_2}][\tau] \frac{\Delta_0 D }{ 2 (D-1)^2 \Omega_D R^{2D} } \\
        \Bigg\{ \del_i \del'_j \sum_{\ell} \frac{e^{-|\tau -\tau'|  \omega_{\ell} } }{R \omega_{\ell} } \frac{ (D+2\ell -2)}{ (D-2) \Omega_D} C^{ \frac{D }{2} -1}_{\ell} (\n \cdot \n')  \, \delta_{\ell_2 \ell_1}\delta_{m_2 m_1} + \frac{ \partial_i Y_{\ell_2 m_2}^* (\n) \, \partial_j' Y_{\ell_1 m_1} (\n') }{ R \sqrt{ \omega_{ \ell_2} \omega_{ \ell_1} }  e^{ - (\tau - \tau') \omega_{ \ell_1}} } \\
        + \frac{ \partial_j' Y_{\ell_2 m_2}^* (\n') \, \partial_i Y_{\ell_1 m_1} (\n) }{ R \sqrt{ \omega_{ \ell_2} \omega_{ \ell_1} }  e^{ (\tau - \tau') \omega_{\ell_2}} } \, \Bigg\} .
\end{multline}
This correlator is symmetric under $(\tau, \n, i ) \leftrightarrow (\tau', \n', j )$.
\begin{multline}
        \braket*{\Vpp* T_{\tau i} (\tau, \n)  T_{\tau\tau} (\tau', \n') \Vpp } = - \Anew[\Delta_Q + R \omega_{ \ell_1}][\Delta_Q + R \omega_{ \ell_2}][\tau] \frac{ \Delta_0 D }{2 \Omega_D R^{2D}} \frac{1}{(D-1)} \\
        \Bigg\{ \del_i \sum_{\ell} e^{- |\tau -\tau'| \omega_{\ell} } \frac{ (D+2\ell -2)}{ (D-2) \Omega_D} C^{ \frac{D }{2} -1}_{\ell} (\n \cdot \n') \, \delta_{ \ell_2 \ell_1} \delta_{m_2 m_1} + \sqrt{ \frac{ \omega_{\ell_2} }{ \omega_{\ell_1} } } \frac{ Y_{\ell_2 m_2}^* (\n') \del_i Y_{\ell_1 m_1} (\n) }{ e^{ (\tau -\tau') \omega_{ \ell_2} }}  \\
        - \sqrt{ \frac{ \omega_{\ell_1} }{ \omega_{\ell_2} } } \frac{ \del_i Y_{\ell_2 m_2}^* (\n) \, Y_{\ell_1 m_1} (\n') }{ e^{ -(\tau -\tau') \omega_{\ell_1}}} + \frac{ (D-1) }{ D } \bigg[ \sqrt{ \frac{ \omega_{\ell_2} }{ \omega_{\ell_1} } } Y_{\ell_2 m_2}^* (\n)  \del_i Y_{\ell_1 m_1} (\n) - (1 \leftrightarrow 2 )^* \bigg] \, \Bigg\} .
\end{multline}
Since $T_{\tau i}$ vanishes on the ground-state solution, the correlator solely receives a second-order contribution from the linear terms and the quadratic term of $T_{\tau i}$. Moving to the combination $T_{\tau i} T_{j k}$ one finds
\begin{multline}
    \braket*{\Vpp* T_{\tau i} (\tau, \n)  T_{j k} (\tau', \n') \Vpp } = \Anew[\Delta_Q + R \omega_{\ell_{1}}][\Delta_Q + R \omega_{\ell_{2}}][\tau] \frac{\Delta_0 D }{2 \Omega_D R^{2D} } \frac{ h_{jk} }{ (D-1)^2} \\
    \Bigg\{ \partial_i \sum_{\ell} e^{- |\tau-\tau'| \omega_{\ell}} \frac{ (D+ 2\ell -2) }{ (D-2) \Omega_D} C^{\frac{ D}{2} -1}_\ell (\n \cdot \n') \, \delta_{\ell_2 \ell_1} \delta_{m_2 m_1} + \sqrt{ \frac{ \omega_{ \ell_2} }{ \omega_{ \ell_1}} } \frac{ Y_{\ell_2 m_2}^* (\n') \, \partial_i Y_{\ell_1 m_1} (\n) }{ e^{ (\tau-\tau') \omega_{\ell_2} } } \\
    - \sqrt{ \frac{ \omega_{ \ell_1} }{ \omega_{ \ell_2}} } \frac{ Y_{\ell_1 m_1} (\n') \, \partial_i Y_{\ell_2 m_2}^* (\n) }{ e^{- (\tau-\tau') \omega_{\ell_1} } } + \frac{ (D-1) }{ D} \bigg[ \sqrt{ \frac{ \omega_{ \ell_2} }{ \omega_{ \ell_1} } } Y_{\ell_2 m_2}^* (\n) \del_i Y_{\ell_1 m_1} (\n) - (1 \leftrightarrow 2)^* \bigg] \Bigg\} .
\end{multline}
Again, besides the linear terms only the quadratic term of $T_{\tau i}$ contributes at second order.
In addition, the correlator $\braket*{\Vpp* T_{\tau i} (\tau, \n) \, h^{jk}(\n') T_{j k} (\tau', \n') \Vpp }$ differs by a minus sign from the previous correlator with an insertion of $T_{\tau\tau} (\tau', \n')$, as imposed by conformal invariance.
\begin{multline}
    \braket*{\Vpp* T_{\tau \tau} (\tau, \n)  T_{ij} (\tau', \n') \Vpp } =  - \Anew[\Delta_Q + R \omega_{ \ell_1}][\Delta_Q + R \omega_{ \ell_2}][\tau] \frac{ \Delta_0 }{ \Omega_D^2 R^{2D}} \frac{ h_{ij} }{ (D-1)} \\
    \Bigg\{ \bigg[ \Delta_0 + 2 \Delta_1 + \frac{ D \, \Omega_D }{ 2} \sum_{\ell} R \omega_{\ell} e^{- |\tau - \tau'| \omega_{\ell} } \frac{ (D+ 2\ell -2) }{ (D-2) \Omega_D} C^{ \frac{D }{2} -1}_{\ell} (\n \cdot \n') \bigg] \delta_{\ell_2 \ell_1} \delta_{m_2 m_1} \\
    + \frac{ D \, \Omega_D  }{2} R \sqrt{ \omega_{ \ell_2} \omega_{ \ell_1} } \bigg[  Y_{\ell_2 m_2}^* (\n) Y_{\ell_1 m_1} (\n')  e^{ (\tau - \tau') \omega_{ \ell_1} } + Y_{\ell_2 m_2}^* (\n') Y_{\ell_1 m_1} (\n) e^{-(\tau - \tau') \omega_{ \ell_2} } \bigg] \Bigg\} \\
    - \Anew[\Delta_Q + R \omega_{\ell_{1}}][\Delta_Q + R \omega_{\ell_{2}}][\tau] \frac{ \Delta_0 R \sqrt{ \omega_{ \ell_1} \omega_{ \ell_2} } }{ 2\Omega_D R^{2D}} h_{ij} \Bigg\{ Y_{ \ell_2 m_2}^* (\n) Y_{\ell_1 m_1} (\n) - \frac{(D-3) }{ (D-1)^2} \frac{ \del_i Y_{\ell_2 m_2}^* (\n) \del_i Y_{\ell_1 m_1} (\n) }{ R^2 { \omega_{ \ell_1} \omega_{\ell_2}}} \Bigg\} \\
    - \Anew[ \Delta_Q + R \omega_{ \ell_1}][ \Delta_Q + R \omega_{ \ell_2}][\tau'] \frac{ \Delta_0 R \sqrt{ \omega_{ \ell_1} \omega_{ \ell_2}} }{2 \Omega_D R^{2D}} \Bigg\{ \, h_{ij} \Big[ \, Y_{ \ell_2 m_2}^* (\n') Y_{\ell_1 m_1} (\n') - \frac{ \del_i Y_{ \ell_2 m_2}^* (\n') \del_i  Y_{\ell_1 m_1} (\n') }{ (D-1) \, R^2 \omega_{ \ell_1 } \omega_{ \ell_2 } } \Big]  \\
    + 2 \frac{ \del_{ ( i} Y_{ \ell_2 m_2 }^* (\n')  \del_{j) }  Y_{\ell_1  m_1} (\n') }{ (D-1) \, R^2 \omega_{ \ell_1} \omega_{ \ell_2}} \Bigg\} .
\end{multline}
This correlator is not symmetric in $(\tau, \n) \leftrightarrow (\tau', \n')$, however, by conformal invariance, the correlator $\braket*{\Vpp* T_{\tau \tau} (\tau, \n) h^{ij}(\n) T_{ij} (\tau', \n') \Vpp }$ is symmetric in $(\tau, \n) \leftrightarrow (\tau', \n')$.\\

\medskip

One can check directly that the above correlators satisfy the Ward identity for $T_{\tau\tau}$ insertions in Eq.~\eqref{eq:WardIdentities}.
For example, for two insertions of $T_{\tau\tau}$ one finds
\begin{equation}
    \int \dd{S(\n)} \braket*{\Vpp* T_{\tau\tau} (\tau,\n) T_{\tau\tau} (\tau',\n') \Vpp }
    = -\frac{1}{R} \pqty*{ \Delta_0 + \Delta_1 + R \omega_{\ell_2}} \braket*{\Vpp* T_{ \tau\tau} (\tau',\n') \Vpp } .
\end{equation}

In the special case $\ell =0$, the above correlators simplify as follows:
\begin{align}
    &\begin{multlined}[][.9\linewidth]
    		\braket{ \Opp* T_{\tau\tau}(\tau, \n) T_{\tau\tau} (\tau', \n') \Opp} =  \Anew  \frac{ \Delta_0 }{\Omega_D^2 R^{2D}} \bigg[ \Delta_0  + 2 \Delta_1 \\
	 + \frac{ D  }{2} \sum_{\ell} e^{-|\tau -\tau'| \omega_{\ell} } R \omega_{\ell} \frac{ (D+2\ell -2)}{ D-2} C^{ \frac{D }{2} -1}_{\ell} (\n \cdot \n') \bigg] ,
    \end{multlined} \\
    &\begin{multlined}[][.9\linewidth]
    \braket{ \Opp* T_{ij}(\tau, \n) T_{kn}(\tau', \n') \Opp } =  \Anew \frac{\Delta_0 }{ \Omega_D^2 R^{2D} } \frac{  h_{ij} h_{kn} }{(D-1)^2} \bigg[ \Delta_0 +  2\Delta_1 \\
     + \frac{ D }{2 } \sum_{\ell} \frac{ R \omega_{\ell} }{ e^{ |\tau -\tau'| \omega_{\ell} } } \frac{ (D+2\ell -2)}{ D-2} C^{ \frac{D }{2} -1}_{\ell} (\n \cdot \n') \bigg] ,
    \end{multlined} \\
    &
    \braket*{\Opp* T_{\tau i} (\tau, \n) T_{\tau j} (\tau', \n') \Opp } = - \frac{ \Anew \Delta_0 D }{ 2 (D-1)^2 \Omega_D^2 R^{2D} } \del_i \del'_j \sum_{\ell}  \frac{ e^{- |\tau -\tau'| \omega_{\ell} } }{ R \omega_{\ell} } \frac{ (D+2\ell -2)}{D-2} C^{ \frac{D }{2} -1}_{\ell} (\n \cdot \n') ,
     \\
    &
    \braket{ \Opp* T_{\tau i}(\tau, \n) T_{\tau\tau}(\tau', \n') \Opp } = - \frac{ \Anew \Delta_0 D }{ 2 (D-1) \Omega_D^2 R^{2D} } \del_i \sum_{\ell} e^{- |\tau -\tau'| \omega_{\ell} } \frac{ (D+2\ell -2)}{ D-2} C^{ \frac{D }{2} -1}_{\ell} (\n \cdot \n') ,
\\
&
    \braket*{\Opp* T_{\tau i} (\tau, \n)  T_{j k} (\tau', \n') \Opp } = \frac{ \Anew \Delta_0 D \, h_{jk} }{2 (D-1)^2 \Omega_D^2 R^{2D} } \partial_i \sum_{\ell} e^{- |\tau-\tau'| \omega_{\ell}} \frac{ (D+ 2\ell -2) }{ D-2} C^{\frac{ D}{2} -1}_\ell (\n \cdot \n') ,
\\
&\begin{multlined}[][.9\linewidth]
    \braket*{\Opp* T_{\tau \tau} (\tau, \n)  T_{ij} (\tau', \n') \Opp } = - \Anew \frac{ \Delta_0 }{ \Omega_D^2 R^{2D}} \frac{ h_{ij} }{ (D-1)} \bigg[ \Delta_0 + 2 \Delta_1 \\
      + \frac{ D  }{ 2} \sum_{\ell} R \omega_{\ell} e^{- |\tau - \tau'| \omega_{\ell} } \frac{ (D+ 2\ell -2) }{ D-2} C^{ \frac{D }{2} -1}_{\ell} (\n \cdot \n') \bigg] .
\end{multlined}
\end{align}
The $\ell_1 = \ell_2 = 0$ correlator with insertions of $T_{\tau i} T_{\tau \tau}$ was computed in the macroscopic limit $R \rightarrow \infty$ in~\cite{Komargodski:2021qp}.

\subsection[\texorpdfstring%
{$\braket{\myatop{Q}{\ell_2 m_2} | TJ  | \myatop{Q}{\ell_1 m_1}}$}%
{<QTJQ>}]%
{$ \braket{\myatop{Q}{\ell_2 m_2} | T J  | \myatop{Q}{\ell_1 m_1} }$ correlators.}%
\label{sec:VTJV}
We now consider correlators with one insertion of the stress-energy tensor and one insertion of the $O(2)$-current respectively at times $\tau > \tau'$ between spinning operators $\Vpp[Q][\ell m]$ at $\tau_1, \tau_2$ such that we have the ordering $\tau_2 > \tau > \tau' > \tau_1$. There are six correlators involving the various components which can be computed as follows:
\begin{align}
\begin{multlined}[][.9\linewidth]
        \braket*{\Vpp* T_{\tau i}(\tau, \n) J_\tau(\tau', \n') \Vpp } =  - i \Anew[ \Delta_Q + R \omega_{ \ell_1}][ \Delta_Q + R \omega_{ \ell_2}][\tau] \frac{ Q }{2 \Omega_D R^{2D-1}} \\
        \Bigg\{  \del_i \sum_{\ell} e^{- |\tau -\tau'| \omega_{\ell} }  \frac{ (D+2\ell -2)}{ (D-2) \Omega_D} C^{ \frac{D }{2} -1}_{\ell} (\n \cdot \n') \, \delta_{\ell_2 \ell_1} \delta_{m_2 m_1}  + \sqrt{ \frac{ \omega_{\ell_2} }{ \omega_{\ell_1} } } \frac{ Y_{\ell_2 m_2}^* (\n') \del_i Y_{\ell_1 m_1} (\n) }{ e^{ (\tau - \tau') \omega_{\ell_2}} } \\
         - \sqrt{ \frac{ \omega_{ \ell_1} }{ \omega_{ \ell_2} } } \frac{ \del_i Y_{\ell_2 m_2}^* (\n) Y_{\ell_1 m_1} (\n') }{ e^{- (\tau - \tau') \omega_{\ell_1}} } + \bigg( \sqrt{ \frac{ \omega_{\ell_2} }{ \omega_{\ell_1} } }   Y_{\ell_2 m_2}^* (\n) \, \del_i Y_{\ell_1 m_1} (\n) - (1 \leftrightarrow 2)^* \bigg) \, \Bigg\} .
\end{multlined}
\end{align}
$T_{\tau i}$ vanishes on the ground state and hence the quadratic contributions only come from the linear terms and the quadratic term of $T_{\tau i}$. The combination $J_i T_{\tau\tau}$ instead leads to
\begin{align}
\begin{multlined}[][.9\linewidth]
        \braket*{\Vpp* J_i (\tau, \n) T_{\tau\tau} (\tau', \n') \Vpp} = - i \Anew[\Delta_Q + R \omega_{ \ell_1}][\Delta_Q + R \omega_{ \ell_2}][\tau] \frac{ Q }{2 \Omega_D R^{2D-1}} \\
        \Bigg[ \delta_{\ell_2 \ell_1} \delta_{m_2 m_1} \del_i \sum_{\ell} e^{- |\tau -\tau'| \omega_{\ell} } \frac{ (D+2\ell -2)}{ (D-2) \Omega_D} C^{ \frac{D }{2} -1}_{\ell} (\n \cdot \n') + \sqrt{ \frac{ \omega_{\ell_2} }{ \omega_{\ell_1} } }  \frac{ Y_{\ell_2 m_2}^* (\n') \, \del_i Y_{\ell_1 m_1} (\n) }{ e^{ (\tau -\tau') \omega_{\ell_2}} } \\
        - \sqrt{ \frac{ \omega_{\ell_1} }{ \omega_{ \ell_2} }} \frac{ \del_i Y_{\ell_2 m_2}^* (\n) \, Y_{\ell_1 m_1} (\n') }{ e^{ -(\tau -\tau') \omega_{ \ell_1}} } + \frac{ (D-2)}{ D} \bigg( \sqrt{ \frac{ \omega_{ \ell_2} }{ \omega_{ \ell_1} } } Y_{\ell_2 m_2}^* (\n) \del_i Y_{\ell_1 m_1} (\n) - (1 \leftrightarrow 2)^* \bigg) \Bigg] .
\end{multlined}
\end{align}
This correlator is related to the previous one.
From the expansions in Eq.~\eqref{eq:ExpansionsOfTsAndQsInTheField} it is clear that this has to be the case.
\begin{align}\label{eq:TtautauJtauCorrelator}
\begin{multlined}[][.9\linewidth]
        \braket*{\Vpp* T_{\tau \tau}(\tau, \n) J_\tau(\tau', \n') \Vpp } = i \Anew[\Delta_Q + R \omega_{\ell_{1}}][\Delta_Q + R \omega_{ \ell_{2}} ][\tau] \frac{ Q (D-1) }{2 \Omega_D R^{2D -1}} R \sqrt{\omega_{ \ell_2} \omega_{ \ell_1} } \\
        \Bigg\{ \, \bigg[ \frac{ Y_{\ell_2 m_2}^* (\n') Y_{\ell_1 m_1} (\n) }{ e^{ (\tau-\tau') \omega_{\ell_2}} }+ \frac{ Y_{\ell_2 m_2}^* (\n) Y_{\ell_1 m_1} (\n') }{ e^{- (\tau-\tau') \omega_{\ell_1}} } \bigg] + \sum_{\ell} \frac{ R \omega_\ell \, e^{- |\tau - \tau'| \omega_\ell}}{  R \sqrt{\omega_{ \ell_2} \omega_{ \ell_1} }}   \frac{ (D+ 2\ell -2) }{ (D-2) \Omega_D} C_\ell^{ \frac{ D}{2} -1} (\n \cdot \n') \\
        +  \frac{ (D-2)}{D} \Big[ Y_{\ell_2 m_2}^* (\n') Y_{\ell_1 m_1} (\n') - \frac{ \partial_i Y_{\ell_2 m_2}^* (\n') \, \partial_i Y_{\ell_1 m_1} (\n') }{ R^2 (D-1) \omega_{\ell_2} \omega_{\ell_1} } \Big] 
        +\frac{2 }{ (D-1)} \Bigg[ \frac{1 }{\Omega_D} \Big( \frac{ \Delta_0 + \Delta_1 }{R\sqrt{ \omega_{ \ell_1} \omega_{ \ell_2} }} \Big) \delta_{\ell_2 \ell_1} \delta_{m_2 m_1} \\
        + \frac{ 1}{2} \bigg( (D-1) Y_{ \ell_2 m_2}^* (\n) Y_{\ell_1 m_1} (\n) 
        - \frac{(D-3) }{ (D-1)} \frac{ \del_i Y_{\ell_2 m_2}^* (\n) \del_i Y_{\ell_1 m_1} (\n) }{ R^2 { \omega_{ \ell_1} \omega_{\ell_2}}} \bigg) \Bigg]  \Bigg\} .
\end{multlined}
\end{align}
Here, the quadratic term from $J_\tau$ vanishes after integration over $\n'$, whereas the quadratic term from $T_{\tau\tau}$ remains finite after integration over $\n$. This is so because it has to correct the energy by $R\omega_{\ell_2}$, in accordance with the Ward identities \eqref{eq:WardIdentities}.

\begin{align}
\begin{multlined}[][.9\linewidth]
        \braket*{\Vpp* T_{\tau i}(\tau, \n) J_j (\tau', \n') \Vpp } = -i \Anew[\Delta_Q + R \omega_{\ell_{1}}][\Delta_Q + R \omega_{ \ell_{2}} ][\tau] \frac{ Q  }{2 \Omega_D R^{2D-1} } \frac{ 1}{ (D-1)} \\
        \Bigg\{ \partial_i \partial'_j \sum_{\ell} \frac{ e^{- |\tau -\tau'| \omega_\ell} }{ R \omega_{\ell} } \frac{ (D+2\ell -2) }{ (D-2) \Omega_D } C_\ell^{ \frac{D}{2} -1} (\n\cdot \n') \, \delta_{\ell_2 \ell_1} \delta_{m_2 m_1} + \frac{ \partial'_j Y_{\ell_2 m_2}^* (\n') \partial_i Y_{\ell_1 m_1} (\n) }{ R \sqrt{ \omega_{ \ell_1} \omega_{ \ell_2}} \, e^{(\tau -\tau') \omega_{ \ell_2}} }\\
        + \frac{ \partial_i Y_{\ell_2 m_2}^* (\n) \partial'_j Y_{\ell_1 m_1} (\n') }{ R \sqrt{ \omega_{ \ell_1} \omega_{ \ell_2}} \, e^{-(\tau -\tau') \omega_{ \ell_1}} } \Bigg\} .
\end{multlined}
\end{align}
Both $T_{\tau i}$ and $J_i$ vanish on the ground state and hence the only quadratic contribution comes from the two linear terms.
\begin{align}
\begin{multlined}[][.9\linewidth]
        \braket*{\Vpp* T_{ij}(\tau, \n) J_\tau (\tau', \n') \Vpp } = - i \Anew[\Delta_Q + R \omega_{ \ell_{1}} ][ \Delta_Q + R \omega_{ \ell_{2}} ][\tau] \frac{ Q }{ \Omega_D R^{2D-1} }  \\
        \Bigg\{ h_{ij} \bigg[ \frac{ \left( \Delta_0 + \Delta_1 \right) }{ \Omega_D (D-1)} \delta_{ \ell_2 \ell_1} \delta_{m_2 m_1} + \frac{ 1}{2} \sum_{\ell}  e^{- |\tau -\tau'| \omega_\ell}  R \omega_{\ell}  \frac{ (D+2\ell -2) }{ (D-2) \Omega_D } C_\ell^{\frac{D}{2} -1} (\n\cdot \n') \, \delta_{\ell_2 \ell_1} \delta_{m_2 m_1} \\
        + \frac{ R \sqrt{\omega_{\ell_2} \omega_{\ell_2}} }{ 2} \bigg( \frac{Y_{\ell_2 m_2}^* (\n') Y_{\ell_1 m_1} (\n)}{e^{(\tau - \tau') \omega_{\ell_2} } } + \frac{ Y_{\ell_2 m_2}^* (\n) Y_{\ell_1 m_1} (\n') }{ e^{- (\tau - \tau') \omega_{\ell_1} } } + \Big[ 1 + \frac{ (D-2)}{D} \Big] Y_{ \ell_2 m_2}^* (\n) Y_{\ell_1 m_1} (\n) \\
        - \Big[ 1 + \frac{ (D-2)}{D} \Big] \frac{ \del_i Y_{ \ell_2 m_2}^* (\n) \del_i  Y_{\ell_1 m_1} (\n) }{ (D-1) \, R^2 \omega_{ \ell_1 } \omega_{ \ell_2 } } \, \bigg)  \bigg] \, + \frac{ R\sqrt{ \omega_{ \ell_1} \omega_{ \ell_2} } }{ (D-1)} \frac{ \del_{ ( i} Y_{ \ell_2 m_2 }^* (\n)  \del_{j) }  Y_{\ell_1  m_1} (\n) }{ R^2 \omega_{ \ell_1} \omega_{ \ell_2}}  \Bigg\} .
\end{multlined}
\end{align}
This correlator is related to the $TJ$ correlator in Eq.~\eqref{eq:TtautauJtauCorrelator} due to the fact that $h^{ij} T_{ij} = -T_{\tau\tau}$, which holds by virtue of conformal invariance.
\begin{align}
\begin{multlined}
        \braket*{\Vpp* J_i (\tau, \n) T_{jk} (\tau', \n') \Vpp } = i \Anew[ \Delta_Q + R \omega_{ \ell_{1}} ][ \Delta_Q + R \omega_{ \ell_{2}} ][ \tau]  \frac{ Q }{2 \Omega_D R^{2D-1} } \frac{ h_{jk} }{ (D-1)}  \\
        \Bigg\{\partial_i \sum_{\ell} e^{- |\tau-\tau'| \omega_{\ell}} \frac{ (D+ 2\ell -2)}{ (D-2) \Omega_D} C_{\ell}^{\frac{D}{2} -1} (\n \cdot \n') \, \delta_{\ell_2 \ell_1} \delta_{m_2 m_1} + \sqrt{ \frac{ \omega_{\ell_2} }{ \omega_{\ell_1} } } \frac{ Y_{\ell_2 m_2}^* (\n') \, \partial_i Y_{\ell_1 m_1} (\n) }{ e^{ (\tau-\tau') \omega_{\ell_2}} }  \\
         - \sqrt{ \frac{ \omega_{\ell_1} }{ \omega_{\ell_2} } } \frac{ Y_{\ell_1 m_1} (\n') \, \partial_i Y_{\ell_2 m_2}^*  (\n) }{ e^{- (\tau-\tau') \omega_{\ell_1}} } + \frac{ (D-2) }{ D} \bigg[ \sqrt{ \frac{ \omega_{ \ell_2} }{ \omega_{ \ell_1} } } Y_{\ell_2 m_2}^* (\n) \del_i Y_{\ell_1 m_1} (\n) - (1 \leftrightarrow 2 )^* \, \bigg] \, \Bigg\} .
\end{multlined}
\end{align}
This correlator is proportional to $h_{jk}$ since the quadratic term in the expansion of $T_{jk}$ only appears at cubic order in the correlator. This is no longer the case once one includes higher-order corrections.

\medskip

In the special case $\ell=0$ the $TJ$ correlators simplify significantly:
\begin{align}
        &
        \braket{\Opp*   T_{\tau i}(\tau, \n) J_\tau(\tau', \n')  \Opp} = -i \frac{ Q \Anew }{2 \Omega_D^2 R^{2D-1} }  \del_i \sum_{\ell} e^{-(\tau -\tau') \omega_{\ell} } \frac{ (D+2\ell -2)}{D-2} C^{ \frac{D }{2} -1}_{\ell} (\n \cdot \n')  , 
       \\
        &
        \braket{\Opp*  J_i (\tau, \n) T_{\tau \tau} (\tau', \n') \Opp} = -i \frac{ Q \Anew }{2 \Omega_D^2 R^{2D-1} } \del_i \sum_{\ell} e^{-(\tau -\tau') \omega_{\ell} } \frac{ (D+2\ell -2)}{D-2} C^{ \frac{D }{2} -1}_{\ell} (\n \cdot \n')  , 
       \\
        &\begin{multlined}[][.9\linewidth]
        \braket*{\Vpp* T_{\tau \tau}(\tau, \n) J_\tau(\tau', \n') \Vpp } = i \Anew \frac{ Q (D-1) }{2 \Omega_D R^{2D -1}} \\
        \times \Bigg\{ \, \sum_{\ell}  R \omega_\ell \, e^{-(\tau - \tau') \omega_\ell}  \frac{ (D+ 2\ell -2) }{D-2} C_\ell^{ \frac{ D}{2} -1} (\n \cdot \n') + \frac{ 2 \big( \Delta_0 + \Delta_1 \big) }{ (D-1) \Omega_D } \delta_{ \ell_2 \ell_1} \delta_{m_2 m_1}  \Bigg\} ,
    \end{multlined}\\
    &
        \braket*{\Vpp* T_{\tau i}(\tau, \n) J_j (\tau', \n') \Vpp } = - i \frac{ Q \Anew  }{2(D-1) \Omega_D^2 R^{2D-1} } \partial_i \partial'_j \sum_{\ell} \frac{ (D+2\ell -2) }{D-2 } \frac{ C_\ell^{ \frac{D}{2} -1} (\n\cdot \n') }{ e^{ (\tau -\tau') \omega_\ell} R \omega_{\ell} }  ,
   \\
    &\begin{multlined}[][.9\linewidth]
        \braket*{\Vpp* T_{ij}(\tau, \n) J_\tau (\tau', \n') \Vpp } = - i \frac{ Q \Anew }{ \Omega_D^2 R^{2D-1} } h_{ij} \\
        \times \Bigg\{ \frac{ \left( \Delta_0 + \Delta_1 \right) }{ D-1}  + \frac{ 1}{2} \sum_{\ell}  e^{- (\tau -\tau') \omega_\ell}  R \omega_{\ell}  \frac{ (D+2\ell -2) }{D-2 } C_\ell^{\frac{D}{2} -1} (\n\cdot \n')   \Bigg\} ,
    \end{multlined}\\
    &
        \braket*{\Vpp* J_i (\tau, \n) T_{jk} (\tau', \n') \Vpp } = i  \frac{ Q \Anew }{2 \Omega_D^2 R^{2D-1} } \frac{ h_{jk} }{ (D-1)} \partial_i \sum_{\ell}  \frac{ (D+ 2\ell -2)}{D-2} \frac{ C_{\ell}^{\frac{D}{2} -1} (\n \cdot \n') }{ e^{ (\tau -\tau') \omega_{\ell}} }  .
\end{align}
The correlators $J_i T_{\tau \tau}$ and $T_{\tau i} J_\tau$ in the special case $\ell_1 = \ell_2 = 0$ has appeared in the macroscopic limit $R \rightarrow \infty$ in~\cite{Komargodski:2021qp}.

\section{Heavy--light--heavy correlators}\label{sec:HLH}

The \ac{eft} for the large-$Q$ sector of the $O(2)$ \ac{cft} can also be used to compute correlators with insertion of small charge (spinning) primaries $\mathscr{O}^q$ with $q \ll Q$.
 In this case, the small-charge operators act as probes around the same large-charge saddle studied before.
 For the sake of completeness, we review the computation for some of these correlators which have appeared in~\cite{Cuomo:2020thesis,Jafferis:2017zna,Monin:2016jmo}.

The starting observation is that in the \ac{eft} every operator must be written in terms of the Goldstone field.
Matching the quantum numbers, an operator of charge \(q\), dimension \(\Delta\) that transforms in a representation of spin \(\ell\) which at leading order in the charge $Q$ must take the form
\footnote{The next term in the large-charge expansion must take the form \(k^{(2)}_{\Delta,q} (\del \chi)^{\Delta-2} \left( \mathcal{R} + \dots \right) e^{iq\chi} \) where we have neglected higher-order terms necessary to obtain a Weyl-invariant quantity.}
\begin{equation}
	\Opp[q][\Delta][\ell m] = k^{(1)}_{\Delta,\ell,q} \Proj\indices{^{\nu_1 \dots \nu_\ell}_{\ell m}} \del_{\nu_1} \chi \dots \del_{\nu_\ell} \chi   \left( \del \chi \right)^{\Delta-\ell} e^{iq\chi} + \dots ,
\end{equation}
and for simplicity
\begin{align}
  \Opp[q][\Delta] &= \Opp[q][\Delta][0 0] , & k^{(1)}_{\Delta,q} &= k^{(1)}_{\Delta,0,q} ,
\end{align}
where  $k^{(1)}_{\Delta,\ell,q}$ is a charge-independent Wilsonian coefficient which cannot be determined in the \ac{eft}. \(\Proj^{\nu_1 \dots \nu_\ell}_{\ell m}{}\) is the change of basis to spherical tensors given in Eq.~\eqref{eq:ProjectorToSphericalBasis}.

\subsection[\texorpdfstring%
{$\braket{ \Vpp[-Q-q][\ell_2 m_2]  \Opp[q][\Delta] \Vpp }$}%
{<QqQ>}]%
{The $\braket{ \Vpp[-Q-q][\ell_2 m_2]  \Opp[q][\Delta] \Vpp }$ correlator}%
\label{sec:VqV}

We first want to compute the correlator
\begin{equation}
    \braket{ \Vpp[-Q-q][\ell_2 m_2] (\tau_2)  \Opp[q][\Delta](\tau_c, \n_c) \Vpp (\tau_1) },
\end{equation}
where
\begin{equation}
    \left[ \Opp*[q][\Delta] (\tau , \n) \right]^\dagger = \Opp[q][\Delta] (-\tau , \n) .
\end{equation}
From general symmetry considerations the classical (homogeneous) contribution to the three-point function must have the form
\begin{equation}
        \braket{ \Vpp[-Q-q][\ell_2 m_2] (\tau_2) \,  \Opp[q][\Delta](\tau_c, \n_c) \, \Vpp (\tau_1) } = \mathcal{C}^\Delta_{Q +q, q, Q} \delta_{\ell_1\ell_2} \delta_{m_1 m_2}   e^{- \omega_{ \ell_2} (\tau_2 - \tau_1)} e^{- \Delta_{Q +q} \frac{ (\tau_2 -\tau_c) }{R} } e^{- \Delta_{ Q} \frac{ ( \tau_c -\tau_1) }{R} } ,
\end{equation}
which is consistent with the general formula for three-point functions given in Eq.~\eqref{eq:limit3pointCFT}.
By dimensional analysis we can extract a factor of $R^\Delta$ from the three-point coefficient, which is coming from the insertion of $\Opp[q][\Delta]$:
\begin{equation}
R^{-\Delta} \, \tilde{ \mathcal{C }}^\Delta_{ Q +q ,q ,Q} = \mathcal{C }^\Delta_{ Q +q ,q ,Q} .
\end{equation}
We want to reproduce this result from a semiclassical large-charge expansion of the path integral.

\medskip

The difference with respect to the saddle that we have studied so far is that the insertion of an operator of charge $q$ introduces a source term in the action and changes the \ac{eom} that now read
\begin{align}
    \nabla_\mu J^\mu &= \nabla_\mu \frac{\delta S}{\delta (\del_\mu \chi )} = \frac{i q   \delta (\tau - \tau_c) \delta (\n - \n_c) }{ \sqrt{g} }, &\begin{dcases} J^\mu(\tau_1= -\infty, \n_1) = \frac{ \delta^{\mu}_0 Q }{ R^{D-1} \Omega} \\
    J^\mu(\tau_2= +\infty, \n_2) = \frac{ \delta^{\mu}_0 ( Q + q) }{ R^{D-1} \Omega}
    \end{dcases} .
\end{align}
On general grounds the solution of this partial differential equation can be written as a sum of the homogeneous solution $\pi$ (in the sense of \textsc{pde}, \emph{i.e.} for $q=0$) and a particular solution $q p(\tau, \n)$.
Based on the form of the equation the particular solution is basically identical to the propagator, except that the boundary conditions will alter the form of the $l=0$ term:
\begin{equation}
\begin{aligned}
        \chi(\tau, \n) ={}& -i \mu \tau + \pi (\tau, \n) + q p(\tau, \n) , %
    \end{aligned}
\end{equation}
where
\begin{equation}
    \mu \propto Q^{\frac{1}{(D-1)}} .
\end{equation}
At $\tau_1 \to -\infty$ and $\tau_2 \to \infty$, the effect of the source appears only in the normalization of the fields, which take the same form as in the unperturbed case of Eq.~\eqref{eq:FieldDecompositionInLadders}.
So we can identify
\begin{align}
	\bra{0}  \Vpp[-Q-q][\ell_2 m_2](\infty) & = \bra*{\myatop{Q+q}{\ell_2 m_2}} = \bra{Q+q}  a_{\ell_2 m_2}\\
	\Vpp[Q][\ell_1 m_1](-\infty)\ket{0} &= \ket*{\myatop{Q}{\ell_1 m_1}} =  a^\dagger_{\ell_1 m_1}\ket{Q}.
\end{align}
We can now compute the correlator (in the limit of large separations), at least to leading order:
\begin{equation}
    \begin{aligned}
    \braket{\Vpp[-Q-q][\ell_2 m_2] (\tau_2) \Opp[q][\Delta](\tau_c, \n_c) \Vpp (\tau_1)} &= \bra{Q+q}   a_{\ell_2 m_2} (\tau_2) \Opp[q][\Delta](\tau_c, \n_c)    a_{\ell_2 m_2}^{\dagger} (\tau_1) \ket{Q}  \\
    = C^{(1)}_{\Delta,q} \mu^\Delta  e^{\mu q \tau_c} &\bra{Q+q}   a_{\ell_2 m_2} (\tau_2)  e^{iq   \pi (\tau_c, \n_c) + iq^2   p(\tau_c, \n_c) }    a_{\ell_2 m_2}^{\dagger} (\tau_1) \ket{Q} + \dots \\
        & = (R\mu)^\Delta \frac{ C^{(1)}_{\Delta,q} }{ R^\Delta } \delta_{\ell_2 \ell_1} \delta_{m_2 m_1} \Anew[\Delta_Q \, + \omega_{\ell}][\Delta_{Q + q } \, + \omega_{\ell} ][\tau_c] + \dots 
    \end{aligned}
\end{equation}
More in general if we insert an operator in a representation of spin \(\ell\) we find
\begin{multline}    
        \braket{Q +q |   a_{\ell_2 m_2} \Opp[q][\Delta][\ell m] (\tau_c, \n_c)   a_{\ell_1 m_1 }^\dagger |Q} = \frac{ (R\mu)^\Delta C^{(1)}_{\Delta,\ell,q} }{ R^\Delta } \braket{ \ell_2 m_2  ; \ell, m | \ell_1 m_1 } \\
        \times  \Anew[\Delta_Q + \omega_{ \ell_1}][\Delta_{Q + q }+ \omega_{ \ell_2}][\tau_c]  + \dots
\end{multline}
where \( \braket{\ell_2 m_2 ; \ell, m | \ell_1 m_1 } \) is the appropriate Clebsch--Gordan coefficient.
The quantum corrections to the above expressions are given in~\cite{Jafferis:2017zna}.
\bigskip

If we specialize the above general result to $\ell_1=\ell_2=0$, we have the correlator of \( \Opp[q][\Delta][\ell m] \) inserted between two large-charge scalars.
By charge conservation one of the scalars must have charge \(-Q -q\), and by rotational invariance only insertions of \(\ell = 0\) operators can lead to non-vanishing results:%
\begin{equation}
\begin{aligned}
   \braket{Q+q | \Opp[q][\Delta][\ell m] (\tau_c , \n_c) | Q} \propto  \delta_{\ell,0} \, \mu^\Delta e^{- \Delta_{Q} \frac{ (\tau_{2} -\tau_{1}) }{R} } e^{ \mu q ( \tau_c -\tau_2)} = \mu^\Delta \Anew[\Delta_Q ][\Delta_{Q+q} ][\tau_c] \delta_{\ell 0}.
\end{aligned}
\end{equation}
This correlator has been presented originally in~\cite{Monin:2016jmo,Cuomo:2020rgt}.
The final result is that the three-point coefficient is given by
\begin{equation}
   \tilde{\mathcal{C }}^\Delta_{ Q +q ,q ,Q} =  C^{(1)}_{\Delta,q} (R\mu )^\Delta \left[ 1 - \frac{ \frac{q^2}{2 } \sum_{ \ell ,m} \frac{ Y_{\ell m}^* (\n_c) Y_{\ell m} (\n_c) }{ R \omega_\ell } }{ c_1 D (D-1) (R\mu)^{D-2} } + \dots \right] + \dots 
\end{equation}
For $D=3$ the first correction is only suppressed by a factor of $\mu \sim \sqrt{Q}$ and is dominant.
It is computable for example via dimensional regularization, or in zeta function regularization using the fact that
\begin{equation}
   \sum_{\ell,m} \frac{ Y_{\ell m}^* (\n_c) Y_{\ell m} (\n_c) }{ \omega_\ell } = \frac{1}{\Omega_D} \sum_{\ell} \frac{ M_{\ell} }{ \omega_{ \ell}} = \frac{\sqrt{D-1} R \zeta_{S^{D-1}}(1/2)}{\Omega_D}.    
\end{equation}
In $D=3$, the \ac{ope} coefficient reads~\cite{Cuomo:2020rgt}
\begin{equation}
\tilde{\mathcal{C }}^\Delta_{ Q +q ,q ,Q} \propto 
\left(Q \right)^{\frac\Delta2} \left[ 1 + 0.0164523 \times \frac{ q^2 \sqrt{12\pi} }{\sqrt{c_1 Q} } + \dots \right] + \dots .
\end{equation}

\subsection[\texorpdfstring%
{$\braket{\Vpp* \Opp[-q][\Delta] \Opp[q][\Delta] \Vpp }$}%
{<QqqQ>}]%
{The $\braket{\Vpp* \Opp[-q][\Delta] \Opp[q][\Delta] \Vpp }$ correlator}%
\label{sec:VqqV}

As a slight generalization we compute the four-point correlator
\begin{equation}
    \braket{ \Vpp[-Q-q_d-q_c][\ell_2 m_2] (\tau_2) \, \Opp[q_d ][\Delta_d] (\tau_d, \n_d) \, \Opp[q_c][\Delta_c] (\tau_c, \n_c) \, \Vpp[ Q][\ell_1 m_1] (\tau_1) } ,
\end{equation}
with $q_d \sim q_c \ll Q$. Computations of four-point functions of this form have first appeared in~\cite{Jafferis:2017zna}.
The \ac{eom} are closely related to the ones of the three-point function with a single scalar insertion:
\begin{align}
    \nabla_\mu J^\mu &= \frac{i q_d \delta (\tau - \tau_d) \delta (\n - \n_d) }{ \sqrt{g}} + \frac{i q_c  \delta (\tau - \tau_c) \delta (\n - \n_c) }{ \sqrt{g}} ,
     &\begin{cases}
     J^\mu(\infty, \n) = \frac{ \delta^{\mu}_0 (Q + q_d + q_c) }{ R^{D-1} \Omega} , \\ J^\mu(-\infty, \n) = \frac{ \delta^{\mu}_0 Q }{ R^{D-1} \Omega} .
     \end{cases}
\end{align}
To leading order, the correlator is 
\begin{equation}
    \begin{multlined}
    \braket{ \Vpp[-Q- q_d- q_c][\ell_2 m_2]  \, \Opp[q_d][\Delta_d]  \, \Opp[q_c][\Delta_c]  \, \Vpp[Q][\ell_1 m_1] } %
    = (R\mu)^{\Delta_d +\Delta_c} \frac{ C^{(1)}_{\Delta_d,q_d} C^{(1)}_{\Delta_c,q_c} }{ R^{ \Delta_d +\Delta_c} }  \delta_{\ell_2 \ell_1} \delta_{m_2 m_1} e^{ - (\tau_2-\tau_1) \omega_{ \ell_2} } \\
    \times e^{- \Delta_{Q +q_d+q_c} \frac{ (\tau_2 -\tau_c) }{R} - \Delta_{Q+q_c} \frac{ (\tau_d -\tau_c) }{R} - \Delta_{Q } \frac{ (\tau_c -\tau_1) }{R} } + \dots
    \end{multlined}
\end{equation}
Setting $q=q_c=-q_d$ the correlator becomes
\begin{equation}
    \begin{multlined}
    \braket{ \Vpp[-Q][\ell_2 m_2]  \,\Opp[-q][\Delta] \, \Opp[q ][\Delta]\, \Vpp[Q][\ell_1 m_1] } = (R\mu)^{2 \Delta} \frac{ |C^{(1)}_{\Delta ,q} |^2 }{R^{2\Delta} } e^{- \Delta_{Q} \frac{ (\tau_2 - \tau_1) }{R} - q \frac{ \del \Delta_Q }{\del Q }  \frac{ (\tau_d - \tau_c) }{R} - (\tau_2 -\tau_1) \omega_{ \ell_2} } \delta_{ \ell_2 \ell_1} \delta_{m_2 m_1} + \dots 
    \end{multlined}
\end{equation}
The next-to-leading order of this correlator has appeared in~\cite{Jafferis:2017zna}.

\newpage

\section{Conclusions}\label{sec:Conclusions}

In this note, we have systematically collected three- and four-point correlators of a \ac{cft} with a global $O(2)$ symmetry using the large-charge expansion. We have studied in particular correlators with current insertions of $J$ and $T$ sandwiched between either the scalar large-charge ground state or higher phonon states with spin. 
The correlators involving spinning states constitute the main result of this article as in the literature, chiefly correlators involving the scalar ground state have appeared to date. We mostly focus on one-phonon excitations but also give selected correlators involving two phonons. The form of the correlators generalizes readily to higher-phonon excitations.

The general structure of our correlators contains contributions with positive $Q$-scaling coming from the tree-level \ac{eft} Lagrangian plus quantum corrections starting at order $Q^0$ which are independent of the Wilsonian coefficients in odd dimensions. A non-trivial position dependence in our correlators must always be due to the quantum corrections since the ground state is homogeneous.
While we have used the large-charge \ac{eft} description in our computations and have not made use directly of conformal invariance, the correlators have the form of conformal correlators on the cylinder at large separation, see Appendix~\ref{sec:constraints}.

We have also studied higher-order loop corrections to the scaling dimension of $\Op^Q$ and found that in even dimensions, the $l$-loop corrections contain terms of the form $Q^{- \frac{(l-1)D}{D-1}} \left( \alpha_0 + \alpha_1 \log Q + ... + \alpha_l (\log Q)^l \right)$. The appearance of the log-terms distinguishes the loop corrections from the tree-level contributions.

We have furthermore collected H--L--$\dots$--H correlators in which operators with a parametrically small charge $q$ are inserted between large-charge operators.
While in the cylinder picture it is easily argued that H--L--$\dots$--H correlators lie within the validity of the \ac{eft}, the same is not \emph{a priori} clear for correlators involving more than two heavy operators.
In the former case, the \ac{eft} correlators in the cylinder picture are constructed by having two heavy states at the in- and out spatial slices, with the possible addition of conserved currents or small charge operators with much smaller scaling dimensions, whose insertions can be thought of as small perturbations of the heavy states. The \ac{eft} might in principle break down very close to the insertions of the small charge operators, but actually accurately captures the behavior.
As discussed in~\cite{Cuomo:2021ygt}, if one instead includes another heavy operator in between the in- and out states, this state can then no longer be considered as small perturbations of the heavy states.
At first sight, the \ac{eft} description seems no longer justified, at least in the cylinder frame.
Each heavy operator with $Q \gg 1$ creates a superfluid state around its insertion point. %
By charge conservation, the three-point function describes the transition between three different superfluids and is now associated to a new classical trajectory in the path integral.
However, the key observation is that also in this case the radial field is locally gapped around all the superfluid states and hence its corresponding modes are not excited by the new classical profile.
It is thus possible to compute $n$-point functions of large charge operators with H--H--H insertions within the \ac{eft}.
The result for the three-point function was obtained using a numerical solution in~\cite{Cuomo:2021ygt}.
In this note, we have not dealt with this type of correlator, but it might be of interest to study such correlators involving phonon states.

\bigskip

The results we have presented here hold for the $O(2)$ model in $D$ dimensions at large charge or the homogeneous $O(2)$ sector of a \ac{cft} with a larger global symmetry group. 
Once we want to discuss the full non-Abelian structure, things become more complicated.
The first observation is that the correct quantity to fix is not a set of charges, but a representation.
The immediate generalization of the \(O(2)\) case, \emph{i.e.} studying the homogeneous sector~\cite{Alvarez-Gaume:2016vff} corresponds to fixing the completely symmetric representation. Other representations can be obtained in two ways:
by exciting type-II Goldstones~\cite{Nielsen:1975hm,Watanabe:2013uya,Watanabe:2019xul} that are charged under the global symmetry or by starting from an inhomogeneous ground state corresponding to a different saddle point.
The two approaches must give the same result in the appropriate limit, but have their own technical complications.
Type-II Goldstones contribute at order \(1/\mu\)~\cite{Alvarez-Gaume:2016vff}.
For this reason they do not play a role in the computations in the present work, but in order to be studied consistently they require the addition of new subleading terms to the \ac{eft}~\cite{Gaume:2020bmp}.
The inhomogeneous saddle is to date only known for the simple case of the \(O(4)\) model and an analytic expression is available only in a special limit~\cite{Banerjee:2019jpw}.
Moreover, since it breaks the \(SO(D)\) rotational invariance that we have used intensively in our present computations, one expects that computing correlation functions using both the tree-level and the quantum corrections will be more technically challenging.
We leave this class of problems for future investigation.

\subsection*{Acknowledgments}

{\dosserif

We would like to thank Gabriel Cuomo, Simeon Hellerman, Vito Pellizzani and Donald Youmans for enlightening discussions.

The work of  N.D., R.M. and S.R. is supported by the Swiss National Science Foundation under grant number 200021 192137.
D.O. acknowledges partial support by the NCCR 51NF40--141869 ``The Mathematics of Physics'' (SwissMAP).
}

\newpage

\appendix

\section{Hyperspherical harmonics and their properties}
\label{sec:Ylm-identities}

In this appendix we collect some useful formulas related to spherical harmonics in \(D\) dimensions, see also~\cite{Avery:2018}.

The hyperspherical harmonic \(Y_{\ell m}\) is an eigenfunction of the Laplacian on \(S^{D-1}\)
\begin{equation}
  -\Laplacian_{S^{D-1}} Y_{\ell m} (\n) = \ell ( \ell + D - 2 ) Y_{\ell m} (\n) ,
\end{equation}
where \(\ell = 0, 1, \dots \) and \(m \) is a vector of \( D - 2\) components satisfying
\begin{equation}
l \ge m_1 \ge m_2 \ge \dots \ge m_{D-3} \ge \abs{m_{D-2}}.	
\end{equation}
The lowest component \(m_{D-2}\) is associated to the standard \(SO(3)\) quantum number.
This is the only component that can have a negative sign.
We denote with \(m^*\) the vector with the sign of \(m_{D-2}\) flipped.
This appears in the conjugation property
\begin{equation}
	Y_{\ell m }^* = (-1)^{m_{D-2}} Y_{\ell m^*}  .
\end{equation}
The eigenvalue does not depend on \(m\), and it appears with multiplicity
\begin{equation}\label{eq:laplacian_degenerancy}
  M_{\ell} = \frac{(D + 2 \ell - 2) \Gamma(D + \ell - 2)}{\Gamma(\ell + 1) \Gamma(D - 1)} \, .
\end{equation}

Since the Laplacian is self-adjoint, the \(Y_{\ell m}\) form an orthonormal basis for \(L^2(S^{D - 1})\)
\begin{equation}
  \scalar{ Y_{\ell m}, Y_{\ell' m'}} = \int_{S^{D -1}} \dd{\Omega} Y_{\ell m}(\n) Y^{*}_{\ell'  m' }(\n) = \delta_{\ell \ell'} \delta_{m m'} \, .
\end{equation}
We defined the rescaled versions of the volume element on the sphere and of the eigenvalues of the Laplacian
\begin{align}
  \dd{S} &= R^{D-1} \dd{\Omega}, & \omega_{\ell}^2 &= \frac{\ell (\ell + D - 2)}{(D-1) R^2}. 
\end{align}
Some useful identities obtained by summing over the indices \(m\) are
\begin{align}\label{eq:Y_orthogonality}
  \sum_m Y_{\ell m}(\n) Y^{*}_{\ell m}(\n) &= \frac{M_{\ell}}{\Omega_D} \, ,\\
  \sum_m Y_{\ell m}(\n) \del_i Y^{*}_{\ell m}(\n) &= 0 \, ,\\
  \sum_m \del_i Y_{\ell m}(\n) \del_j Y^{*}_{\ell m}(\n) &= \frac{M_{\ell}}{\Omega_D} ( R \omega_{\ell} )^2 h_{ij}(\n) \, ,
\end{align}
where $\Omega_D = \frac{2 \pi^{D/2}}{\Gamma(D/2)} $ is the volume of the \(D - 1\) sphere. Sums involving the eigenvalues $\omega_\ell$ can be written in term of the sum \eqref{eq:regulated_sphere_sum}
\begin{equation}
  \sum_{\ell, m} \omega_{\ell}^s Y_{\ell m}(\n) Y^{*}_{\ell m}(\n)  = \frac{\Sigma(s)}{\Omega_D R^s}= \frac{\zeta( \sfrac{s}{2} \mid S^{D-1} ) }{(D - 1)^{s/2} R^s \Omega_D} \, ,
\end{equation}
where the $\Lambda$-independent part of $\Sigma(s)$ is related to the sphere zeta-function~\cite{Monin:2016bwf}. 
\begin{equation}
  \zeta(s \mid S^{D-1}) = \Tr*[ (-\Laplacian_{S^{D-1}}{})^s] \, .
\end{equation}

In the special case of \(s =1\) we recover the Casimir energy of a free scalar computed in Appendix~\ref{sec:Casimir_appendix}
\begin{equation}
  \sum_{\ell, m} \omega_{\ell} Y_{\ell m}(\n) Y^{*}_{\ell m}(\n) = \frac{\Sigma(1)}{\Omega_D R} = \frac{2 \Delta_1}{\Omega_D R} \ .
\end{equation}

Similarly, sums featuring open derivative indexes can be computed
\begin{equation}
  \sum_{\ell, m} \omega_{\ell}^s \del_i Y_{\ell m}(\n) \del_j Y^{*}_{\ell m}(\n) = \frac{\Sigma(s+2)}{\Omega_D R^s} h_{ij} =  \frac{\zeta(\sfrac{s}{2} + 1 \mid S^{D - 1})}{(D-1)^{s/2 + 1}  R^s \Omega_D} h_{ij} \, ,
\end{equation}
and in this case, for \(s = -1\)
\begin{equation}
  \sum_{\ell,m} \frac{1}{\omega_{\ell}} \del_i Y_{\ell m}(\n) \del_j Y^{*}_{\ell m}(\n) = \frac{R \Sigma(1)}{\Omega_D} h_{ij} = \frac{2 R \Delta_1}{\Omega_D} h_{ij} \, .
\end{equation}

In loop computations we will make use of properties of the Gegenbauer polynomials, defined from the hyperspherical harmonics as follows
\begin{equation}\label{eq:DefGegenbauerPolynomials}
 C^{D/2-1}_{\ell}(\n\cdot \n') = \frac{(D-2)\Omega_D}{D+2\ell-2} \sum_{m} Y_{\ell m }^* (\n) Y_{\ell m} (\n') .
\end{equation}
Monomials can be decomposed in terms of Gegenbauer polynomials
\begin{equation}
    (\n \cdot \n')^\ell = \frac{ \ell!}{2^\ell} \sum_{s=0}^{\floor{\frac{\ell}{2}}} \frac{ \left( \frac{D}{2} - 1 + \ell -2s\right) \Gamma \left( \frac{D}{2} -1 \right)}{ s! \, \Gamma \left( \frac{D}{2} + \ell -s \right)} C^{D/2 -1}_{\ell-2s} (\n \cdot \n') .
\end{equation}
In addition, the Gegenbauer polynomials satisfy an addition property of the form
\begin{equation}\label{eq:Gegenbauer_addition}
C_{\ell_a}^{D/2-1}(\n\cdot \n') C_{\ell_b}^{D/2-1}(\n\cdot \n') =\sum_{k=0}^{\min(\ell_a, \ell_b)} \braket{k| \ell_a \ell_b} C^{D/2-1}_{\ell_a + \ell_b -2k}(\n \cdot \n') ,
\end{equation}
where the coefficients $\braket{ k | \ell_a \ell_b}$  are given by the expression
\begin{multline}
 \braket{k| \ell_a \ell_b } = \left( \frac{D}{2} -1 - 2 k + \ell_a + \ell_b \right) \frac{\Gamma(\ell_a + \ell_b +1 -2k ) }{\Gamma\left( \frac{D}{2} -1 \right)^2 \Gamma(\ell_a + \ell_b -2k + D-2)} \\
\times \frac{\Gamma\left( \frac{D}{2} + k -1 \right) 
\Gamma(\ell_a + \ell_b - k + D-2) \Gamma\left( \ell_a - k + \frac{D}{2} -1 \right) \Gamma\left( \ell_b - k + \frac{D}{2} -1 \right) }{ \Gamma(k+1)  \Gamma\left( \ell_a + \ell_b -k +\frac{D}{2} \right) \Gamma(\ell_a - k +1 ).\Gamma(\ell_b -k +1) } .
\end{multline}
This is an $SO(D)$ generalization of angular momentum addition in $D=3$.
\section{Constraints from conformal symmetry}
\label{sec:constraints}

Conformal invariance strongly constrains the form of the correlators.
In order to use the state-operator correspondence we have been working in the cylinder frame in the limit of large separation, where two of the insertions are taken to be at $\tau = \pm \infty$.
For spinful operators, it is moreover most convenient to work in the spherical tensor basis.

An object transforming in an irreducible representation of \(SO(D)\) is in the standard basis in Cartesian coordinates represented by a completely symmetric traceless tensor \(T_{\nu_1 \dots \nu_{\ell}}\).
In the spherical basis, the same object is labeled by a pair \(\ell, m\).
To pass from one basis to the other  we use the operator
\(\Proj\indices{^{\nu_1 \dots \nu_\ell}_{\ell m}}{}\), written as the integral on the unit sphere
\begin{equation}\label{eq:ProjectorToSphericalBasis}
  \Proj\indices{^{\nu_1 \dots \nu_\ell}_{\ell m}}{} = k_{D, \ell}  \int \dd{\Omega} n^{\nu_1} \dots n^{\nu_{\ell}} Y^*_{\ell m}(\n)  \, ,
\end{equation}
where \(k_{D,\ell}\) is a normalization obtained by requiring that \(\Proj{}\) squares to one,
\begin{equation}
  \abs{\Proj_{\ell m}{}}^2 = \delta_{\mu_1 \nu_1} \dots \delta_{\mu_{\ell} \nu_{\ell}} \pqty{ \Proj\indices{^{\nu_1 \dots \nu_\ell}_{\ell m}}}^{*} \Proj\indices{^{\mu_1 \dots \mu_\ell}_{\ell m}} = 1,
\end{equation}
which reads
\begin{equation}
  k_{D,\ell} = \sqrt{\frac{2^{\ell} }{ \Omega_D} \frac{ \Gamma \left( \frac{D}{2} + \ell\right) }{ \ell! \,\, \Gamma \left( \frac{D}{2} \right) } } \, .
\end{equation}

The simplest non-trivial example is a vector \(V_{\mu}\) in \(D =3 \), which is mapped to \(V_{1m}\) with components
\begin{equation}
  \begin{pmatrix}
    V_{1,-1} \\
    V_{1,0} \\
    V_{1,1}
  \end{pmatrix}
  = 
  \begin{pmatrix}
    - \frac{1}{\sqrt{2}} \pqty*{ V_1 + i V_2 } \\
    V_3 \\
    \frac{1}{\sqrt{2}} \pqty*{ V_1 - i V_2 }
  \end{pmatrix}.
\end{equation}

\bigskip

The two-point function of two primary operators is non-vanishing only if they have the same dimension \(\Delta\), and they transform in conjugate representations.
On the cylinder, in the limit \(\tau_2 - \tau_1 \gg 1\) we have, up to an arbitrary normalization
\begin{equation}\label{eq:TwoPointCylinder}
  \braket{ \Opp*[q][\Delta][\ell \bar m](\tau_2, \n_2)  \Opp[q][\Delta][\ell m](\tau_1, \n_1) } = e^{-(\tau_2 - \tau_1) \Delta/R}  I^{\ell}_{m \bar m}(\n_2) \coloneq \Anew I^{\ell}_{m \bar m}(\n_2),
\end{equation}
where we have used that in the large-separation limit the unit vector in the direction of the separation between the two insertions is
\begin{equation}
  \n = \frac{x-y}{\abs{x-y}} = \frac{  e^{\tau_2/R} \n_2 - e^{\tau_1/R} \n_1 }{|e^{\tau_2/R}\n_2 - e^{\tau_1/R} \n_1|} \overset{\tau_{2,1} \to \pm \infty}{\longrightarrow} \n_2 .
\end{equation}
Here \(I^{\ell}_{m \bar m}\) is the intertwiner between the two representations.

\bigskip
Similarly, the three-point function of scalar primaries is fixed up to a constant.
From the flat space form
\begin{equation}
    \braket{ \Opp[2](x_2) \Opp[c](x) \Opp[1](x_1) } = \frac{\mathcal{C}_{1c2}}{\abs{x_2 - x}^{2 \Delta_{2 c \mid 1}} \abs{x_2 - x_1}^{2 \Delta_{2 1 \mid c}} \abs{x - x_1}^{2 \Delta_{c 1 \mid 2}}}  ,
\end{equation}
where $\Delta_{ij \mid k} = \Delta_i + \Delta_j - \Delta_k $, we find that in the limit of large separation $\tau_{1,2} \to \mp \infty$ on the cylinder the result does not depend on the scaling dimension of the middle operator \(\Delta_c\)
\begin{equation}\label{eq:limit3pointCFT}
         \braket{ \Opp[2][\Delta_2] \Opp[c][\Delta] \Opp[1][\Delta_1] } \longrightarrow \mathcal{C}_{1c2} e^{- \Delta_2 ( \tau_2 - \tau)} e^{-\Delta_1(\tau - \tau_1)} =  \Anew[\Delta_1][\Delta_2][\tau] \mathcal{C}_{1c2}.
\end{equation}
The correlator becomes independent of \(\tau\) in the special case where \(\Delta_1 = \Delta_2  = \Delta \)
\begin{equation}
  \braket{ \Opp[2][\Delta] \Opp[c] \Opp[1][\Delta] } \longrightarrow  \mathcal{C}_{1c2} e^{-\Delta (\tau_2 - \tau_1)/R} = \Anew[\Delta] \mathcal{C}_{1c2} \, .
\end{equation}

The result is similar for the four-point function of scalar operators, which depends on the cross ratios
\begin{equation}
  \braket{\Opp[2](x_2) \Opp[c](x) \Opp[c'](x') \Opp[1](x_1)} = f\pqty*{\frac{x_{2c} x_{c'1}}{x_{2 c'} x_{c 1}}, \frac{x_{2c} x_{c'1}}{x_{2 1} x_{c c' }}} \prod_{i < j} x_{ij}^{\sum_{k} \Delta_k/3 - \Delta_i - \Delta_j}.
\end{equation}
In the large-separation limit on the cylinder it can be rewritten in terms of the function $\Anew$
 \begin{equation}
 \begin{multlined}
   \braket{\Opp[2](x_2) \Opp[c](x) \Opp[c'](x') \Opp[1](x_1)} = e^{- \Delta_2 ( \tau_2 - \tau)} e^{-\Delta_1(\tau - \tau_1)}  f_c(\tau' - \tau , \n \cdot \n')\\ = \Anew[\Delta_1][\Delta_2][\tau] f_c(\tau' - \tau , \n \cdot \n') .
    \end{multlined}
 \end{equation}

 \bigskip
 
For correlators involving spinful operators, the scalar part remains the same as above and is supplemented by an appropriate tensor structure~\cite{Osborn:1993cr,Costa:2011mg}.
For example, in the case of a scalar--scalar--spin-\(\ell\) correlator one needs to multiply by \((V^{(ijk)} \cdot t)^{\ell}\), where
\begin{equation}
    V^{(ijk)} = \frac{\abs{x_{ki}} \abs{x_{kj}}}{\abs{x_{ij}}} \pqty*{ \frac{x_{ki}}{\abs{x_{ki}}^2} - \frac{x_{kj}}{\abs{x_{kj}}^2} } ,
\end{equation}
and  $t$ is an auxiliary vector which squares to zero, $t^2=0$, to ensure the tracelessness of \(V^{(ijk)}\).
This object has a particularly simple expression as a spherical tensor.
First we observe that \(\Proj\indices{^{\nu_1 \dots \nu_\ell}_{\ell m}}{}\) is antisymmetric and traceless by construction, so we do not need to subtract any traces and just need to compute one integral
\begin{multline}
    V^{(ijk)}_{\ell m} = \Proj\indices{^{\mu_1 \dots \mu_\ell}_{\ell m}} {} V^{(ijk)}_{\mu_1} \dots V^{(ijk)}_{\mu_\ell} = k_{\ell , D} \int \dd{\Omega} Y^*_{\ell m}(\n) \pqty*{\n \cdot V^{(ijk)}}^\ell \\
    = \frac{1}{k_{\ell, D}} \frac{ \abs{ \abs{x_{kj}}^2 x_{ki} - \abs{x_{ki}}^2 x_{kj}}^{\ell }}{ \abs{x_{ij}}^{\ell} \abs{x_{ki}}^{\ell} \abs{x_{kj}}^{\ell} }  Y^*_{\ell m} \pqty*{ \frac{  \abs{x_{kj}}^2 x_{ki} - \abs{x_{ki}}^2 x_{kj}  }{  \abs{ \abs{x_{kj}}^2 x_{ki} - \abs{x_{ki}}^2 x_{kj}} } } .
\end{multline}
In the limit of large separation, if we write \(x_i = R e^{\tau_2/R} \n_2\), \(x_j = R e^{\tau_1/R} \n_1\),  \(x_k = R e^{\tau/R} \n\),  we find that
\begin{equation}
    V^{(ijk)}_{\ell m} =  \frac{1}{k_{\ell, D}} Y^{*}_{\ell m}(\n) \pqty*{1 + \order{e^{-(\tau_2 - \tau )/R}}} ,
\end{equation}
as expected from representation theory.

\section{Casimir energy in various dimensions}
\label{sec:Casimir_appendix}
The Casimir energy for the \ac{eft} on $\mathbb{R} \times S^{D-1}$ gives the first correction to the scaling dimension of the primary $\Opp$ as follows
\begin{equation}
    \Delta_1 = \frac{1}{2\sqrt{D-1}} \sum_{\ell =1}^\infty M_\ell \sqrt{\ell(\ell + D-2)}.
    \label{eq:casimir}
\end{equation}
The \ac{eft} can describe only phonon states $a_\ell^\dagger \ket{Q}$ with $\ell \ll \Lambda$, where $ \Lambda \sim R \mu \sim Q^{\frac{1}{D-1}}$
which provides a natural regularization procedure for Eq.~\eqref{eq:casimir}. We can regulate it employing a smooth cutoff~\cite{Monin:2016bwf}
\begin{equation}
   \Sigma(1/2) =  \sum_{\ell=1}^\infty \text{deg}_D(\ell) \sqrt{\ell(\ell + D-2)} e^{-\ell(\ell + D-2)/\Lambda^2}.
\end{equation}
This regulated sum can be computed as an asymptotic series for $\Lambda \rightarrow \infty$ as follows. First, note that for large $\ell$ the summand of the original series can be expanded in powers of $1/\ell$ starting at $\ell^{D-1}$
\begin{align}
	\text{deg}(\ell) \sqrt{\ell (\ell + D-2)} \xrightarrow{\ell \rightarrow \infty}  \sum_{k=1} a_{k}(D) \ell^{D-k}.
\end{align}
The regulated sum can then be split as follows
\begin{equation}
	\begin{aligned}
\sum_{\ell = 0}^\infty \text{deg}(\ell) \sqrt{\ell (\ell+D-2)} e^{- \ell(\ell+D-2)/\Lambda^2} ={}& \Sigma_{\text{div.}} + \Sigma_{\text log} + \Sigma_{\text{conv.}},
\end{aligned}
\end{equation}
where we defined 
\begin{equation}
	\begin{aligned}
\Sigma_{\text{conv.}} ={}& \sum_{\ell=0}^\infty \left( \text{deg}(\ell) \sqrt{\ell (\ell+D-2)} - \sum_{k=1}^{D+1} a_k(D) \ell^{D-k} \right) e^{-\ell(\ell +D-2) /\Lambda^2},\\
\Sigma_{\text log} ={}& a_{D+1}(D) \sum_{\ell =0}^\infty \frac{1}{\ell} e^{-\ell(\ell +D-2) /\Lambda^2}, \\
\Sigma_{\text{div.}} ={}& \sum_{\ell=0}^\infty \left(\sum_{k=1}^{D} a_k(D) \ell^{D-k}\right) e^{-\ell(\ell+D-2)/\Lambda^2} .
\end{aligned}
\label{eq:various_sums}
\end{equation}
Let us discuss these three sums separately. By construction, the $\Sigma_{\text{conv.}}$ term contains a convergent series
\begin{align}
	\text{deg}(\ell) \sqrt{\ell (\ell+D-2)} - \sum_{k=1}^{D+1} a_k(D) \ell^{D-k} &\sim  \order*{\frac{1}{\ell^2}}  & \text{as} && \ell \rightarrow \infty.
\end{align}
This means that no further regulation is needed and $\Sigma_{\text{conv.}} = \text{const.} + \order{1/\Lambda}$. 
The $\Sigma_{\text log}$ term is only present when $D$ is even. This is due to the fact that the coefficients $a_k(D)$ satisfy
\begin{align}
	a_{2k} &\propto (D-2k+1), & \forall k &\in \mathbb{N},
\end{align}
while $a_{2k+1}$ does not have zeroes for any integer dimension $D>2$ for all $k \in \mathbb{N}$. This term leads to a $\log \Lambda$ first found in this context in~\cite{Cuomo:2020rgt}. The asymptotic expansion of this term for $\Lambda \rightarrow \infty$ can be found by a straightforward application of the Euler--Maclaurin formula
\begin{equation}
\begin{aligned}
	\sum_{\ell=1}^\infty\frac{1}{\ell} e^{-\ell(\ell+D-2)/\Lambda^2} &\sim \int_0^\infty \frac{\dd x}{x} e^{-x(x+D-2)/\Lambda^2} + \frac{1}{2} e^{-(D-1)/\Lambda^2} - \sum_{k=1}^\infty \frac{B_{2k}}{(2k)!} \left. \left( \frac{\del}{\del x} \right)^{2k-1} \right|_{1}\frac{1}{x} e^{-x(x+D-2)/\Lambda^2} \\
	&\sim \frac{1}{2} \left( \gamma + \log \Lambda^2 \right) + \order{1/\Lambda}.
\end{aligned}	
\end{equation}
This term produces the following term in the 1-loop scaling dimension in even dimension
\begin{equation}
\Delta_1|_{\text{$D=$ even}} \supset \frac{a_{D+1}}{2(D-1)\sqrt{D-1}} \log Q.
\end{equation}
As this cannot be corrected by any other classical or loop correction, it is a universal prediction that is independent of the details of the underlying strongly-coupled \ac{cft}, depeding only on its global symmetry group.

In odd dimensions the universal term is instead proportional to $Q^0$ and is computed from the terms $(\Sigma_{\text{div.}} + \Sigma_{\text{conv.}})|_{\text{const.}}$. The term $\Sigma_{\text{div.}}$ contains also positive powers of the cutoff $\Lambda$ which can be absorbed into the Wilsonian coefficients $c_i$. Using a symmetry-preserving regulator will always guarantee that to be the case. In the present case, to estimate which powers are going to appear one can use 
\begin{equation}
\sum_{\ell=0}^{\infty} \ell^{\alpha} e^{- \ell(\ell+D-2)/\Lambda^2} \sim \frac{1}{2} \Gamma\left( \frac{\alpha + 1}{2} \right) \Lambda^{\alpha +1} + \sum_{k=1}^\infty a_k(\alpha) \Lambda^{\alpha - k},
\end{equation}
valid for $\alpha \in \mathbb{N}$, with coefficients $a_k(\alpha)$ computable order by order. For example, one finds 
\begin{align}
  \eval[\Big]{ \Sigma_{\text{div.}}}_{D=3} &= \frac{\sqrt{\pi}}{2} \Lambda^3 - \frac{1}{4} + \order{\Lambda^{-1}}, \\
  \eval[\Big]{\Sigma_{\text{div.}}}_{D=4} &= \frac{1}{2} \Lambda^4 + \frac{1}{4} \Lambda^{2} - \frac{9}{20} + \order{\Lambda^{-1}}, \\
  \eval[\Big]{\Sigma_{\text{div.}}}_{D=5} &= \frac{\sqrt{\pi}}{8} \Lambda^5 + \frac{\sqrt{\pi}}{6} \Lambda^{3} - \frac{21}{64} + \order{\Lambda^{-1}}, \\
  \eval[\Big]{\Sigma_{\text{div.}}}_{D=6} &= \frac{1}{12} \Lambda^6 + \frac{5}{24} \Lambda^4 + \frac{1}{6} \Lambda^2 - \frac{18553}{30240} + \order{\Lambda^{-1}}.
\end{align}
The relevant contributions to the conformal dimensions $\Delta_1$ are summarized in~\autoref{tab:casimir} for different spacetime dimensions $D$. 

\begin{table}
  \renewcommand{\arraystretch}{1.5}
  \centering
  \begin{tabular}{p{0.5cm}p{3cm}p{3cm}Sp{3cm} }
 \toprule
  $D$ & $\Sigma_{\text{div.}}|_{\text{const.}}$ & $\Sigma_{\text{log}}$ & \mcL{$\Sigma_{\text{conv.}}$} & $\Delta_{1,\text{univ.}}$ \\
  \midrule
 \(3\) & $ - \frac{1}{4}$ & $0$ & -0.01509 & $-0.09372 \times Q^0$\\
 \(4\) & $- \frac{9}{20}$ & $- \frac{1}{8} \left( \frac{\gamma}{2} + \log \Lambda \right)$ & 0.1106 & $- \frac{1}{48\sqrt{3}} \log Q$\\
 \(5\) & $ - \frac{21}{64}$ & $0$ & -0.1035 & $-0.1079 \times Q^0$\\
 \(6\) & $- \frac{18553}{30240}$ & $-\frac{1}{6} \left( \frac{\gamma}{2} +\log \Lambda \right)$ & 0.1990 &  $-\frac{1}{60\sqrt{5}} \log Q$ \\
 \(7\) & $- \frac{4735}{12288}$ & $0$ & -0.1684 & $-0.1130 \times Q^0$\\
 \(8\) & $- \frac{534983}{725760}$ & $-\frac{981}{5120}\left( \frac{\gamma}{2} +\log \Lambda \right)$ & 0.2655 & $- \frac{981}{71680\sqrt{7}} \log Q$ \\
 \(9\) & $- \frac{1273741}{2949120}$ & $0$ & -0.2203 & $-0.1153  \times Q^0$\\
 \(10\) & $- \frac{10420037}{12418560}$ & $-\frac{22}{105}\left( \frac{\gamma}{2} + \log \Lambda \right)$ & 0.3192 & $ -\frac{11}{2835} \log Q$ \\
 \(11\) & $-\frac{277116003}{587202560}$ & $0$ & -0.2641 & $-0.1163 \times Q^0$\\
 \bottomrule
\end{tabular}
\caption{Relevant values for the sums in~\eqref{eq:various_sums} in different spacetime dimension $D$. We indicate with $\Delta_{1, \text{univ.}}$ the universal contribution to the scaling dimension.}
\label{tab:casimir}
\end{table}

\section{Details of the loop computations}
\label{sec:Loop_appendix}

The subleading terms in the large-$Q$ expansion involve contribution from loop corrections arising from the $\ac{eft}$. The perturbation theory is set up on $S^1_\beta \times S^{D-1}$, and in this Appendix we summarize the relevant technology for 2-loop computations.  

\subsection{Matsubara sums}

Summations over Matsubara frequencies can be performed using the formula
\begin{equation}
\sum_{n \in \mathbb{Z}} f \pqty*{\tfrac{2\pi i n}{\beta}} = \beta \int \frac{\dd k }{2\pi} \left( \frac{f(i k) + f(-i k)}{2} \right) + \order{e^{-\beta}},
\end{equation}
where corrections are neglected in the $\beta \rightarrow \infty$ limit.

There are three relevant Matsubara sums appearing in our loop computations. These are readily computed using the formula above and result in
\begin{align}
\sum_{n_a} D_{n_a \ell_a} &= \frac{\beta}{2 \omega_{\ell_{a}}} , \\
  \sum_{n_a + n_b = n} D_{n_a \ell_a} D_{n_b \ell_b} &= \frac{\beta }{2 } \left[ \frac{1}{\omega_{\ell_a}} + \frac{1}{\omega_{\ell_b}} \right] \frac{1}{\omega_n^2 + (\omega_{\ell_a} + \omega_{\ell_{b}} )^2 },\\
\sum_{n} D_{n \ell} \sum_{n_a + n_b = n} D_{n_a \ell_a} D_{n_b \ell_b} 
&= \frac{\beta^2}{4} \frac{1}{\omega_{\ell_a}\omega_{\ell_b}\omega_{\ell} } \frac{1}{\omega_\ell + \omega_{\ell_a} + \omega_{\ell_b}},
\end{align}
where the propagator is $D_{n_a \ell_a} = (\nu^2_{n_a} + \omega_{\ell_a}^2)^{-1}$.
Related sums with powers of $\omega_{n_a} ,\, \omega_{n_b}$ at the numerator can be expressed in terms of the sums above using the linearity property of the smooth regularizsation~\eqref{eq:reg_lin}. This is analog to the reduction to scalar master integrals in multi-loop computations~\cite{Grozin:2003ak,Grozin:2005yg}.
Due to the derivative interactions, this reduction procedure may produce a divergent Matsubara sum. In our smooth cutoff regularization these are computed as follows in the limit $\beta \rightarrow \infty$
\begin{align}
\sum_{n_a} 1 \quad\quad \rightarrow \quad\quad \sum_{n_a} e^{-\frac{(2\pi n_a)^2}{\beta^2 \Lambda^2}} = \frac{\beta \Lambda}{2\sqrt{\pi}} + \order{\beta^{-1}} + \dots
\end{align}
\subsection{Kinematic vertex factors}

In perturbation theory on $S^{D-1}$ one needs to compute vertex factors proportional to multiple integrals of hyperspherical harmonics $Y_{\ell m} $. In the diagram arising from the four-point vertex~\eqref{eq:fourpoint_diag} one finds
\begin{align}
T^{0\del}(1,2,3,4) &=\int_{S^{D-1}}  Y_{\ell_1 m_1}   Y_{\ell_2 m_2}  Y_{\ell_3 m_3}  Y_{\ell_4 m_4} , \\
T^{2\del}(1,2,3,4) &=\int_{S^{D-1}} Y_{\ell_1 m_1}  Y_{\ell_2 m_2} \del_j Y_{\ell_3 m_3} \del_j Y_{\ell_4 m_4}  ,\\
T^{4\del}(1,2,3,4) &=\int_{S^{D-1}} \del_i Y_{\ell_1 m_1}  \del_i Y_{\ell_2 m_2} \del_j Y_{\ell_3 m_3} \del_j Y_{\ell_4 m_4}  .
\end{align}
These integrals are rather non-trivial in general, but in loop diagrams it is sufficient to compute their contraction in the $m$-type indices using~\eqref{eq:Y_orthogonality}:
\begin{align}
\sum_{m_a,m_b} T^{0\del} (a,-a,b,-b) &= \frac{R^{D-1}}{\Omega_D} M_{\ell_a} M_{\ell_b},  \\
\sum_{m_a,m_b} T^{2\del}(a,-a,b,-b) &= \frac{R^{D-1}}{\Omega_D} M_{\ell_a} M_{\ell_b} \frac{\lambda_{\ell_b}}{R^2}, \\
\sum_{m_a, m_b} T^{2\del}(a,b,-a,-b) &= 0, \\
\sum_{m_a,m_b} T^{4\del} (a,-a,b,-b) &= \frac{R^{D-1}}{\Omega_D}M_{\ell_a} M_{\ell_b} \frac{\lambda_{\ell_a}}{R^2} \frac{\lambda_{\ell_b}}{R^2} ,\\
\sum_{m_a,m_b} T^{4\del}(a,b,-a,-b) &= \frac{R^{D-1}}{\Omega_D} \frac{1}{D-1} M_{\ell_a} M_{\ell_b} \frac{\lambda_{\ell_a}}{R^2} \frac{\lambda_{\ell_b}}{R^2}.
\end{align}
Pure two-loop topologies appear in diagrams with three-point vertices~\eqref{eq:threepoint_diag}. In that case the following structures appear:
\begin{align}
T^{0\del}(1,2,3 | 4,5,6) &= \int_{S^{D-1}}  Y_{\ell_1 m_1} Y_{\ell_2 m_2}   Y_{\ell_3 m_3} \int_{S^{D-1}}  Y_{\ell_4 m_4}   Y_{\ell_5 m_5}   Y_{\ell_6 m_6},  \\
T^{2\del}(1,2,3|4,5,6) &= \int_{S^{D-1}}  Y_{\ell_1 m_1}  Y_{\ell_2 m_2}   Y_{\ell_3 m_3} \int_{S^{D-1}}  Y_{\ell_4 m_4} \del_i  Y_{\ell_5 m_5}  \del_i Y_{\ell_6 m_6} , \\
T^{4\del}(1,2,3|4,5,6) &= \int_{S^{D-1}}  Y_{\ell_1 m_1}  \del_j Y_{\ell_2 m_2}   \del_j Y_{\ell_3 m_3} \int_{S^{D-1}}  Y_{\ell_4 m_4} \del_i  Y_{\ell_5 m_5}  \del_i Y_{\ell_6 m_6} ,
\end{align}
with $\lambda_\ell$ eigenvalue of $-\Delta_{S^{D-1}}$.
The last two structures can be expressed in terms of the first one via integration-by-parts relations
\begin{align}
    T^{4\del}(1,2,3,4,5,6) &= \frac{1}{4} \Big\{ \lambda_{\ell_3} + \lambda_{\ell_2} - \lambda_{\ell_1} \Big\} \Big\{ \lambda_{\ell_6} + \lambda_{\ell_5} - \lambda_{\ell_4} \Big\} T^{0\del}(1,2,3,4,5,6),\\
    T^{2\del}(1,2,3,4,5,6) &= \frac{1}{2} \Big\{ \lambda_{\ell_6} + \lambda_{\ell_5} - \lambda_{\ell_4} \Big\} T^{0\del}(1,2,3,4,5,6).
\end{align}
The expression for $T^{0\partial}$ is not as simple as its flat space counterpart, due to the fact that momentum conservation on $S^{D-1}$ is replaced by $SO(D)$-angular momentum addition. Its $m$-index contraction can be computed using the properties of Gegenbauer polynomials summarized in \autoref{sec:Ylm-identities}
\begin{multline}
\sum_{m_a , m_b, m_c} T^{0\del}(a,b,c|a,b,c)= \frac{1}{3}
 \frac{R^{2D-2}}{(D-2)\Omega_D} \frac{(D+2\ell_a -2)(D+2\ell_b -2)}{(D+2 \ell_c +2)} M_{\ell_c}  \\
\times \sum_{k=0}^{\min(\ell_a \ell_b)}  \braket{ k | \ell_a \ell_b } \delta_{\ell_c - \ell_a -\ell_b +2k}  + ( \text{2 perm. in $\ell_a \ell_b \ell_c$} ) ,
\end{multline}
where the permutations in $\ell_a , \ell_b, \ell_c$ are included to make the permutation symmetry manifest, and correspond to the different ways of applying the Gegenbauer addition formula~\eqref{eq:Gegenbauer_addition}. The summation appearing above can be computed as
\begin{multline}
\sum_{k=0}^{\min(\ell_a ,\ell_b)}	\braket{ k| \ell_a \ell_b} \delta_{\ell_c - \ell_a - \ell_b + 2k} = \vartri_{\ell_a \ell_b \ell_c}   \frac{(D+2 \ell_c-2)\Gamma(\ell_c+1)}{2 \Gamma\left( \frac{D}{2} -1 \right)^2\Gamma(\ell_c+D-2)}  \\
	 \times \frac{\Gamma\left( \frac{\ell_{abc}}{2} + \frac{D-2}{2} \right)}{\Gamma\left( \frac{\ell_{abc}}{2} +1 \right)} \frac{\Gamma\left( \frac{\ell_{cab}}{2} + \frac{D-2}{2} \right)}{\Gamma\left( \frac{\ell_{cab}}{2} +1 \right)} \frac{\Gamma\left( \frac{\ell_{bca}}{2} + \frac{D-2}{2} \right)}{\Gamma\left( \frac{\ell_{bca}}{2} +1 \right)} \frac{\Gamma\left( \frac{\ell_a + \ell_b +\ell_c}{2} + \frac{2D-4}{2} \right)}{\Gamma\left(\frac{\ell_a+\ell_b+\ell_c}{2} + \frac{D}{2} \right)} ,
\end{multline}
where we introduced the notation $\ell_{abc} = \ell_a +\ell_b - \ell_c$ and the symbol $\vartri$ imposing a triangle inequality
\begin{equation}
      \vartri_{\ell_a \ell_b \ell_c} = \begin{cases} 1 & \text{if} \ \abs{\ell_b - \ell_a}  \leq \ell_c \leq \ell_b+\ell_a \quad \text{and} \quad \ell_c - \ell_a - \ell_b \ \text{even}, \\
    0 & \text{otherwise.} \end{cases}
\end{equation}
Putting these result together and using that $\vartri$ is a fully symmetric symbol, one finds
\begin{equation}\label{eq:def_Sabc}
  \begin{aligned}
\sum_{m_a  m_b m_c} T^{0\del}(a,b,c \mid a,b,c) ={}& \vartri_{\ell_a \ell_b \ell_c}
 \frac{R^{2D-2}}{(D-2)\Omega_D} \frac{(D+2\ell_a -2)(D+2\ell_b -2)(D+2\ell_c -2)}{2\Gamma(D-1) \Gamma\left( \frac{D}{2}-1\right)^2 } \\
	& \times \frac{\Gamma\left( \frac{\ell_{abc}}{2} + \frac{D-2}{2} \right)}{\Gamma\left( \frac{\ell_{abc}}{2} +1 \right)} \frac{\Gamma\left( \frac{\ell_{cab}}{2} + \frac{D-2}{2} \right)}{\Gamma\left( \frac{\ell_{cab}}{2} +1 \right)} \frac{\Gamma\left( \frac{\ell_{bca}}{2} + \frac{D-2}{2} \right)}{\Gamma\left( \frac{\ell_{bca}}{2} +1 \right)} \frac{\Gamma\left( \frac{\ell_a + \ell_b +\ell_c}{2} + \frac{2D-4}{2} \right)}{\Gamma\left(\frac{\ell_a+\ell_b+\ell_c}{2} + \frac{D}{2} \right)} \\
	 \eqcolon{}& \frac{R^{2D-2}}{(D-2)\Omega_D} S_{\ell_a \ell_b \ell_c} ,
  \end{aligned}
\end{equation}
where we defined the fully symmetric structure $S_{\ell_a \ell_b \ell_c}$ which appears in Eq.~\eqref{eq:sum-of-two-loop-graphs}. 

This is useful to get fully symmetric amplitudes in intermediate steps. Sums involving the $\vartriangle$ symbol are computed as
\begin{equation}
\sum_{\ell_a \ell_b \ell_c =1}^\infty \vartriangle_{\ell_a \ell_b \ell_c} f(\ell_a, \ell_b, \ell_c)= \sum_{\ell_a \ell_b =1}^\infty \sum_{k=0}^{\min(\ell_a \ell_b)} f(\ell_a, \ell_b ,\ell_a + \ell_b -2k) - \sum_{\ell_a \ell_b = 1}^\infty\delta_{\ell_a \ell_b} f(\ell_a , \ell_a, 0),
\end{equation}
where the second term comes from the exclusion of the $\ell_c = 0$ term. In particular, some sums involving $S_{\ell_a \ell_b \ell_c}$ can be computed as follows
\begin{align}
    \sum_{\ell_a \ell_b \ell_c = 1}^\infty \vartriangle_{\ell_a \ell_b \ell_c} S_{\ell_a \ell_b \ell_c} &= (D-2) \left[  \sum_{\ell_a \ell_b} M_{\ell_a} M_{\ell_b} - \sum_{\ell_a} M_{\ell_a} \right],\\
    \sum_{\ell_a \ell_b \ell_c=1}^\infty \vartriangle_{\ell_a \ell_b \ell_c} S_{\ell_a \ell_b \ell_c} \omega_{\ell_c} &= (D-2) \left[  \sum_{\ell_a \ell_b} M_{\ell_a} M_{\ell_b} \omega_{\ell_b}  - \sum_{\ell_a} M_{\ell_a}\omega_{\ell_a} \right],\\
    \sum_{\ell_a \ell_b \ell_c=1}^\infty  \vartriangle_{\ell_a \ell_b \ell_c}  S_{\ell_a \ell_b \ell_c} \omega_{\ell_c}^2 &= 2(D-2) \sum_{\ell_a \ell_b} M_{\ell_a} M_{\ell_b}  \omega_{\ell_b}^2.
\end{align}

\subsection[\texorpdfstring%
{Graphs for $\Delta_2$}%
{Two-loop graphs}]%
{Graphs for $\Delta_2$}%

Using the notation of the previous section, the graphs appearing in the 4-point vertex contribution of $\Delta_2$ evaluate to
\begin{align}
  &\begin{aligned}
   \vcenter{\hbox{\includegraphics{quartic-1}}} \,  &= \beta \left( \frac{\beta}{R} \right)^2 \sum_{\{n_i , \ell_i , m_i\}} T^{4\del}(1,2,3,4) \times (\text{3 contractions} )  \\
    &= \frac{D+1}{D-1} \frac{C^2 R^{D-3}}{\beta \Omega_D} \left[\sum_{n \ell}  D_{n \ell} \lambda_{\ell}  M_{\ell} \right]^2 ,
  \end{aligned} \\
  & \begin{aligned}
  \vcenter{\hbox{\includegraphics{quartic-2}}} \, &=- \beta \left( \frac{\beta}{R} \right)^2 \sum_{\{n_i , \ell_i , m_i\}} \omega_{n_1} \omega_{n_2}  T^{2\del}(1,2,3,4) \times (\text{3 contractions} )  \\
    &= \frac{C^2 R^{D-3}}{\beta \Omega_D} \left[\sum_{n \ell}  D_{n \ell} \lambda_{\ell}  M_{\ell} \right] \left[ \sum_{n \ell} D_{n \ell} \nu_n^2 \right] ,
  \end{aligned} \\
  &\begin{aligned}
   \vcenter{\hbox{\includegraphics{quartic-3}}} \,
   &= \beta \left( \frac{\beta}{R} \right)^2\sum_{\{n_i , \ell_i , m_i\}} \delta_{\sum_i n_i} \left[ \prod_{i=1}^4 \nu_{n_i} \right] T^{0\del}(1,2,3,4) \times (\text{3 contractions} )  \\
   &= 3\frac{C^2 R^{D-3}}{\beta \Omega_D} \left[ \sum_{n \ell} D_{n \ell} \nu_n^2 M_\ell \right]^2.
  \end{aligned}
\end{align}
Similarly, the graphs appearing using three-point vertices give
\begin{align}
  &\begin{aligned}
   \vcenter{\hbox{\includegraphics{cubic-1}}} \, &=- \beta^2 \left( \frac{\beta}{R} \right)^3 \sum_{\{n_i , \ell_i , m_i\}}^{(\sum_i n_i = 0)} \sum_{\{n_j , \ell_j , m_j\}}^{(\sum_j n_j = 0)} \left[ \prod_{i=1}^6 \nu_{n_i} \right] T^{0\del}(1,2,3,4,5,6) \times (\text{6 contractions} ) \\
   &= 6\frac{C^3 R^{2D-5}}{\beta} \sum_{n_a + n_b + n_c = 0} \nu_{n_a}^2 \nu_{n_b}^2 \nu_{n_c}^2\sum_{\ell_a \ell_b \ell_c} D_{n_a \ell_a} D_{n_b \ell_b} D_{n_c \ell_c}  S_{\ell_a \ell_b \ell_c} ,
  \end{aligned} \\
  &\begin{aligned}
   \vcenter{\hbox{\includegraphics{cubic-2}}} \, &= \beta^2 \left( \frac{\beta}{R} \right)^3 \sum_{\{n_i , \ell_i , m_i\}}^{(\sum_i n_i=0)} \sum_{\{n_j , \ell_j , m_j\}}^{(\sum_j n_j=0)} \nu_{n_1} \nu_{n_2} \nu_{n_3} \nu_{n_4} T^{2\del}(1,2,3,4,5,6) \times (\text{6 contractions} )  \\
   &=- 6\frac{C^3 R^{2D-5}}{2\beta} \sum_{n_a + n_b + n_c = 0} \nu_{n_a}^2 \nu_{n_b} \nu_{n_c}  \sum_{\ell_a \ell_b \ell_c} D_{n_a \ell_a} D_{n_b \ell_b} D_{n_c \ell_c} (\omega_{\ell_c}^2+\omega_{\ell_b}^2 - \omega_{\ell_a}^2 ) S_{\ell_a\ell_b  \ell_c} ,
  \end{aligned} \\
  &\begin{aligned}
   \vcenter{\hbox{\includegraphics{cubic-3}}} \, ={}& -\beta^2 \left( \frac{\beta}{R} \right)^3 \sum_{\{n_i , \ell_i , m_i\}}^{(\sum_i n_i=0)} \sum_{\{n_j , \ell_j , m_j\}}^{(\sum_j n_j=0)} \nu_{n_1} \nu_{n_2} T^{4\del}(1,2,3,4,5,6) \times (\text{6 contractions} ) \\
   ={}& \frac{C^3 R^{2D-5}}{4\beta} \sum_{n_a + n_b + n_c = 0} \sum_{\ell_a \ell_b \ell_c} D_{n_a \ell_a} D_{n_b \ell_b} D_{n_c \ell_c} S_{\ell_a \ell_b \ell_c} \\
   &\times \Big[ 2 \nu_{n_a}^2 (\omega_{\ell_c}^2 + \omega_{\ell_b}^2 - \omega_{\ell_a}^2) - 4 \nu_{n_a} \nu_{n_b} (\omega_{\ell_c}^2 + \omega_{\ell_a}^2 - \omega_{\ell_b}^2) \Big] (\omega_{\ell_c}^2 + \omega_{\ell_b}^2 - \omega_{\ell_a}^2) .
  \end{aligned}
\end{align}

\section[\texorpdfstring%
{Methods and details for the computations in Section~\refstring{sec:ConformalAlgebraAndChargeCorrelators}}%
{Methods and details for the computations in Section four}]%
{Methods and details for the computations in Section~\refstring{sec:ConformalAlgebraAndChargeCorrelators}}%
\label{sec:methods}

\subsection[\texorpdfstring%
{Computing the $\braket{\Opp*  T_{\tau\tau} \Opp}$ correlator}%
{<QTQ>}]%
{Computing the $\braket{\Opp*  T_{\tau\tau} \Opp} $ correlator}
Using the property of $\ket{Q}$ as a vacuum~\eqref{eq:QisVacuumOfTheAs}, the field decomposition in terms of creation and annihilation operators~\eqref{eq:FieldDecompositionInLadders} and the expansions of $T$ and $Q$ in Eq.~\eqref{eq:ExpansionsOfTsAndQsInTheField} one can compute the tree level results for the correlators in \eqref{sec:ConformalAlgebraAndChargeCorrelators}.
We demonstrate the computation of $\braket{ \Opp* T_{\tau\tau} \Opp }$
\begin{multline}
    \braket*{\Opp* T_{\tau\tau}(\tau,\n)\Opp} =\\
    - \frac{ \Delta_0 }{ R^D \Omega_D} \braket*{Q | 1 + i  \frac{D}{\mu} \dot\pi  - \frac{ D}{2\mu^{2}} \left( (D-1) \dot \pi^2 -  \frac{ (D-3)}{ R^2 (D-1)} \pi\Delta\pi + \frac{ (D-3)}{ R^2 (D-1)} \del^i (\pi \del_i \pi) \right) | Q} .
\end{multline}
We ignore the total derivative term for now and show later that it vanishes. 
Setting $A={c_1 \Omega_D R^{D-1} D(D-1) \mu^{D-2} }$ we find up to quadratic order in the fields
\begin{multline}
        \braket{ \Opp* T_{\tau\tau} (\tau,\n)  \Opp} =  - \frac{ \Delta_0 }{ R^D \Omega_D } \braket*{Q | \bigg[ 1 + i  \frac{D}{\mu} \dot\pi - \frac{ D}{2\mu^{2}} \Big[ (D-1) \dot \pi^2 - \frac{ (D-3)}{R^2 (D-1)} \pi\Delta\pi \Big] \bigg]_{(\tau,\n)}  |Q}  \\
        = - {}\frac{ \Delta_0 }{R^D \Omega_D} \Bigg\langle Q   \Bigg|   e^{- \frac{ (\tau_2 -\tau) }{R}   D} \bigg[ 1 + i  \frac{D}{\mu} \bigg[ - \frac{i \Pi_0}{A } + \sqrt{\frac{\Omega_D}{A}} \sum_{m,l} \sqrt{\frac{\omega_l}{2}}\Big( a^\dagger_{lm} Y_{l m}^* (\n) - {a_{lm}} Y_{l m} (\n) \Big) \bigg] \\
         + \frac{ D (D-1)}{2\mu^{2}} {\frac{\Omega_D}{2 A}}  \sum_{\substack{m, l\\ m', l'}} \sqrt{ \omega_l \omega_{l'} } \bigg[  a^\dagger_{lm} a_{l'm'}  Y_{l m}^* (\n) Y_{l' m'} (\n) + a_{lm} a^\dagger_{l'm'} Y_{l m} ( \n) Y_{l' m'}^* (\n) \bigg] \\
         - \frac{ \Pi^2_0}{A^2 } - 2 \frac{i \Pi_0}{A} \sqrt{ \frac{ \Omega_D }{ A}} \sum_{m,l} \sqrt{ \frac{ \omega_l }{2}} \Big[ - {a_{lm}} Y_{l m} (\n) + {a^\dagger_{lm}} Y_{l m}^* (\n) \Big] \\
        - \frac{ D}{ \mu^{2}}   \frac{ \Omega_D }{2A} \frac{ (D-3)}{2 (D-1)} \sum_{ \substack{ m, l \\ m', l'}} \frac{ (D-1) \omega_{l'}^2 }{ \sqrt{\omega_l \omega_{l'}}} \bigg[ {a^\dagger_{lm}} {a_{l'm'}} Y_{l m}^* (\n) Y_{l' m'} (\n) + {a_{lm}}  a^\dagger_{l'm'} Y_{l m} (\n) Y_{l' m'}^* (\n) \bigg] \\
         + \frac{(D-3) }{ (D-1)} \pi_0 \sqrt{ \frac{ \Omega_D }{A} } \sum_{m, l} \frac{l(l+D-2)}{ R^2 \sqrt{2\omega_{l}}} \Big[ {a_{l m}}  Y_{l m}(\n) + {a^\dagger_{lm}} Y_{l m}^* (\n) \Big]   \bigg]  e^{- \frac{ (\tau -\tau_1) }{R}   D}   \Bigg| Q \Bigg \rangle .
\end{multline}        
Terms linear in the $a_{lm}$-operators and terms directly proportional to $\Pi_0$ are directly zero and can be ignored.
We have
\begin{multline}
        \braket*{\Opp* T_{\tau\tau}(\tau,\n) \Opp} = - \frac{ \Delta_0 }{R^D \Omega_D} \bra{Q} e^{- \frac{ (\tau_2 -\tau)   D }{R}} \bigg[ 1 + \frac{ D (D-1)\, \Omega_D }{ 4\mu^{2} A} \sum_{\substack{m, l\\ m', l'}} \sqrt{ \omega_l \omega_{l'} }   [ a_{lm} , a^\dagger_{l'm'}] Y_{l m} (\n) Y_{l' m'}^* (\n) \\
        - \frac{(D-3)}{(D-1)}  \frac{ D }{ 2\mu^{2}} \frac{ \Omega_D }{2A} \sum_{ \substack{ m, l \\ m', l'}} \frac{ (D-1) \omega_{l'}^2 }{ \sqrt{\omega_l \omega_{l'}}} [ {a_{lm}} , a^\dagger_{l'm'} ] Y_{l m} (\n) Y_{l' m'}^* (\n) \bigg] e^{- \frac{ (\tau -\tau_1)   D }{R} } \ket{Q} \\
        = - \Anew \bigg[ \frac{ \Delta_0 }{R^D \Omega_D} + \frac{ c_1 (D-1) \mu^D D \Omega_D \Big[ (D-1) - (D-3)\Big]}{4\mu^{2} c_1 \Omega_D R^{D-1}  D(D-1) \mu^{D-2} } \sum_{m, l} \omega_l Y_{l m} (\n) Y_{l' m'}^* (\n) \bigg] \\
        = - \Anew \bigg[ \frac{ \Delta_0 }{R^D \Omega_D} + \frac{ 1}{2} \frac{1}{ R^{D-1}} \sum_{m,l} \omega_l Y_{l m} (\n) Y_{l' m'}^* (\n)  \bigg] .
\end{multline}
The sum over the product of spherical harmonics is evaluated in Appendix~\ref{sec:Loop_appendix}.\\

We still need to show that the total derivative term vanishes
\begin{equation}
    \begin{aligned}
        \frac{ \Delta_0 }{R^D \Omega_D}\frac{ D}{2\mu^{2}} \bra{Q}  \del^i \Big( \pi ( \tau, \n) \, \del_i \pi(\tau ,\n) \Big) \ket{Q} &= - \frac{\Anew }{4  R^{D-1}}   \sum_{m,l}    \frac{\del^i \Big( Y_{l m} (\n) \del_i Y_{l m}^* (\n) \Big)}{ \omega_l } \\
        &= - \frac{\Anew }{4  R^{D-1}} \sum_{l} \frac{ 1}{ \omega_l} \frac{1 }{2} \frac{\del^i \del_i }{ 2 \omega_l} \frac{ M_{\ell} }{ \Omega_D} = 0 .
    \end{aligned}
\end{equation}
Putting everything together, the final result is
\begin{equation}
  \braket*{\Opp* T_{\tau\tau}(\tau,\n) \Opp} = - \Anew \frac{ \Delta_0 +\Delta_1 }{R^D \Omega_D} + \order{\mu^{D-3}}  .
\end{equation}
\subsection[\texorpdfstring%
{Computing the $\braket{ \Vpp* T_{\tau\tau} T_{\tau\tau} \Vpp }$ correlator}%
{<QTTQ>}]%
{Computing the $\braket{ \Vpp* T_{\tau\tau} T_{\tau\tau} \Vpp }$ correlator}%
\label{sec:VTTVcorrelator}

We compute the correlator $\braket*{\Vpp* T_{\tau\tau}(\tau,x) T_{\tau\tau}(\tau',x')\Vpp }$.
We use the expansion of $T_{\tau\tau}$ in terms of the Goldstone field in \eqref{eq:ExpansionsOfTsAndQsInTheField} and the explicit form of the leading order energy density $c_1 (D-1) \mu^D = \Delta_0/(R^D \Omega_D)$.
Expanded to second order we have
\begin{multline}
        T_{\tau\tau} (\tau, \n) T_{\tau\tau} (\tau', \n') = \frac{ \Delta_0^2 }{R^{2D} \Omega_D^2} - \frac{ \Delta_0^2 }{R^{2 D} \Omega_D^2} \frac{D^2}{\mu^2} \dot\pi(\tau, \n) \dot\pi(\tau', \n') - \frac{ \Delta_0^2 (D-1) }{R^{2D} \Omega_D^2} \frac{ D }{2 \mu^2} \Big[ \dot{\pi}^2 + \frac{(D-3) (\del_i \pi)^2 }{R^2 (D-1)^2} \Big]_{(\tau, \n)} \\
        - \frac{ \Delta_0^2 (D-1) }{R^{2D} \Omega_D^2} \frac{D}{2\mu^2} \Big[ \dot{\pi}^2 + \frac{(D-3) (\del_i \pi)^2 }{R^2 (D-1)^2} \Big]_{(\tau', \n')} + \order{\mu^{2D-3}}.
\end{multline}
The correlator becomes
\begin{multline}
        \bra{Q} a_{\ell_2 m_2} T_{\tau\tau} (\tau, \n) T_{\tau\tau} (\tau', \n') a_{\ell_1 m_1}^\dagger \ket{Q} = \frac{ \Delta_0^2 }{R^{2D} \Omega_D^2} e^{-\omega_\ell (\tau_2-\tau_1)} \Anew \delta_{m_1m_2} \delta_{\ell_1 \ell_2} \\
        - \frac{ \Delta_0^2 }{R^{2D} \Omega_D^2} \frac{D^2}{\mu^2} \bra{Q} a_{\ell_2 m_2} \dot\pi (\tau, \n) \dot\pi (\tau', \n') a_{\ell_1 m_1}^\dagger \ket{Q} \\
        - \frac{ \Delta_0^2 }{R^{2D} \Omega_D^2} \frac{ D}{2\mu^{2}} \bra{Q} a_{\ell_2 m_2} \Big[(D-1) \dot \pi^2 (\tau, \n) +  \frac{ (D-3) }{ R^2 (D-1)} \big( \del_i \pi (\tau, \n) \big)^2 \Big] a_{\ell_1 m_1}^\dagger \ket{Q} \\
        - \frac{ \Delta_0^2 }{R^{2D} \Omega_D^2} \frac{ D}{2\mu^{2}} \bra{Q} a_{\ell_2 m_2} \Big[  (D-1) \dot \pi^2(\tau', \n') + \frac{ (D-3) }{R^2 (D-1)} \big( \del_i \pi (\tau', \n') \big)^2 \Big] a_{\ell_1 m_1}^\dagger \ket{Q} .
\end{multline}
The last two terms are equivalent, already appear in the $\braket*{ \Vpp* T_{\tau\tau} (\tau, \n) \Vpp }$ correlator and are computed as follows (where $A={\Delta_0 D R^{-1} \mu^{-2} }$)
\begin{multline}
        - \frac{\Delta_0 (D-1)}{R^D \Omega_D} \frac{ D}{2\mu^{2}} \bra{Q} a_{\ell_2 m_2} \Big[ \frac{ (D-3) \big( \del_i \pi \big)^2 }{ R^2 (D-1)^2 } + \dot \pi^2 \Big]_{ (\tau, \n)} \, a_{\ell_1 m_1}^\dagger \ket{Q} = - \frac{\Delta_0 D}{2 \mu^{2} R^D \Omega_D} \bra{Q} a_{\ell_2 m_2} e^{-(\tau_2-\tau)   H} \dots \\
        \shoveright{ \dots \bigg[ \frac{(D-3) }{ R^2 (D-1)} \big( \del_i \pi \big)^2 + (D-1) \dot \pi^2 \Big] e^{- (\tau -\tau_1)   H} a_{\ell_1 m_1}^\dagger \ket{Q} } \\
        = \frac{ \Delta_0 D (D-1)}{2\mu^{2} R^D \Omega_D}  \frac{ \Omega_D }{2 A} \bigg( \sum_{ \substack{m, l\\ m', l'}} \sqrt{ \omega_l \omega_{l'}} \bigg[ \bra{Q} a_{\ell_2 m_2} e^{-(\tau_2-\tau)   H} a^\dagger_{lm} a_{l'm'} e^{- (\tau -\tau_1)   H} a_{\ell_1 m_1}^\dagger \ket{Q} Y_{l m}^* (\n) Y_{l' m'} (\n) \\
        + \bra{Q} a_{\ell_2 m_2} e^{-(\tau_2-\tau)   H}a^\dagger_{l'm'} a_{lm} e^{- (\tau -\tau_1)   H} a_{\ell_1 m_1}^\dagger \ket{Q} Y_{l m} (\n) Y_{l' m'}^* (\n) \\
        + \bra{Q} a_{\ell_2 m_2} e^{-(\tau_2-\tau)   H} [a_{lm} , a^\dagger_{l' m'} ] e^{- (\tau -\tau_1)   H} a_{\ell_1 m_1}^\dagger \ket{Q} Y_{l m} (\n) Y_{l' m'}^* (\n) \bigg]   \\
        - \frac{ (D-3) }{ R^2 (D-1)^2 } \sum_{ \substack{ m, l \\ m', l'}} \frac{ 1 }{ \sqrt{ \omega_l \omega_{l'} } } \bigg[ \bra{Q} a_{\ell_2 m_2} e^{-(\tau_2-\tau)   H} {a^\dagger_{lm}} {a_{l'm'}} e^{- (\tau -\tau_1)   H} a_{\ell_1 m_1}^\dagger \ket{Q} \del_i Y_{l m}^* (\n) \del_i Y_{l' m'} (\n) \\
        + \bra{Q} a_{\ell_2 m_2} e^{-(\tau_2-\tau)   H} a^\dagger_{l'm'} {a_{lm}} e^{- (\tau -\tau_1)   H} a_{\ell_1 m_1}^\dagger \ket{Q} \del_i Y_{l m} (\n) \del_i Y_{l' m'}^* (\n) \\
        + \bra{Q} a_{\ell_2 m_2} e^{-(\tau_2-\tau)   H} [{a_{lm}} , a^\dagger_{l' m'}] e^{- (\tau -\tau_1)   H} a_{\ell_1 m_1}^\dagger \ket{Q} \del_i Y_{l m} (\n) \del_i Y_{l' m'}^* (\n) \bigg] \bigg)  \\
        = \frac{ \Omega_D }{ 2 R^{D} \Omega_D } \Anew[\Delta_Q + R \omega_{ \ell_2}] \bigg( \sum_{m, l} R \omega_l Y_{l m} (\n) Y_{l m}^* (\n) \, \delta_{ \ell_2 \ell_1} \delta_{m_2 m_1} \\
        + (D-1) R \sqrt{ \omega_{ \ell_1} \omega_{ \ell_2} } \, \frac{ Y_{\ell_2 m_2}^* (\n) Y_{\ell_1 m_1} (\n) }{ e^{- ( \tau -\tau_1) ( \omega_{ \ell_1} - \omega_{ \ell_2} ) } } - \frac{ 1 }{\sqrt{ \omega_{ \ell_1} \omega_{ \ell_2} } } \frac{ (D-3) }{R (D-1) } \frac{ \del_i Y_{\ell_2 m_2}^* (\n) \del_i Y_{\ell_1 m_1} (\n) }{ e^{- ( \tau -\tau_1) ( \omega_{ \ell_1} - \omega_{ \ell_2} ) } } \bigg) .
    \end{multline}
In the above computation we have used that
\begin{equation}
        \sum_{m,l} \frac{\del_i  Y_{l m} (\n) \del_i Y_{l m}^* (\n) }{ R \omega_l} = \sum_{m,l} \frac{ Y_{l m} (\n) \, \big( - \Delta \big) Y_{l m}^* (\n) }{ R \omega_l} + \sum_{m,l}\frac{ \del_i \Big( Y_{l m} (\n) \del_i Y_{l m}^* (\n) \Big) }{ R \omega_l} ,  
\end{equation}
where $\sum_{m} \del_i \Big( Y_{l m} (\n) \del_i Y_{l m}^* (\n)\Big) = 0$ .\\
The Laplacian acting on the hyperspherical harmonics is easily evaluated using \eqref{eq:DispersionRelation}:
\begin{equation}
    - \Delta Y_{l m}^* (\n) = l(l+D-2) Y_{l m}^* (\n) = R^2 (D-1) \omega_l^2  Y_{l m}^* (\n) . 
\end{equation}
We are left with computing a single term, 
\begin{multline}
        \frac{\Delta_0^2 }{ R^{2D} \Omega_D^2} \frac{D^2}{\mu^2} \bra{Q} a_{\ell_2 m_2} \dot\pi (\tau, \n) \dot\pi (\tau', \n') a_{\ell_1 m_1}^\dagger \ket{Q} = - \frac{\Delta_0 \Omega_D D }{ 2 R^{2D} \Omega_D^2} \sum_{ \substack{m',l' \\ m,l}} R \sqrt{ \omega_{l'} \omega_{l} } \bra{Q} a_{\ell_2 m_2} e^{- (\tau_2 -\tau)   H} \dots  \\
        \dots\Big( {a^\dagger_{lm}} Y_{l m}^* (\n) - {a_{lm}} Y_{l m} (\n) \Big) e^{- (\tau-\tau')   H} \Big( {a^\dagger_{l'm'}} Y_{l' m'}^* (\n') - {a_{l'm'}} Y_{l' m'} (\n') \Big) e^{- (\tau' -\tau_1)   H}  a_{\ell_1 m_1}^\dagger \ket{Q} \\
        = - \frac{\Delta_0 \Omega_D D }{ 2 R^{2D} \Omega_D^2} \sum_{\substack{m',l' \\ m,l}} R \sqrt{ \omega_{l'} \omega_{l} } \bigg( Y_{l' m'} (x') Y_{l m}^* (\n) \bra{Q} a_{\ell_2 m_2} e^{- (\tau_2 -\tau)   H} {a^\dagger_{lm}}  e^{- (\tau -\tau')   H} {a_{l'm'}}  e^{- (\tau' -\tau_1)   H} \dots \\
        \dots a_{\ell_1 m_1}^\dagger \ket{Q} + Y_{l m} (\n) Y_{l' m'}^* (\n') \bra{Q} a_{\ell_2 m_2} e^{- (\tau_2 -\tau)   H} a_{lm}  e^{- (\tau -\tau')   H} a_{l'm'}^\dagger  e^{- (\tau' -\tau_1)   H}  a_{\ell_1 m_1}^\dagger \ket{Q} \bigg)\\
        = - \frac{\Delta_0 \Omega_D D \, \Anew }{ 2 R^{2D} \Omega_D^2 e^{ \omega_{\ell_2} (\tau_2-\tau_1)} } \sum_{ \substack{m',l'\\ m,l}} R \sqrt{ \omega_{l'} \omega_{l} } \bigg( \frac{ Y_{l'm'} (\n') Y_{l m}^* (\n) }{ e^{ - (\tau -\tau_1) \omega_{l} + (\tau' -\tau_1) \omega_{l'} } }  \delta_{\ell_2 l} \delta_{\ell_1 l'}  \delta_{m_2 m} \delta_{m_1 m'} \\
        + Y_{l m} (\n) Y_{l' m'}^* (\n') e^{- (\tau -\tau') \omega_{l} } \Big[ e^{- (\tau' -\tau_1 )( \omega_{l} -\omega_{l'} ) } \delta_{\ell_1 l} \delta_{\ell_2 l'}  \delta_{m_1 m} \delta_{m_2 m'} + \delta_{l' l} \delta_{m' m} \delta_{ \ell_2 \ell_1} \delta_{m_2 m_1} \Big] \bigg) \\
        = - \frac{\Delta_0 \Omega_D D }{ R^{2D} \Omega_D^2 } \Anew[\Delta_Q + R \omega_{ \ell_2}] \bigg( \delta_{\ell_2 \ell_1} \delta_{m_2 m_1} \sum_{m,l} R \omega_{l} Y_{l m} (\n) Y_{l m}^* (\n') e^{- (\tau -\tau') \omega_{l} } \\
        + \frac{ R \sqrt{ \omega_{ \ell_1} \omega_{ \ell_2} } }{ e^{ (\tau -\tau_1) ( \omega_{ \ell_1} - \omega_{ \ell_2} ) } } \Big[ Y_{\ell_2 m_2 }^* (\n) Y_{\ell_1 m_1} (\n')  e^{ (\tau -\tau') \omega_{ \ell_1} } + Y_{\ell_2 m_2}^* (\n') Y_{\ell_1 m_1} (\n) e^{- (\tau -\tau') \omega_{ \ell_2} } \Big] \bigg) .
\end{multline}
After appropriately evaluating the sums appearing in these expressions the total correlator becomes 
\begin{multline}
        \braket*{\Vpp* T_{\tau\tau} (\tau, \n) T_{\tau\tau} (\tau', \n')\Vpp } = \frac{ \Delta_0 }{ R S_D} \Anew[ \Delta_Q + R \omega_{ \ell_2}] \left[ \frac{\Delta_0}{ R^D \Omega_D} + \frac{ 2\Delta_1 }{ R^{2D} \Omega_D^2 } \right] \delta_{m_1 m_2} \delta_{\ell_1 \ell_2}  \\
        + \frac{\Delta_0 \Omega_D D }{ 2 R^{2D} \Omega_D^2 } \Anew[ \Delta_Q + R \omega_{ \ell_2}] \bigg[ \delta_{\ell_2 \ell_1} \delta_{m_2 m_1} \braket*{\n'| \Laplacian{} e^{-(\tau - \tau') \Laplacian{}}| \n} \\
        + \frac{ R \sqrt{ \omega_{ \ell_1} \omega_{ \ell_2} } }{ e^{ (\tau -\tau_1) ( \omega_{ \ell_1} -\omega_{ \ell_2} ) } } \Big(  Y_{ \ell_2 m_2}^* (\n) Y_{\ell_1 m_1} (\n')  e^{ (\tau -\tau') \omega_{ \ell_1} } + Y_{\ell_2 m_2}^* (\n') Y_{\ell_1 m_1} (\n) e^{- (\tau -\tau') \omega_{ \ell_2} } \Big) \bigg] \\
        + \frac{ \Delta_0 \Omega_D }{2 R^{2D} \Omega_D^2 } \frac{ \Anew[ \Delta_Q + R \omega_{ \ell_2}] }{ R \sqrt{ \omega_{ \ell_1} \omega_{ \ell_2}} } \bigg[ R^2 \omega_{ \ell_1} \omega_{ \ell_2} \, (D-1)  \frac{ Y_{\ell_2 m_2}^* (\n) Y_{\ell_1 m_1} (\n) }{ e^{ (\tau -\tau_1) ( \omega_{ \ell_1} - \omega_{ \ell_2} ) } } - \frac{ (D-3) }{(D-1)}  \frac{ \del_i Y_{\ell_2 m_2}^* (\n) \del_i Y_{\ell_1 m_1} (\n) }{ e^{ (\tau - \tau_1) (\omega_1 - \omega_2) }  }  \bigg] \\
        + \frac{\Delta_0 \Omega_D }{ 2 R^{2D} \Omega_D^2 } \frac{ \Anew[ \Delta_Q + R \omega_{ \ell_2}] }{ R\sqrt{ \omega_{ \ell_1} \omega_{ \ell_2} } } \bigg[ R^2 \omega_{ \ell_1} \omega_{ \ell_2} \, (D-1) \frac{ Y_{\ell_2 m_2}^* (\n') Y_{\ell_1 m_1} (\n') }{ e^{ ( \tau' -\tau_1) ( \omega_{ \ell_1} - \omega_{ \ell_2})}} - \frac{ (D-3) }{ (D-1)} \frac{ \del_i Y_{ \ell_2 m_2}^* (\n') \del_i Y_{\ell_1 m_1} (\n') }{ e^{ (\tau' - \tau_1) (\omega_1 - \omega_2) }  } \bigg] .
\end{multline}

\setstretch{1}

\printbibliography{}
\end{document}